\newcount\mgnf\newcount\tipi\newcount\tipoformule\newcount\greco

\tipi=2          
\tipoformule=0   

\global\newcount\numsec\global\newcount\numfor
\global\newcount\numapp\global\newcount\numcap
\global\newcount\numfig\global\newcount\numpag
\global\newcount\numnf

\def\SIA #1,#2,#3 {\senondefinito{#1#2}%
\expandafter\xdef\csname #1#2\endcsname{#3}\else
\write16{???? ma #1,#2 e' gia' stato definito !!!!} \fi}

\def \FU(#1)#2{\SIA fu,#1,#2 }

\def\etichetta(#1){(\veroparagrafo.\veraformula)%
\SIA e,#1,(\veroparagrafo.\veraformula) %
\global\advance\numfor by 1%
\write15{\string\FU (#1){\equ(#1)}}%
\write16{ EQ #1 ==> \equ(#1)  }}
\def\etichettaa(#1){(A\veraappendice.\veraformula)
 \SIA e,#1,(A\veraappendice.\veraformula)
 \global\advance\numfor by 1
 \write15{\string\FU (#1){\equ(#1)}}
 \write16{ EQ #1 ==> \equ(#1) }}
\def\getichetta(#1){Fig. \verafigura
 \SIA g,#1,{\verafigura}
 \global\advance\numfig by 1
 \write15{\string\FU (#1){\graf(#1)}}
 \write16{ Fig. #1 ==> \graf(#1) }}
\def\retichetta(#1){\numpag=\pgn\SIA r,#1,{\verapagina}
 \write15{\string\FU (#1){\rif(#1)}}
 \write16{\rif(#1) ha simbolo  #1  }}
\def\etichettan(#1){(n\verocapitolo.\veranformula)
 \SIA e,#1,(n\verocapitolo.\veranformula)
 \global\advance\numnf by 1
\write16{\equ(#1) <= #1  }}

\newdimen\gwidth
\gdef\profonditastruttura{\dp\strutbox}
\def\senondefinito#1{\expandafter\ifx\csname#1\endcsname\relax}
\def\BOZZA{
\def\alato(##1){
 {\vtop to \profonditastruttura{\baselineskip
 \profonditastruttura\vss
 \rlap{\kern-\hsize\kern-1.2truecm{$\scriptstyle##1$}}}}}
\def\galato(##1){ \gwidth=\hsize \divide\gwidth by 2
 {\vtop to \profonditastruttura{\baselineskip
 \profonditastruttura\vss
 \rlap{\kern-\gwidth\kern-1.2truecm{$\scriptstyle##1$}}}}}
\def\verapagina{
{\romannumeral\number\numcap}.\number\numsec.\number\numpag}}

\def\alato(#1){}
\def\galato(#1){}
\def\veroparagrafo{\number\numsec}\def\veraformula{\number\numfor}
\def\veraappendice{\number\numapp}
\def\verapagina{\number\pageno}\def\veranformula{\number\numnf}
\def\verafigura{{\romannumeral\number\numcap}.\number\numfig}
\def\verocapitolo{\number\numcap}\def\veranformula{\number\numnf}
\def\Eqn(#1){\eqno{\etichettan(#1)\alato(#1)}}
\def\eqn(#1){\etichettan(#1)\alato(#1)}
\def\ver{\veroparagrafo}
\def\Eq(#1){\eqno{\etichetta(#1)\alato(#1)}}
\def\eq(#1){\etichetta(#1)\alato(#1)}
\def\Eqa(#1){\eqno{\etichettaa(#1)\alato(#1)}}
\def\eqa(#1){\etichettaa(#1)\alato(#1)}
\def\dgraf(#1){\getichetta(#1)\galato(#1)}
\def\drif(#1){\retichetta(#1)}

\def\eqv(#1){\senondefinito{fu#1}$\clubsuit$#1\else\csname fu#1\endcsname\fi}
\def\equ(#1){\senondefinito{e#1}\eqv(#1)\else\csname e#1\endcsname\fi}
\def\graf(#1){\senondefinito{g#1}\eqv(#1)\else\csname g#1\endcsname\fi}
\def\rif(#1){\senondefinito{r#1}\eqv(#1)\else\csname r#1\endcsname\fi}
\def\bib[#1]{[#1]\numpag=\pgn
\write13{\string[#1],\verapagina}}

\def\include#1{
\openin13=#1.aux \ifeof13 \relax \else
\input #1.aux \closein13 \fi}

\openin14=\jobname.aux \ifeof14 \relax \else
\input \jobname.aux \closein14 \fi
\openout15=\jobname.aux
\openout13=\jobname.bib


\ifnum\tipoformule=1\let\Eq=\eqno\def\eq{}\let\Eqa=\eqno\def\eqa{}
\def\equ{}\fi


{\count255=\time\divide\count255 by 60 \xdef\hourmin{\number\count255}
        \multiply\count255 by-60\advance\count255 by\time
   \xdef\hourmin{\hourmin:\ifnum\count255<10 0\fi\the\count255}}

\def\oramin{\hourmin }

\def\data{\number\day/\ifcase\month\or january \or february \or march \or
april \or may \or june \or july \or august \or september
\or october \or november \or december \fi/\number\year;\ \oramin}

\setbox200\hbox{$\scriptscriptstyle \data $}

\newcount\pgn \pgn=1
\def\foglio{\number\numsec:\number\pgn
\global\advance\pgn by 1}
\def\foglioa{A\number\numsec:\number\pgn
\global\advance\pgn by 1}

\footline={\rlap{\hbox{\copy200}}\hss\tenrm\folio\hss}

\def\TIPIO{
\font\setterm=amr7 
\def \settepunti{\def\rm{\fam0\setterm}
\textfont0=\setterm   
\normalbaselineskip=9pt\normalbaselines\rm
}\let\nota=\settepunti}

\def\TIPITOT{
\font\twelverm=cmr12
\font\twelvei=cmmi12
\font\twelvesy=cmsy10 scaled\magstep1
\font\twelveex=cmex10 scaled\magstep1
\font\twelveit=cmti12
\font\twelvett=cmtt12
\font\twelvebf=cmbx12
\font\twelvesl=cmsl12
\font\ninerm=cmr9
\font\ninesy=cmsy9
\font\eightrm=cmr8
\font\eighti=cmmi8
\font\eightsy=cmsy8
\font\eightbf=cmbx8
\font\eighttt=cmtt8
\font\eightsl=cmsl8
\font\eightit=cmti8
\font\sixrm=cmr6
\font\sixbf=cmbx6
\font\sixi=cmmi6
\font\sixsy=cmsy6
\font\twelvetruecmr=cmr10 scaled\magstep1
\font\twelvetruecmsy=cmsy10 scaled\magstep1
\font\tentruecmr=cmr10
\font\tentruecmsy=cmsy10
\font\eighttruecmr=cmr8
\font\eighttruecmsy=cmsy8
\font\seventruecmr=cmr7
\font\seventruecmsy=cmsy7
\font\sixtruecmr=cmr6
\font\sixtruecmsy=cmsy6
\font\fivetruecmr=cmr5
\font\fivetruecmsy=cmsy5
\textfont\truecmr=\tentruecmr
\scriptfont\truecmr=\seventruecmr
\scriptscriptfont\truecmr=\fivetruecmr
\textfont\truecmsy=\tentruecmsy
\scriptfont\truecmsy=\seventruecmsy
\scriptscriptfont\truecmr=\fivetruecmr
\scriptscriptfont\truecmsy=\fivetruecmsy
\def \eightpoint{\def\rm{\fam0\eightrm}
\textfont0=\eightrm \scriptfont0=\sixrm \scriptscriptfont0=\fiverm
\textfont1=\eighti \scriptfont1=\sixi   \scriptscriptfont1=\fivei
\textfont2=\eightsy \scriptfont2=\sixsy   \scriptscriptfont2=\fivesy
\textfont3=\tenex \scriptfont3=\tenex   \scriptscriptfont3=\tenex
\textfont\itfam=\eightit  \def\it{\fam\itfam\eightit}%
\textfont\slfam=\eightsl  \def\sl{\fam\slfam\eightsl}%
\textfont\ttfam=\eighttt  \def\tt{\fam\ttfam\eighttt}%
\textfont\bffam=\eightbf  \scriptfont\bffam=\sixbf
\scriptscriptfont\bffam=\fivebf  \def\bf{\fam\bffam\eightbf}%
\tt \ttglue=.5em plus.25em minus.15em
\setbox\strutbox=\hbox{\vrule height7pt depth2pt width0pt}%
\normalbaselineskip=9pt
\let\sc=\sixrm  \let\big=\eightbig  \normalbaselines\rm
\textfont\truecmr=\eighttruecmr
\scriptfont\truecmr=\sixtruecmr
\scriptscriptfont\truecmr=\fivetruecmr
\textfont\truecmsy=\eighttruecmsy
\scriptfont\truecmsy=\sixtruecmsy
}\let\nota=\eightpoint}

\newfam\msbfam   
\newfam\truecmr  
\newfam\truecmsy 
\newskip\ttglue
\ifnum\tipi=0\TIPIO \else\ifnum\tipi=1 \TIPI\else \TIPITOT\fi\fi

\global\newcount\numpunt


\def\a{\alpha}
\def\b{\beta}
\def\d{\delta}
\def\e{\epsilon}

\def\g{\gamma}

\def\l{\lambda}

\def\s{\sigma}
\def\t{\tau}
\def\th{\theta}

\def\o{\omega}
\def\D{\Delta}
\def\L{\Lambda}
\def\G{\Gamma}
\def\O{\Omega}

\def\del #1{\frac{\partial^{#1}}{\partial\l^{#1}}}

\def\1{{1\kern-.25em\hbox{\rm I}}}
\def\eu{{1\kern-.25em\hbox{\sm I}}}

\def\R{{\Bbb R}}  
\def\N{{\Bbb N}}  
\def\P{{\Bbb P}}  
\def\Z{{\Bbb Z}}  
\def\E{{\Bbb E}}  
\def\M{{\Bbb M}}  

\def\del{\partial}


\def\AA{{\cal A}}
\def\BB{{\cal B}}
\def\CC{{\cal C}}
\def\DD{{\cal D}}

\def\FF{{\cal F}}

\def\SS{{\cal S}}

\def\NN{{\cal N}}

\def\WW{{\cal W}}

\def\LL{{\cal L}}

\def\QQ{{\cal Q}}
\def\A{{\cal A}}

\def\sign{\,\hbox{sign}\,}

\def\wt{\widetilde}


\newcount\foot
\foot=1
\def\note#1{\footnote{${}^{\number\foot}$}{\ftn  #1}\advance\foot by 1}

\def\frac#1#2{{#1\over #2}}
\def\sfrac#1#2{{\textstyle{#1\over #2}}}
\def\text#1{\quad{\hbox{#1}}\quad}
\def\newpage{\vfill\eject}
\font\pr=cmbxsl10


\font\ftn=cmr8
\font\smit=cmti8

\font\it=cmti10
\font\bf=cmbx10
\font\sm=cmr7


\magnification=1200  
\input amssym.def  

\catcode`\@=12

\voffset=2.5truepc
\hoffset=2.5truepc
\hsize=5.1truein
\vsize=8.4truein 

\font\ninerm=cmr9 at 9pt

\font\twelveit=cmti12  


\tolerance=6000
\parskip=0pt
\parindent=18pt

 \abovedisplayskip=12pt plus3pt minus2pt
 \belowdisplayskip=12pt plus3pt minus2pt 
 \abovedisplayshortskip=12pt plus3pt minus2pt
 \belowdisplayshortskip=12pt plus3pt minus2pt

 \def\lskipamount{12pt}
 \def\lskip{\vskip\lskipamount plus3pt minus2pt}
 \def\lbreak{\par \ifdim\lastskip<\lskipamount
  \removelastskip \penalty-200 \lskip \fi}

 \def\lnobreak{\par \ifdim\lastskip<\lskipamount
  \removelastskip \penalty200 \lskip \fi}

\font\sectionfont=cmbx10 at 10pt


    \def\tit#1{{\titrm\centerline{#1}}\vskip 2truepc}


\def\sec#1{\vskip 1.5truepc\centerline{\hbox {{\sectionfont #1}}}
\vskip1truepc\noindent}
\def\chap #1#2{\sec {#1}\numsec=#2\numfor=1} 

\def\newsubsec#1{\vskip 1truepc\line{\hbox{{\sectionfont #1}}\hfill}\vskip 3pt}
\def\theo#1#2{\medskip
  {\noindent\bf   Theorem #1.} {\it #2}\medskip}
\def\lem#1#2{\medskip{\noindent \bf  Lemma #1.}
  {\it #2}\medskip}
\def\cor#1#2{\medskip{\noindent \bf  Corollary #1.}
  {\it #2}\medskip}
\def\prop#1#2{\medskip{\noindent \bf  Proposition #1.} {\it #2}\medskip}
\def\defi#1#2{\medskip {\noindent \bf Definition #1}. {#2}\medskip }

\def\proof#1{\medskip {\noindent \bf  Proof.} {\rm #1} \medskip}
\def\proofof#1#2{\medskip{ \noindent \bf Proof of #1.}{\rm #2}\medskip}
\def\blacksquare{{ \vrule height7pt width7pt depth0pt}}
\def\qed{\line{\hfil$\blacksquare$}\vskip 1.5truepc}

{\headline={\ifodd\pageno\rightheadline \else \leftheadline \fi}}
\def\rightheadline{\it  {}\hfil\tenrm\folio}
\def\leftheadline{\tenrm \folio \hfil\it  {}}

\def\proposition{\prop}

\def\lemma{\lem} 

\def\definition{\defi}
\def\corollary{\cor}
\def\proofof #1{{\noindent\pr Proof of #1. }}
\def\endproof{\par\qed}
\def\remark{\noindent{\bf Remark. }}
\def\thanks{\noindent{\bf Acknowledgements. }}

\def\wt{\tilde}

\font\tit=cmbx12
\font\aut=cmbx12
\font\aff=cmsl12
\def\s{\char'31}
\nopagenumbers
{$  $}
\vskip1.5truecm
\centerline{\tit HOPFIELD MODELS AS  GENERALIZED RANDOM MEAN FIELD   }
\vskip.2truecm
\centerline{\tit MODELS
\footnote{${}^\#$}{\ftn Work
partially supported by the Commission of the European Union
under contract  CHRX-CT93-0411}\footnote{${}^\dagger$}{\ftn 
To appear in ``Mathematics of spin glasses and neural networks'', 
A. Bovier and P. Picco, eds, 
``Progress in Probability'', Birkh\"auser, 1997}}
\vskip1.5truecm
\centerline{\aut Anton Bovier
\footnote{${}^1$}{\ftn e-mail:
bovier@wias-berlin.de}}
\vskip.1truecm
\centerline{\aff Weierstra\s {}--Institut}
\centerline{\aff f\"ur Angewandte Analysis und Stochastik}
\centerline{\aff Mohrenstra\s e 39, D-10117 Berlin, Germany}
\vskip.5truecm
\centerline{\aut  V\'eronique Gayrard\footnote{${}^2$}{\ftn
e-mail: gayrard@cpt.univ-mrs.fr}}
\vskip.1truecm
\centerline{\aff Centre de Physique Th\'eorique - CNRS}
\centerline{\aff Luminy, Case 907}
\centerline{\aff F-13288 Marseille Cedex 9, France}
\vskip1truecm\rm
\def\s{\sigma}
\noindent {\bf Abstract:} We give a comprehensive self-contained review
on the rigorous analysis of the thermodynamics of a class of random 
spin systems of mean field type whose most prominent example is the 
Hopfield model. We focus on the low temperature phase and the analysis of the 
Gibbs measures with large deviation techniques.  There is a very detailed and 
complete picture in  the regime of ``small $\a$''; a particularly satisfactory
result concerns a non-trivial regime of parameters in which we prove 
1) the convergence of the local ``mean fields'' to gaussian random variables 
with constant variance and random mean; the random means are from site to site
independent gaussians themselves; 2) ``propagation of chaos'', i.e. 
factorization of the extremal infinite volume Gibbs measures, and 3)
the correctness of the ``replica symmetric solution'' of Amit, Gutfreund 
and Sompolinsky [AGS]. This last result was first proven by
M. Talagrand [T4], using different techniques.


 \vfill
\eject

\count0=1
\footline={\hss\tenrm\folio\hss}
\overfullrule=0pt

\chap{Table of Contents}0

\vskip2.0truecm

\line{1. \hskip0.5em Introduction \dotfill     2}
\line{2. \hskip0.5em Generalized random  mean field models \dotfill 5}
\line{\hskip.4cm 2.1\hskip0.5em Derivation of the Hopfield model as a
mean field spin glass \dotfill     5}
\line{\hskip.4cm 2.2\hskip0.5em The Hopfield model as an autoassociative
memory \dotfill     7}
\line{\hskip.4cm 2.3\hskip0.5em Definition of generalized  random mean
field models \dotfill 8}
\line{\hskip.4cm 2.4\hskip0.5em Thermodynamic limits for 
generalized random mean field models \dotfill     10}
\line{\hskip.4cm 2.5. Convergence, laws of large numbers and propagation
of chaos \dotfill     12}
\line{\hskip.4cm 2.6\hskip0.5em Examples  \dotfill     13}
\line{3.\hskip0.5em  Large deviation estimates and 
Transfer principle
\dotfill     16}
\line{\hskip.4cm 3.1\hskip0.5em Large deviation estimates \dotfill     16}
\line{\hskip.4cm 3.2\hskip0.5em Transfer principle \dotfill   20}
\line{4.\hskip0.5em Bounds on the norm of random matrices
\dotfill     24}
\line{5.\hskip0.5em Properties of the induced measures \dotfill     27}
\line{\hskip.4cm 5.1\hskip0.5em  Localization of the induced measures 
\dotfill 28}
\line{\hskip.4cm 5.2\hskip0.5em Symmetry and concentration of measure \dotfill
27}
\line{6.\hskip0.5em Global estimates on the free energy function \dotfill 33}
\line{7.\hskip0.5em Local analysis of $\Phi$ \dotfill   47}
\line{\hskip.4cm 7.1 The case $p\geq 3$
\dotfill     46}
\line{\hskip.4cm7.2 The case $p=2$ \dotfill  48}
\line{8.\hskip0.5em Convexity, the replica symmetric solution, and limiting Gibs measures \dotfill   54}
\line{References \dotfill     70}

\newpage

\line{\hfill\sl L'intuition ne peut nous  donner  la rigeur,}
\vskip-2pt
\line{\hfill \sl ni m\^eme la certitude, on s'en est aper\c{c}u de 
plus en plus.}
\line{\hfill \ftn Henri Poincar\'e,}
\vskip-6pt
\line{\hfill \smit ``La Valeur de La Science''}

\chap{I. Introduction}1

Twenty years ago, Pastur and Figotin [FP1,FP2] first introduced and studied
what has become known to be the Hopfield model and which turned out,
over the years, as one of the more  successful and important models of a 
disordered system. This  is also reflected in the fact that several 
contributions in this book are devoted to it. The Hopfield model is quite 
versatile and models various situations: Pastur and Figotin introduced it 
as a simple model for a spin glass, while Hopfield, in 1982, independently
considered it as a model for associative memory. The first viewpoint naturally
 put it in the context of equilibrium statistical mechanics, while Hopfield's
main interest was its dynamics. But the great success of what became known
as the Hopfield model came from the realization, mainly in the work 
of Amit, Gutfreund, and Sompolinsky [AGS] that a more complicated 
version of this model is reminiscent 
to a spin glass, and that the (then) recently developed methods of 
spin-glass theory, in particular the replica trick and Parisi's 
replica symmetry breaking scheme could be adapted to this model and allowed a 
``complete'' analysis of the  equilibrium statistical mechanics of the 
model and  to recover some of the most prominent 
``experimentally'' observed features of the model like the ``storage 
capacity'', and ``loss of memory'' in a precise analytical way. 
This observation sparked a surge of interest by theoretical physicists
into neural network theory in general  that  has led to considerable 
progress in the field (the literature on the subject is extremely rich, and 
there are a great number of good review papers. See for example 
[A,HKP,GM,MR,DHS]). We will not review this development here. 
In spite of their success, the methods used in the analysis by theoretical 
physicist were of heuristic nature and involved mathematically 
unjustified procedures and it may not be too unfair to say that they do not
really provide a deeper understanding for what is really going on in these 
systems. Mathematicians and mathematical physicists were only late entering
this field; as a matter of fact, spin glass theory was (and is) considered 
a field difficult, if not impossible, to access by rigorous mathematical 
techniques. 

As is demonstrated in this book, in the course of the last decade the 
attitude of at least some mathematicians and mathematical physicists 
towards this field has changed, and some now consider it as a major challenge
to be faced rather than a nuisance to be avoided. And already, 
substantial progress in  a rigorous 
mathematical sense has begun to be made. The Hopfield model has been
for us the focal point of attention in this respect over the last five years
and in this article we will review the results obtained by us in this spirit.
Our approach to the model may be called ``generalized random mean field
models'', and is in spirit close to  large deviation theory. We will 
give a precise outlay of this general setting in the next section. 
Historically, our basic approach can be traced back even to the original papers
by Pastur and Figotin. In this setting, the ``number of patterns'', $M$, or
rather its relation to the system size $N$, is a crucial parameter 
and the larger it is, the more difficult  things are getting. The case where
$M$ is is strictly bounded could be termed ``standard disordered mean field'', 
and it is this type of models that were studied by Pastur and Figotin in 1977,
the case of two patterns having been introduced by Luttinger [Lut] 
shortly before that. Such ``site-disorder'' models were studied again 
intensely some years later by a number of people, emphasizing applications of 
large deviation methods [vHvEC,vH1,GK,vHGHK,vH2,AGS2,JK,vEvHP]. 
A general large deviation theory for such systems was 
obtained by Comets [Co] somewhat later.
This was far from the ``physically'' interesting case where the 
ratio between $M$ and $N$, traditionally called $\a$, is a finite positive 
number [Ho, AGS]. 
 The approach  of Grensing and K\"uhn [GK],
that could be described as the most straightforward generalization of 
the large deviation analysis of the Curie-Weiss model by combinatorial 
computation of the entropy (see Ellis' book [El] for a detailed exposition),
was the first to be generalized to unbounded $M$ by Koch and Piasko [KP]
(but see also [vHvE]).
Although their condition on $M$, namely $M< \frac{\ln N}{\ln 2}$, 
was quite strong, 
 until 1992 this remained the only rigorous result on 
the thermodynamics of the model with an unbounded number of patterns 
and their analysis involved for the first time a non-trivial 
control on fluctuations of a free energy functional. Within their framework,  
however, the barrier $\ln N$ appeared unsurmountable, and some crucial 
new ideas were needed. They came in two almost simultaneous papers by 
Shcherbina and Tirozzi [ST] and Koch [K]. They proved 
that the free energy of the Hopfield model in the thermodynamic limit
is equal to that of the Curie-Weiss model, provided only 
that $\lim_{N\uparrow\infty} \frac MN=0$, without condition on the speed
of convergence. In their proof this fact was linked to the convergence in norm
of a certain random matrix constructed from the patterns to the identity 
matrix. Control on this matrix proved one key element in further progress. 
Building on this observation, in  a  paper with Picco [BGP1] we were able to 
give a construction of the extremal Gibbs states under the same hypothesis,
and even get first results on the Gibbs states in the case $\frac MN=\a\ll 1$.
Further progress in this latter case, however, required  yet another key
idea: the use of exponential concentration of measure estimates. Variance 
estimates based on the Yurinskii martingale construction had already 
appeared in [ST] where they were used to prove self-averaging of the 
free energy.  With Picco [BGP3] we proved exponential estimates on 
``local'' free
energies and  used this to show that 
disjoint Gibbs states corresponding to all patterns
 can be constructed for small enough $\a$. 
A considerable refinement of this analysis that included a 
detailed analysis of the local minima near the Mattis states [Ma] was given 
in a later paper by the present authors [BG5]. 
The  result is a fairly complete and rigorous picture 
of the Gibbs states and even metastable states in the small $\a$ regime,
which is in good agreement with the heuristic results of [AGS]. 
During the preparation of this manuscript, a
remarkable breakthrough was obtained by Michel Talagrand [T4]. He succeeded
in proving that in a certain (nontrivial) range of the 
parameters $\b$ and $\a$,
the validity of the ``replica symmetric solution'' of [AGS] can be rigorously 
justified. It turns out that a result obtained in [BG5] can be used to 
give an alternative proof of that also yields some 
complementary information and in particular allows to analyse the 
convergence properties of the Gibbs measures in that regime. 
We find it particularly pleasant that, 10 years after the paper by Amit et al.,
we can present this development in this review.

In the 
present paper we will give a fairly complete and streamlined version of our
approach, emphasizing generalizations beyond the standard Hopfield model,
even though we will not work out all the details at every point.
We have tried to give proofs that are either simpler or more systematic
than the original ones and believe to have succeeded to some extent. At some 
places technical proofs that we were not able to improve 
substantially are omitted and reference is made to the original papers.
 In Section 
2 we present a derivation of the Hopfield model as a mean field spin glass,
 introduce the concept of {\it generalized random  mean field models} and
discuss
 the thermodynamic formalism for such systems. We point out some popular
variants of the Hopfield model and place them in this general framework.
Section 3 discusses some necessary background on large deviations, 
emphasizing calculational aspects. This section is quite general and 
can be regarded as completely independent from particular models. Section 
4 brings the last proof on exponential estimates on maximal and minimal 
eigenvalues of some matrices that are used throughout in the sequel. 
In Section 5 we show how large deviation estimates lead to estimates on Gibbs 
measures. Here the theme of {\it concentration of measure} appears 
in a crucial way. Section 6 as well as Section 7 are devoted to 
the study of the function $\Phi$ that emerged from Section 3 as a crucial 
instrument to control large deviations. Section 8, finally gives a 
rigorous derivation of the replica symmetric solution of [AGS] 
in an appropriate range of parameters, and the comstruction of the 
limiting distribution of the Gibbs measures (the ``metastate'' in the 
language of [NS]).

 There are a number of other results on the Hopfield model that we
do not discuss. We never talk here about the 
high temperature phase, and we also exclude the study of the zero temperature 
case. Also  we do not speak about the case $\a=0$ but will always assume 
$\a>0$. However, all  proofs work also when $\frac MN\downarrow 0$, 
with some trivial modifications necessary when $M(N)$ remains bounded or 
grows slowly. In this situation some more refined results, like large 
deviation principles [BG4] and central limit theorems [G1] can be obtained.
Such results will be covered in other contributions to this volume.

\thanks We are grateful to Michel Talagrand for sending us copies of his 
work, in particular [T4] prior to publication. This inspired most of Section 8.
We also are indebted to Dima Ioffe for suggesting at the right moment that the 
inequalities in [BL] could be the right tool to make use of Theorem 8.1.
This proved a key idea. We thank Aernout van Enter for a careful reading of the
manuscript and numerous helpful comments.

\newpage


\chap{2. Generalized random mean field models}2

This section introduces the general setup of our approach, including a 
definition of the concept of ``generalized random mean field model''
and the corresponding thermodynamic formalism. But before giving
formal definitions, we will show how such a class of models and the 
Hopfield model in particular arises naturally in the attempt to 
construct mean field models for spin glasses, or to 
construct models of autoassociative memory.

\newsubsec{ \ver.1. The Hopfield model as a mean 
field spin glass.}

The derivation we are going to present does not follow the historical 
development. In fact, what is generally considered ``the'' mean field 
spin glass model, the {\it Sherrington-Kirkpatrick model} [SK], is different
(although, as we will see, related) and not even, according to the definition 
we will use, a mean field model (a fact which may explain why 
it is so much harder to analyse than its inventors apparently expected, 
and which in many ways makes it much more interesting). 
What do we mean by ``mean field model''?  A spin system on a lattice is, 
roughly, given by a lattice, typically $\Z^d$, a  local spin space 
$\SS$, which could be some Polish space but which for the present we can think
of as the discrete set $\SS=\{-1,+1\}$, the {\it configuration space}
$\SS_\infty\equiv \SS^{\Z^d}$ and its finite volume subspaces $\SS_\L\equiv 
\SS^\L$ for any finite $\L\subset \Z^d$, and a Hamiltonian function $H$
that for any finite $\L$ gives the energy of a configuration $\s\in 
\SS_\infty$ in the volume $\L$, 
as $H_\L(\s)$. We will 
say that a spin system is a {\it mean field model} 
if its Hamiltonian depends on $\s$ only through 
a set of so-called {\it macroscopic functions} or {\it order parameters}.
By this we mean typically spatial averages of local functions of the 
configuration. If the mean field model is supposed to describe reasonably well
a given spin system, a set of such functions 
should be used so that their equilibrium values 
suffice to characterize completely the {\it phase diagram} of the model. For 
instance, for a ferromagnetic spin system it suffices to consider the 
{\it total magnetization} in a volume $\L$, 
$m_\L(\s)\equiv \frac 1{|\L|} \sum_{i\in \L}\s_i$ as order parameter. A mean 
field Hamiltonian for a ferromagnet is then $H_\L^{fm}(\s)=-|\L| E(m_\L(\s))$; 
the physically most natural choice $E(m)=\frac 12 m^2$  gives the 
{\it Curie-Weiss model}.
 Note that 
$$
H_\L^{fm}(\s)=-\sum_{i\in \L} \s_i\left[\frac { E(m_\L(\s))}{m_\L(\s)}\right]
\Eq(G.1)
$$
which makes manifest the idea that in this model the spins $\s_i$ at the site 
$i$ interact only with the (non-local) {\it mean-field}
$\frac { E(m_\L(\s))}{m_\L(\s)}$. In the Curie-Weiss case this mean field 
is of course the mean magnetization itself. Note that the order parameter 
$m_\L(\s)$ measures how close the spin configuration in $\L$ is to the 
ferromagnetic ground states $\s_i\equiv +1$, resp. $
\s_i\equiv -1$. If we wanted to model an antiferromagnet,
the corresponding order parameter would be the {\it staggered magnetization}
$m_\L(\s)\equiv \frac 1{|\L|}\sum_{i\in \L} (-1)^{\sum_{\g=1}^d i_\g}\s_i$.

In general, a natural choice for a set of order parameters will be given by 
the projections of the 
spin configurations to the {\it ground states}   of the system. 
By ground states we mean configurations $\s$  that for all
$\L$  minimize the Hamiltonian $H_\L$
in the sense that  $H_\L(\s)$ cannot be made smaller by 
changing $\s$ only within $\L$\note{We are somewhat too simplistic here. The 
notion of ground states should in general  not only  be applied to 
individual  configurations but rather
to measures on configuration space (mainly to avoid the problem of local 
degeneracy); however, we will ignore such 
complications here.} .
So if 
$\xi^1,  \dots, \xi^M $ 
are the ground states of our system, we should define the $M$ order 
parameters
$m^1_\L(\s)=\frac 1{|\L|}\sum_{i\in \L} \xi^1_i\s_i,\dots,
m^M_\L(\s)=\frac 1{|\L|}\sum_{i\in \L} \xi^M_i\s_i$
and take as a Hamiltonian a function 
$H^{mf}_\L(\s)=-|\L| E\left(m_\L^1(\s),\dots, m^M_\L(\s)\right)$. 
For consistency, 
one should of course choose $E$ in such a way that $\xi^1,\dots,\xi^M$ are 
ground states of the so defined  $H^{mf}_\L(\s)$. 
We see that in this spirit, the construction of a mean field model 
departs from assumptions on the ground states of  the real model. 

Next we should say what we mean by ``spin glass''. This is a more complicated 
issue. The generally accepted model for a lattice spin-glass is the 
Edwards-Anderson model [EA] in which Ising spins on a lattice $\Z^d$ interact
via nearest-neighbour couplings $J_{ij}$ that are independent random variables
with zero mean. Little is known about the low-temperature properties of this 
model on a rigorous level, and even on the heuristic level there are 
conflicting opinions, and it will be difficult to find
consensus 
within a reasonably large crowd of experts on what should be reasonable 
assumptions on the nature of ground states in a spin glass. But there will be 
some that would agree on the two following features which should hold 
in high enough dimension\note{For arguments in favour of this, see e.g. 
[BF,vE], for a different  view e.g.  [FH].}

\item{(1)} The ground states are ``disordered''.
\item{(2)} The number of ground states is infinite.

Moreover, the most ``relevant'' ground states should be stationary random 
fields, although not much more can be said a priori on their distribution.
Starting from these assumptions, we should choose some function 
$M(\L)$ that tends to infinity as $\L\uparrow\Z^d$ and $M(\L)$ random 
vectors $\xi^\mu$, defined on some probability space 
$(\O,\FF,\P)$ and taking values in 
$\SS_\infty$ and define, for all $\o\in \O$, a $M(\L)$-dimensional vector of 
order parameters with components,
$$
m^\mu_\L[\o](\s)\equiv \frac 1{|\L|}\sum_{i\in \L}\xi_i^\mu[\o]\s_i
\Eq(G.2)
$$
and finally choosing the Hamiltonian as some function of this vector. The most
natural choice in many ways is
$$
\eqalign{
H_\L[\o](\s)&=-\frac {|\L|}2 \left\|m_\L[\o](\s)\right\|_2^2\cr
             &=-\frac {|\L|}2 \sum_{\mu=1}^{M(\L)}\left[m_\L[\o](\s)\right]^2
\cr
             &=-\frac 1{2|\L|} \sum_{i,j\in \L}\sum_{\mu=1}^{M(\L)}
                 \xi_i^\mu[\o] \xi_j^\mu[\o]\s_i\s_j
}
\Eq(G.3)
$$

If we make the additional assumption that 
the random variables $\xi_i^\mu$ are independent and identically distributed
with 
$\P[\xi_i^\mu=\pm 1]=\frac 12$ we have obtained exactly the Hopfield model
[Ho] in its most standard form\note{Observe that the lattice structure of the
set $\scriptstyle\L$ plays no r\^ole anymore and we can consider it 
simply as a set of points}.
Note that at this point we can replace without any loss $\L$ by the set 
$\{1,\dots,N\}$. Note also that many of the most common variants of the 
Hopfield model are  simply obtained by a different choice of the function 
$E(m)$ or by different assumptions on the distribution of $\xi$. 

In the light of what we said before we should check whether this choice was 
consistent, 
i.e. whether the ground states of the Hamiltonian \eqv(G.3) 
are indeed the vectors $\xi^\mu$, at least with probability tending to one. 
This will depend on the behavior of the  function $M(N)$. From what is known 
today, in a strict sense this is true only if $M(N)\leq c\frac N{\ln N}$
[McE,Mar] whereas under a mild relaxation (allowing deviations that are 
invisible on the level of the macroscopic variables $m_N$), this holds 
as long as $\lim_{N\uparrow\infty} \frac {M(N)}N=0$ [BGP1]. It does not hold
for faster growing $M(N)$ [Lu].
On the contrary, one might ask whether for given $M(\L)$ consistency can be 
reached by the choice of a different distribution $\P$. This seems an
 interesting,  and to our knowledge completely uninvestigated question.

\newsubsec{\ver.2 The Hopfield model as an autoassociative memory.}

Hopfield's purpose when deriving his model was not to model spin glasses, 
but to describe the capability of a neural network to act as a memory. 
In fact, 
the type of interaction for him was more or less dictated by assumptions on 
neural functioning. Let us, however, give another, fake, 
derivation of his model.
By an {\it autoassociative memory} we will understand an algorithm that is 
capable of associating input data to a preselected set of learned {\it 
patterns}. Such an algorithm may be deterministic or stochastic. 
We will generally only be interested in {\it complex} data, i.e. a pattern
should contain a large amount of information. 
A pattern is thus naturally described as an element of a set $\SS^N$,
and a reasonable description of any possible datum $\s\in \SS^N$ 
within that set
in relation to the {\it stored} patterns $\xi^1,\dots \xi^M$ is in terms 
of its similarity to these patterns that is expressed in terms of the 
 vector of {\it overlap parameters} $m(\s)$ whose components are
$m^\mu(\s)=\frac 1N\sum_{i=1}^N\xi_i^\mu\s_i$. 
If we agree that this should be all the information we 
care about, it is natural to construct an algorithm that can be expressed in
 terms of these variables only. A most natural candidate for such an 
algorithm  is  a Glauber 
dynamics with respect to a mean field Hamiltonian like \eqv(G.3). 
Functioning of the memory is then naturally interpreted by the existence of 
equilibrium measures corresponding to the stored patterns. Here the assumptions
on the distribution of the patterns are dictated by a priori assumptions on 
the types  of patterns one wants to store, and the maximal 
$M(N)$ for which the memory ``functions'' is called {\it storage capacity}
and should be determined by the theory. In this paper we will not say much 
about this dynamical aspect, mainly because there are almost no mathematical 
results on this. It is clear from all we know about Glauber dynamics,
that a detailed knowledge of the equilibrium distribution is necessary, but 
also ``almost'' sufficient to understand the main features of the 
long time properties of the dynamics. These things are within 
reach of the present 
theory, but only first steps have been carried out (See e.g. [MS]). 

\newsubsec{ \ver.3 Definition of generalized random mean field models.}

Having seen how the Hopfield model emerges naturally in the framework of 
mean field theory, we will now introduce a rather general framework 
that allows to encompass this model as well as numerous generalizations. 
We like to call this framework {\it generalized } random mean field models 
mainly due to the fact that we allow 
an unbounded number of order parameters, rather than a finite 
(independent of $N$) one 
which would fall in the classical setting of mean field theory and for which 
the standard framework of large deviation theory, 
as outlined in Ellis' book [El],
applies immediately. 

A generalized random mean field model needs the following ingredients.

\item {(i)} A single spin space $\SS$ that we will always take to be a 
           subset of some linear space,  equipped with 
            some a priori probability measure $q$. 
\item {(ii)} A state space $\SS^N$ whose elements we denote by $\s$ and call
             {\it spin configurations}, equipped with the product measure
             $\prod_i q(d\s_i)$.  
\item {(iii)} The dual space ${(\SS^N)^*}^M$ of linear maps 
             $\xi_{N,M}^T:\SS^N\rightarrow \R^M$.
\item {(iv)} A mean field potential  which is some real valued function
$E_M:\R^M\rightarrow \R$, that we will assume
\itemitem{(iv.1)} Bounded below (w.l.g. $E_M(m)\geq 0$).
\itemitem{(iv.2)} in most cases, convex and ``essentially smooth'', that 
              is, it has a domain $\DD$ with non-empty interior,
             is differentiable on its domain,  and  
              $\lim_{m\rightarrow \del\DD}|\nabla E_M(m)|=+\infty$ 
              (see [Ro]).  
\item{(v)} An abstract probability space $(\O,\FF, \P)$ and measurable maps
           $\xi^T: \O\rightarrow {(\SS^\N)^*}^\N$. Note that 
           if $\Pi_N$ is the canonical projection $\R^\N\rightarrow \R^N$, 
then $\xi_{M,N}^T[\o]\equiv\Pi_M\xi^T[\o]\circ\Pi_N^{-1}$ are random elements 
of   $ {(\SS^N)^*}^M$. 
\item {(vi)} The random order parameter 
$$
m_{N,M}[\o](\s)\equiv \frac 1N   \xi_{M,N}^T[\o]\s \in \R^M
\Eq(G.4)
$$
\item{(vii)} A random Hamiltonian 
$$
H_{N,M}[\o](\s) \equiv -N E_M\left(m_{N,M}[\o](\s)\right)   
\Eq(G.5)
$$

\remark The formulation above corresponds to what in large deviation theory is
known as ``level 1'', i.e. we consider the Hamiltonian as a function
of order parameters that are functions (``empirical averages'') rather than 
as a function of empirical measures as in a ``level 2'' formulations. In some
cases a level 2 formulation would be more natural, but since in our main 
examples everything can be done on level 1, we prefer to stick to this 
language. 

With these objects we define the {\it finite volume Gibbs measures},
(which more precisely are probability  measure valued random variables)
$\mu_{\b,N,M}$ on 
$(\SS^N, \BB(\SS^N))$ through
$$
\mu_{\b,N,M}[\o](d\s)=\frac {e^{-\b H_{N,M}[\o](\s)}}
{Z_{\b,N,M}[\o]}\prod_{i=1}^N
q(d\s_i)
\Eq(G.6)
$$
where the normalizing factor,  called {\it partition function}, is 
$$
Z_{\b,N,M}[\o]\equiv \E_\s e^{-\b H_{N,M}[\o](\s)}
\Eq(G.7)
$$
where $\E_\s$ stands for the expectation with respect to the a priori product 
measure on $\SS^N$.
 Due to the special feature of these models that
$H_{N,M}[\o]$ depends on $\s$ only through $m_{N,M}[\o](\s)$, the distribution 
of these quantities contains essentially all information on the 
Gibbs measures themselves (i.e. the measures $\mu_{\b,N,M}[\o]$ restricted to 
the level sets of the functions $m_{N,M}[\o]$ are the uniform distribution on 
these sets) and thus 
play a particularly prominent r\^ole. They are measures on $(\R^M,\BB(\R^M))$ 
and we will call them {\it induced measures}
and denote them by
$$
\QQ_{\b,N,M}[\o]\equiv \mu_{\b,N,M}[\o]\circ \left(\frac 1N \xi^T_{N,M}[\o]
\right)^{-1}
\Eq(G.8)
$$

In the classical setting of mean field theory, $N$ would now be considered 
as the large parameter tending to infinity while $M$ would be some constant 
number, independent of $N$.  The main new feature here is that 
both $N$ and $M$ are large parameters and that as $N$ tends to infinity,
we choose $M\equiv M(N)$ as some function of $N$ that tends to infinity as 
well. However, we stress that the entire approach is geared to the case where 
at least $M(N)<N$, and even $M(N)/N\equiv \a $ is small. In fact, the passage
to the induced measures $\QQ$ appears reasonably motivated only in this case,
since only then we work in a space of lower dimension. To study e.g. the
Hopfield model for $\a$ large will require entirely different ideas which we 
do not have. 

It may be worthwhile to make some remarks on randomness and self averaging 
at this point in a somewhat informal way. 
As was pointed out in [BGP1], the distribution $\QQ$ of the 
order parameters can be expected to be much less ``random'' 
than the distribution 
of the spins. This is  to be understood in a rather strong sense: Define
$$
f_{\b,N,M,\rho}[\o](m)\equiv -\frac 1{\b N} \ln  
\QQ_{\b,N,M}[\o]\left(B_\rho(m)\right)
\Eq(G.8bis)
$$
where $B_\rho(m)\subset \R^M$ is the ball of radius $\rho$ centered at $m$. 
Then by strong self-averaging we mean that (for suitably chosen $\rho$)
$f$ as a function of $m$ is everywhere ``close'' to its expectation with 
probability close to one (for $N$ large)). Such a fact holds in a sharp
sense when $M$ is bounded, but it remains ``essentially'' true as long as
$M(N)/N\downarrow 0$ (This statement will be made precise in Section 6). 
This is the reason why under this hypothesis, these systems actually 
behave very much like ordinary mean field models.
When $\a>0$, what ``close'' can mean will depend on $\a$, but for small $\a$
this will be controllable. This is the reason why it will turn out to be possible
to study the situation with $\a$ small as a perturbation of the case $\a=0$.

\newsubsec{\ver.4 Thermodynamic limits
}

Although in some sense ``only finite volume estimates really count'', 
we are interested generally in asymptotic results as $N$ (and $M$) tend to 
infinity, and it is suitable to discuss in a precise way the corresponding 
procedure of {\it thermodynamic limits}.

In standard spin systems with short range interactions there is a well 
established beautiful  procedure  
of constructing infinite volume Gibbs measures
from the set of all finite volume measures (with ``boundary conditions'')
due to Dobrushin, Lanford and Ruelle (for a good exposition see e.g. [Geo]). 
This procedure cannot be applied
in the context of mean field models,  
essentially because the finite volume Hamiltonians are not restrictions 
to finite volume
of some formal infinite volume Hamiltonian, but contain parameters that  
  depend in an explicit way 
on the volume  $N$. It is however still possible to consider so called
{\it limiting Gibbs measures} obtained as accumulation points of sequences
of finite volume measures. This does, however require some discussion.

Observe first that it is of course trivial to 
extend the finite volume Gibbs measures $\mu_{\b,N,M}$ 
to measures on the infinite 
product space $(\SS^\N, \BB(\SS^\N))$, e.g. by tensoring it with 
the a priori measures $q$ on the components $i>N$. 
Similarly, the induced measures can be extended to the space
$(\R^\N, \BB(\R^\N))$ by tensoring with the Dirac measure concentrated on 
$0$. One might now be tempted to define the set of limiting Gibbs measures
as the set of limit points, e.g. 
$$
\CC_\b[\o]\equiv \hbox{clus}_{N\uparrow\infty}\left\{\mu_{\b,N,M(N)}[\o]
\right\}
\Eq(G.9)
$$
where $\hbox{clus}_{N\uparrow\infty} a_N$ denotes the set of limit points
(``cluster set'') of 
the sequence $a_N$.
However, it is easy to see that in general this set is not rich enough
to describe the physical content of the model.  
E.g., if we consider the Curie-Weiss model (c.f. \eqv(G.1)) it is easy to see 
and well known that this cluster set would always consist of a single 
element, namely the measure $\frac 12\left(\prod_{i=1}^\infty
q^{m^*(\b)}+\prod_{i=1}^\infty
q^{-m^*(\b)}\right)$, where $q^a(\s_i)=\frac {e^{\b a\s_i}}{2\cosh(\b a)}$
and where $m^*(\b)$ is the largest solution of the equation 
$$
x=\tanh \b x
\Eq(G.10)
$$
(and which we will have many occasions to meet in the sequel of this article).
If $\b>1$, $m^*(\b)>0$, and the limiting measure is a mixture; we would 
certainly want 
to be allowed to call the two summands limiting Gibbs measures as well, 
and to consider them as {\it extremal}, with {\it all} limiting Gibbs measures 
convex combinations of them. The fact that more than one such extremal 
measure exists  would be the sign of the occurrence of a {\it phase transition}
if $\b>1$. 

The standard way out of this problem is to consider a richer class of 
{\it tilted Gibbs measures}
$$
\mu_{\b,N,M}^h[\o](d\s)\equiv \frac{ e^{-\b  H_{N,M}[\o](\s)+
\b N h\left(m_{N,M}[\o](\s)\right)}}
{Z_{\b,N,M}^h[\o]}\prod_{i=1}^N
q(d\s_i)
\Eq(G.11)
$$
where $h:\R^M\rightarrow \R$ is a {\it small} perturbation that plays 
the r\^ole of a symmetry breaking term. In most cases it suffices to 
choose {\it linear} perturbations, $h\left(m_{N,M}[\o](\s)\right)=\left(
h,m_{N,M}[\o](\s)\right)$, in which case $h$ can be interpreted  as a 
{\it magnetic field.}
Instead of \eqv(G.9) one defines then the set
$$
\wt\CC_\b[\o] \equiv \hbox{clus}_{\|h\|_\infty\downarrow 0, N\uparrow \infty}
\left\{\mu_{\b,N,M(M)}^h[\o]\right\}
\Eq(G.12)
$$
where we first consider the limit points that can be obtained for 
all $h\in \R^\infty$ and then collect all possible limit points that can be 
obtained as $h$ is taken to zero (with respect to the sup-norm). Clearly 
$\CC_\b\subset \wt\CC_\b$. If this inclusion is strict, this means that
the infinite volume Gibbs measures depend in a discontinuous way 
on $h$ at $h=0$, which corresponds to the standard physical definition of
a first order phase transition. We will call $\wt\CC_\b[\o]$ the set of 
{\it limiting Gibbs measures}. 

The set $\wt\CC_\b[\o]$ will  in general {\it not} be a convex set. 
E.g., in the Curie-Weiss case, it consists, for $\b>1$ of three elements, 
$\mu_{\b,\infty}^+,\mu_{\b,\infty}^-$, and 
$\frac 12(\mu_{\b,\infty}^++\mu_{\b,\infty}^-)$. (Exercise: Prove this 
statement!). However, we may still consider the convex closure of this set
and call its extremal points {\it extremal Gibbs measures}. It is likely,
but we are not aware of a proof, that all elements of the convex closure 
can be obtained as limit points if the limits $N\uparrow 0$, 
$\|h\|_\infty\downarrow 0$ are allowed to be taken  jointly
(Exercise: Prove that this is true in the Curie-Weiss model!).

Of course, in the same way we define the tilted induced measures, and the main
aim is to construct, in a more or less explicit  way, the set of limiting
induced measures. We denote these sets by 
$\CC_\b^\QQ[\o]$, and $\wt\CC_\b^\QQ[\o]$, respectively.  
  The techniques used will basically of large deviation type,
with some modifications necessary. We will discuss this formalism briefly in 
Section 3 and 5. 

\newsubsec{\ver.5 Convergence and propagation of
chaos.}
 
Here we would like to discuss a little bit the expected or possible behaviour 
of generalized random mean field models.  Our first remark is that 
all the sets $\CC_\b[\o] $ and $\wt\CC_\b[\o]$ will not be empty if $\SS$ is 
compact. The same holds in most cases for $\CC_\b^\QQ[\o] $ and 
$\wt\CC_\b^\QQ[\o]$, namely when the image of $\SS^N$ under $\xi^T_{N,M}$ is 
compact. This may, however, be misleading. 
Convergence of a sequence of measures $\QQ_{\b,N,M(N)}$ on 
$(\R^\infty,\BB(\R^\infty))$ in the usual weak sense means simply 
convergence of all finite dimensional marginals. 
Now take the sequence $\d_{e^{M(N)}}$, of Dirac-measures concentrated on the 
 $M(N)$-th unit vector in $\R^\infty$. Clearly, this sequence converges to 
the Dirac measure concentrated on zero, and this observation 
obviously misses 
a crucial point about this sequence. Considered rather as a measure on the 
set of unit vectors, this sequence clearly does {\it not} converge. 
For most purposes it thus more appropriate to use a $\ell^2$-topology rather 
than the more conventional product topology. In this sense, the above sequence
of Dirac measures does, of course, not converge weakly, but converges 
vaguely to the zero measure.

It is an interesting question whether one can expect, in a random situation,
that there exist {\it subsequences} of untilted measures converging 
 weakly in the $\ell_2$ topology 
in a phase transition region. Ch. K\"ulske [Ku] recently constructed 
an example in which the answer to this question is negative. He also showed, 
that, as long as $M(N)<\ln N$, in the standard Hopfield model, 
the sets $\CC^\QQ_\b[\o]$ and $\wt\CC^\QQ_\b[\o]$ coincide for almost all $\o$.

In conventional mean field models, the induced measures converge (if properly 
arranged) to 
Dirac measures, implying that in the thermodynamic limit, the macroscopic 
order parameters verify a {\it law of large numbers}. In the case of infinitely
 many order parameters, this is not obviously true, and it may not even seem 
reasonable to expect, if $M(N)$ is not considerably smaller than $N$. Indeed, 
it has been shown in [BGP1] that in the Hopfield model this holds if
$\frac {M(N)}N\downarrow 0$. Another paradigm of mean field theory is 
{\it propagation of chaos} [Sn], i.e. the fact that 
the (extremal) limiting Gibbs measures are product measures, 
i.e. that any finite subset of spins forms a family of independent random 
variables in the 
thermodynamic limit. In fact, both historically and in most standard textbooks 
on statistical mechanics, {\it this} is the starting assumption for the 
derivation of mean field theory, while models such as 
the Curie-Weiss model are just convenient examples where these assumptions
happen to be verified. In the situation of random models, this is a 
rather subtle issue, and we will  come back to this in Section 8 
where we will learn actually a lot about this.

\newsubsec{ \ver.6 Examples.}

Before turning to the study of large deviation techniques, we conclude this 
section by presenting a list of commonly used variants of the Hopfield model
and to show how they fit into the above framework. 

\vskip 0.2cm 
\line{\sl \ver.6.1 The standard Hopfield model.\hfill} 

Here $\SS=\{-1,1\}$, $q$ is the Bernoulli measure $q(1)=q(-1)=\frac 12$. 
$(\SS^N)^*$ may be identified with $\R^N$ and $\xi^T_{N,M}$ are real 
$M\times N$-matrices. 
The mean field potential is $E_M (m)=\frac 12\|m\|_2^2$, where
$\|\cdot\|_2 $ denotes the $2$-norm in $\R^M$. The measure $\P$ is such that 
$\xi_i^\mu$ are independent and  identically distributed with 
$\P[\xi_i^\mu=\pm 1]=\frac 12$. The order parameter is the $M$-dimensional 
vector
$$
m_{N,M}[\o](\s)=\frac 1N\sum_{i=1}^N \xi_i \s_i
\Eq(G.13)
$$
and the Hamiltonian results as the one in \eqv(G.3).

\vskip 0.2cm 
\line{\sl \ver.6.2 Multi-neuron interactions.\hfill}

This model was apparently introduced by Peretto and Niez [PN] and studied 
for instance by Newman [N]. Here all is the same as in the previous case,
except that the mean field potential is
$E_M(m) =\frac 1p\|m\|_p^p$, $p>2$. For (even) integer $p$, 
the Hamiltonian is then
$$
H_{N,M}[\o](\s)=-\frac 1{N^p} \sum_{i_1,\dots,i_p} \s_{i_1}\dots \s_{i_p}
\sum_{\mu=1}^M \xi_{i_1}^\mu\dots \xi_{i_p}^\mu
\Eq(G.14)
$$ 

\vskip 0.2cm 
\line{\sl \ver.6.3 Biased Hopfield model.\hfill} 

Everything the same as in \ver.6.1, but the distribution of $\xi_i^\mu$ 
is supposed to reflect an asymmetry (bias) between $+1$ and $-1$ (e.g. 
to store pictures that are typically more black than white). That is, we
have (e.g.)
$\P[\xi_i^\mu=2x]=(1-x)$ and $\P[\xi_i^\mu=2(1-x)]=x$.           
              One may, of course, consider the model with yet 
different distributions of the $\xi_i^\mu$. 

\vskip 0.2cm 
\line{\sl \ver.6.4 Hopfield model with correlated patterns.\hfill}

In the same context, also the assumption of independence of the $\xi_i^\mu$
is not always reasonable and may be dropped. One speaks of
{\it semantic} correlation, if the components of each vector 
$\xi^\mu$ are independent, while the different vectors are correlated, and of 
{\it spatial} correlation, if the  different vectors $\xi^\mu$ are independent,
but have correlated components $\xi^\mu_i$. Various reasons for considering
such types of patterns can be found in the literature [FZ,Mi].
Other types of correlation considered include the case where $\P$ is the  
distribution of a family of Gibbs random fields [SW].

\vskip 0.2cm 
\line{\sl \ver.6.5 Potts-Hopfield model.\hfill}

Here the space $\SS$ is the set $\{1,2,\dots,p\}$, for some integer $p$, and $q$ is the uniform measure on this set. We again have random patterns $\xi_i^\mu$
that are independent and the marginal distribution of $\P$ coincides with $q$.
The order parameters are defined as
$$
m^\mu_M[\o](\s)=\frac 1N\sum_{i=1}^N \left[\d_{\s_i,\xi^\mu_i}-\sfrac 1p\right]
\Eq(G.14bis)
$$
for $\mu=1,\dots,M$. $E_M$ is the same as in the standard Hopfield model.
Note that the definition of $m_M$ seems not to fit exactly our 
setting. The reader should figure out how this can be fixed.
See also [G1].
A number of other interesting variants of the model really lie outside 
our setting. We mention two of them:

\vskip 0.2cm 

\line{\sl \ver.6.6 The dilute Hopfield model.\hfill}

Here we are in the same setting as in the standard Hopfield model, except that
the Hamiltonian is no longer a function of the order parameter. Instead, 
we need another family of, let us say independent,
 random variables, $J_{ij}$, with 
$(i,j)\in \N\times \N$ with distribution e.g. 
$\P[J_{ij}=1]=x$, $\P[J_{ij}=0]=1-x$, and the Hamiltonian is
$$
H_{N,M}[\o](\s)=-\frac 1{2Nx}\sum_{i,j}\s_i\s_i J_{i,j}[\o]\sum_{\mu=1}^M
\xi_i^\mu\xi_j^\mu
\Eq(G.15)
$$
This model describes a neural network  in which each neuron 
interacts only with a fraction $x$ of the other neurons, with the set of 
a priori connections between neuron described as a random graph [BG1,BG2].
 This is 
certainly a more realistic assumption when one is modelling biological 
neural networks like the brain of a rat. The point here is that, while 
this model is not a generalized mean field model, if we replace the 
Hamiltonian \eqv(G.15) by its average with respect to the 
random variables $J$, we get back the original Hopfield Hamiltonian. On the 
other hand, it is true that 
$$
\sup_{\s\in \SS^N}\left|
H_{N,M}[\o](\s)-\E\left[H_{N,M}[\o](\s)\big|\FF_\xi\right]\right]\leq 
cN\sqrt{\frac M{x N}}
\Eq(G.16)
$$
with overwhelming probability, which implies that in most respects the dilute model 
has the same behaviour as the normal one, provide $\frac M{x N}$ is small.
The estimate \eqv(G.16) has been proven first in [BG2], but a much simpler 
proof can be found in [T4]. 

\vskip 0.2cm
\line{\sl \ver.6.7 The Kac-Hopfield model.\hfill}

This model looks similar to the previous one, but here some non-random 
geometry is introduced. The set $\{1,\dots,N\}$ is replaced by 
$\L\subset \Z^d$,
and the random $J_{ij}$ by some deterministic function
$J_\g(i-j)\equiv \g^d J(\g(i-j))$ with $J(x)$ some function with 
bounded support
(or rapid decay) whose integral equals one. Here $\g$ is a small parameter.
 This model had already been 
introduced by Figotin and Pastur [FP3] but has been investigated more 
thoroughly only recently [BGP2, BGP4]. It shows very interesting features and
an entire article 
in this volume is devoted to it.

\newpage
 
\def\PPP{\Phi}
\def\PSS{\Psi}

\def\ff{E_M}

\chap{3. Large deviation estimates and transfer principle}3

The basic tools to study the models we are interested in are large deviation 
estimates for the induced measures $\QQ_{\b,N,M}$. 
Compared to the standard situations, there are two 
particularities in the setting of generalized random mean field models that 
require some special attention: (i) the dimension $M$ 
of the space on which these measures are defined must be allowed to depend on 
the basic large parameter $N$ and (ii) the measure $\QQ_{\b,N,M}$ is itself
random. A further aspect is maybe even more important. We should be 
able to compute, in a more or less explicit form, the ``rate function'',
or at least be able to identify its minima. In the setting we are in,
this is a difficult task, and we will stress the calculational aspects here. 
We should mention that in the particular case of the Hopfield model with
quadratic interaction, there is a convenient trick, called the 
{\it Hubbard-Stratonovich transformation} [HS] that allows one to circumvent 
the technicalities we discuss here. This trick 
has been used frequently in the past,
and we shall come back to it in Section 8. The techniques we present here 
work in much more generality and give essentially equivalent results. 
The central result that will be used later is Theorem \ver.5.

\newsubsec{\ver.1. Large deviations estimates.}

Let us start with the general large deviation framework adopted to 
our setting. 
Let $M$ and $N$ be two integers. Given a family  $\{\nu_N,\,N\geq 1\}$
of probability measures on $(\R^M,\BB(\R^M))$, and a function $\ff$:
$\R^M\rightarrow\R$ (hypotheses on $\ff$ will be specified later on), 
we define a new  family 
$\{\mu_N,\,N\geq 1\}$ of probability measures on $(\R^M,\BB(\R^M))$ via
$$
\mu_N(\G)\equiv\frac{\int_{\G}e^{N\ff(x)}d\nu_N(x)}
{\int_{\R^M}e^{N\ff(x)}d\nu_N(x)}\,\,\,,\,\,\,\,\G\in\BB(\R^M)
\Eq(T.1)
$$
We are interested in the large deviation properties of this new family. In 
the case  when $M$ is a fixed integer, it follows from Varadhan's lemma on 
the asymptotics of integrals that, if $\{\nu_N,\,N\geq 1\}$ satisfies a 
large deviation principle with good rate function $I(\cdot)$, and if $\ff$ is 
suitably chosen (we refer to [DS], Theorem 2.1.10 and exercise (2.1.24) for a 
detailed presentation of these results in a more general setting) then
$\{\mu_N,\,N\geq 1\}$ satisfies a large deviation principle 
with good rate function $J(x)$ where
$$
J(x)=-[\ff(x)-I(x)]+\sup_{y\in\R^M}[\ff(y)-I(y)]
\Eq(T.2)
$$
Here we address the question of the large deviation behaviour of 
$\{\mu_N,\,N\geq 1\}$
in the case where $M\equiv M(N)$ is an unbounded function of $N$ 
and where the measure $\nu_N$ is defined as follows:

Let $\xi$ be a linear transformation from $\R^N$ to $\R^M$. To avoid 
complications, we assume that $M\leq N$ and $\xi$ is non-degenerate, 
i.e. its image is all $\R^M$.
We will use the same symbol to denote the corresponding $N\times M$
matrix $\xi\equiv\{\xi_{i,\mu}\}_{i=1,\dots N; \mu=1,\dots M}$ and we will
denote by $\xi^{\mu}\equiv(\xi^{\mu}_1,\dots,\xi^{\mu}_N)\in\R^M$, respectively
$\xi_i\equiv(\xi_i^1,\dots,\xi_i^M)\in\R^N$, 
the $\mu$-th row vector and  $i$-th 
column vector. The transposed matrix (and the corresponding adjoint linear 
transformation from $\R^M$ to $\R^N$) is denoted $\xi^T$. Consider a 
probability space $(\R,\BB(\R),{\cal P})$ and its $N$-fold 
power $(\R^N,{\cal P}_N)$ 
where ${\cal P}_N={\cal P}^{\otimes N}$. We set
$$
\nu_N \equiv {\cal P}_N\circ\left(\sfrac{1}{N}\xi^T\right)^{-1}
\Eq(T.3)
$$

In this subsection we will  present upper and lower large 
deviation bounds for fixed $N$. More precisely we set, 
for any $\rho>0$ and $x^*\in\R^M$,
$$
Z_{N,\rho}(x^*)\equiv \int_{B_{\rho}(x^*)} e^{N\ff(x)}d\nu_N(x)
\Eq(T.4)
$$
In the regime where $\lim_{N\rightarrow\infty} \sfrac{M}{N}=0$,
estimates on these quantities provide a starting point to prove a strong 
large deviation principle for $\{\mu_N,\,N\geq 1\}$ in a formulation that 
extends the ``classical'' Cram\`er's formulation. This was done in 
[BG4] in the case of the standard Hopfield model. In the regime where 
$\lim_{N\rightarrow\infty} \sfrac{M}{N}=\a$ with $\a>0$, we cannot 
anymore establish such a LDP. But estimates on $Z_{N,\rho}(x^*)$ will 
be used to 
establish concentration properties for $\QQ_N$ asymptotically as $N$
tends to infinity, as we will see later in  the paper. 

Following the classical procedure, we obtain an upper bound on 
$Z_{N,\rho}(x^*)$ by optimizing on a family of exponential Markov 
inequalities. As is well known, this will require the computation of the 
conjugate of\note{We have chosen to follow Rockafellar's  terminology and 
speak about conjugacy correspondence and conjugate of a (convex) function 
instead of Legendre-Fenchel conjugate, as is often done. 
This will allow us to refer to [Ro] and the classical Legendre transform 
avoiding confusions that might otherwise arise.}
{$ $} the 
{\it logarithmic moment generating function},
defined as 
$$
\LL_{N,M}(t)\equiv \frac{1}{N}\log \int_{\R^M}e^{N(t,x)}\nu_N(dx)\,\,\,, 
t\in\R^M
\Eq(T.5)
$$
In the setting we are in, the computation of this quantity is generally quite 
feasible. A recurrent theme in large deviation theory is that of the \
Legendre transform. To avoid complications that will not arise in our examples,
we restrict the following discussion mainly
to 
the case when the Legendre transform is well
defined (and involutary) which  is essentially the case where the convex 
function is {\it strictly convex} and {\it essentially smooth}. We recall from
[Ro]: 

\definition{\ver.1} {\it  A real valued function $g$ on a convex 
set $C$ is said to be
{\rm strictly convex} on $C$ if 
$$
g((1-\l)x+\l y)<(1-\l)g(x)+\l g(y)\,\,\,\,\,\,\, 0<\l<1
\Eq(T.24)
$$
for any two different points $x$ and $y$ in $C$. It it called {\rm proper} 
if it is not identically equal to $+\infty$. 
\hfill\break\vskip2pt\noindent
An extended-real-valued function $h$ on $R^M$ is 
{\rm essentially smooth} if it satisfies the following three conditions
for $C=\hbox{int}(\hbox{dom} h)$:
\item{(a)} $C$ is non empty;
\item{(b)} $h$ is differentiable throughout $C$;
\item{(c)} $\lim_{i\rightarrow\infty}|\nabla h(x_i)|=+\infty$ whenever 
           $x_1,x_2,\dots,$ is a sequence in $C$ converging to a boundary
           point $x$ of $C$.
}

(Recall that $\hbox{dom}g\equiv\{x\in\R^M\mid g(x)<\infty\}$).
 Note that if a function $\ff$ is essentially smooth, it follows  
(c.f. [RV], 
Theorem A and B and [Ro], pp. 263-272) that
$\ff$  attains a minimum value and the set on which  
this (global) minimum is attained consists of a single point belonging to the 
interior of it's domain. Without loss of generality we will assume in the 
sequel that $\ff(x)\geq 0$ and $\ff(0)=0$.

All through this chapter we adopt the usual approach that consists in 
identifying 
a convex function $g$ on  $\hbox{dom}g$ with the convex function defined 
throughout the space $\R^M$ by setting $g(x)=+\infty$ for $x\notin\hbox{dom}g$.

\definition{\ver.2} {\it Let $g$ be a proper convex function. The function 
$g^*$ defined by
$$
g^*(x^*)=\sup_{x\in\R^M}\left\{(x, x^*)-g(x)\right\}
\Eq(T.7)
$$
is called its {\rm (ordinary) conjugate}.}

\noindent For any set $S$ in  $\R^M$ we denote by $\hbox{int} S$ its interior.
For smooth $g$  we denote by 
$\nabla g(x)\equiv\left(\sfrac{\partial g(x)}
{\partial x^1},\dots,\sfrac{\partial g(x)}{\partial x^{\mu}},\dots,
\sfrac{\partial g(x)}{\partial x^{M}}\right)$,
$\nabla^2 g(x)\equiv\left(
\sfrac{\partial^2 g(x)}{\partial x^{\mu}\partial x^{\nu}}
\right)_{\mu,\nu=1,\dots,N}$ 
and
$\Delta g(x)\equiv\sum_{\mu=1}^{M}
\sfrac{\partial^2 g(x)}{\partial^2 x^{\mu}}$
respectively the 
gradient vector, the Hessian matrix, and the Laplacian of $g$ at $x$.

The following lemma collects some well-known properties of
$\LL_{N,M}$ and its conjugate:

\lemma{\ver.3} {\it
\item{(a)} $\LL_{N,M}$ and $\LL_{N,M}^*$ are proper convex functions
           from   $\R^M$ to $\R\cup\infty$. 
\item{(b)}  $\LL_{N,M}(t)$ is infinitely  differentiable
on }$\hbox{int(dom} \LL_{N,M})${\it.
Defining the measure $\tilde\nu_{N, t}$ via 
$d\tilde\nu_{N, t}(X)\equiv\frac{\exp\{N(t,X)\}}
{\int\exp\{N(t,X)\}d\nu_N(X) }d\nu_N(X)$,
and denoting by 
$\wt\E_{t}(\cdot)$, the expectation 
w.r.t. 
$\tilde\nu_{N,t}$ we have, for any $t$ in }$\hbox{dom} \LL_{N,M}$,{\it 
$$
\eqalign{
\nabla \LL_{N,M}(t)
&= \wt\E_{t}(X)
=\left(\wt\E_{t}(X_{\mu})\right)_{\mu=1,\dots,M}\cr
\sfrac{1}{N}\nabla^2 \LL_{N,M}(t)
&=\left(
\wt\E_{t}(X_{\mu}X_{\nu})
-\wt\E_{t}(X_{\mu})
\wt\E_{t}(X_{\nu})
\right)_{\mu,\nu=1,\dots,M}\cr
}
\Eq(T.8)
$$
and, if $\LL^*$ is  smooth, 
the following three conditions on $x$ are equivalent
$$
\eqalign{
1)&\,\,\,\,\,\nabla \LL_{N,M}(t)=x\cr
2)&\,\,\,\,\,\LL_{N,M}^*(x)=(t,x)- \LL_{N,M}(t)\cr
3)&\,\,\,\,\,(y,x)- \LL_{N,M}(y)\,\,\,\,\,\, \hbox{achieves its supremum over} 
\,\,\,y\,\,\,\hbox{at}\,\,\,y=t
}
\Eq(T.9)
$$
\item{(c)} $\LL_{N,M}^*(x)\geq 0$ and, 
        if $\wt\E_{0}(X)<\infty$, 
            $\LL_{N,M}^*(\wt\E_{0}(X))=0$.
}

\proof {The proofs of statements (a) and (c) can be found in [DZ], as well
as the proof of the differentiability property.  
The formulae \eqv(T.8) are simple algebra.
Finally, the equivalence of the three conditions  \eqv(T.9) is an application
of Theorem 23.5 of [Ro] to the particular case of a differentiable proper 
convex function.
\endproof}  

Setting
$$
\PSS_{N,M}(x)\equiv -\ff(x)+\LL_{N,M}^*(x)\,\,\,, x\in\R^M
\Eq(T.10)
$$
we have

\lemma{\ver.4} {\it 
For any $x^*$ in $\R^M$, define $t^*\equiv t^*(x^*)$ through 
$\LL_{N,M}^*(x^*) =(t^*,x^*)-\LL_{N,M}(t^*)$
if such a  $t^*$ exists while otherwise $\|t^*\|_2\equiv\infty$
(note that $t^*$ need not be unique). We have, for any $\rho>0$,
$$
\frac{1}{N}\log Z_{N,\rho}(x^*)\leq
-\PSS_{N,M}(x^*)+ \sup_{x\in B_{\rho}(x^*)}[\ff(x)-\ff(x^*)]+\rho\|t^*\|_2
\Eq(T.12)
$$
and
$$
\eqalign{
\frac{1}{N}\log Z_{N,\rho}(x^*)\geq
-\PSS_{N,M}(x^*)&+ \inf_{x\in B_{\rho}(x^*)}[\ff(x)-\ff(x^*)]-\rho\|t^*\|_2
\cr
&+\sfrac{1}{N}\log(1-\sfrac{1}{\rho^2N}\Delta\LL_{N,M}(t^*))
}
\Eq(T.13)
$$
}

\proof {Analogous bounds were obtained in [BG4], Lemmata 2.1 and 2.2,
in the special case of an application to the Hopfield model.  The proofs of 
\eqv(T.12) and \eqv(T.13) follow the proofs of these lemmata with only minor 
modifications. We will only recall the main lines of the proof of the lower 
bound: the essential step is to perform an exponential 
change of measure i.e., with the definition of $\tilde\nu_{N, t}$ from
Lemma \ver.4, we have,
$$
\frac{1}{N}\log Z_{N,\rho}(x^*)
=\wt\E_{t^*}\left(e^{N\{\ff(X)-(t^*,X)\}}\1_{\{B_{\rho}(x^*)\}}\right)
\wt\E_{0}\left(e^{N(t^*,X)}\right)
\Eq(T.14)
$$
from which, together with \eqv(T.5) and \eqv(T.9), we easily obtain,
$$
\eqalign{
\frac{1}{N}\log Z_{N,\rho}(x^*)&\geq 
e^{N\left\{-\PSS_{N,M}(x^*)+ \inf_{x\in B_{\rho}(x^*)}[\ff(x)-\ff(x^*)]
-\rho\|t^*\|_2\right\}}\cr
&\times\tilde\nu_{N,t^*}(B_{\rho}(x^*))
}
\Eq(T.15)
$$
When the law of large numbers is not available, as is the case here,
the usual procedure to estimate the term $\tilde\nu_{N,t^*}(B_{\rho}(x^*))$ 
would be to use the upper bound. Here we simply use the Tchebychev
inequality to write
$$
1-\tilde\nu_{N,t^*}(B_{\rho}(x^*))
=\wt\E_{t^*}\left(\1_{\{\|X-x^*\|_2^2>\rho^2\}}\right)
\leq\sfrac{1}{\rho^2}\wt\E_{t^*}\|X-x^*\|_2^2
\Eq(T.16)
$$
Now, by \eqv(T.9), $t^*$ satisfies $\nabla \LL_{N,M}(t^*)=x^*$, and it 
follows from \eqv(T.8) that
$$
\eqalign{
\wt\E_{t^*}\|X-x^*\|_2^2
=\frac{1}{\rho^2}\sum_{\mu=1}^M\left[
\wt\E_{t^*}X_{\mu}^2-\left(\wt\E_{t^*}X_{\mu}\right)^2\right]
=\sfrac{1}{\rho^2N}\Delta\LL_{N,M}(t^*)
}
\Eq(T.17)
$$
Collecting \eqv(T.15), \eqv(T.16) and \eqv(T.17) proves \eqv(T.13).
\endproof}

\remark {The lower bound \eqv(T.13) is meaningful only if 
 $\sfrac{1}{N\rho^2}\Delta\LL_{M,N}(x)<1$. But the Laplacian of 
a function on $\R^M$ has a tendency to be of order $M$. Thus, typically, 
the lower bound will be useful only if $\rho^2\geq O(M/N)$. 
We see that if $\lim_{N\uparrow\infty} \frac MN=0$, 
one may shrink $\rho$ to $0$ and get 
upper and lower bounds 
that are asymptotically the same (provided $\ff$ is continuous), provided 
the norm of $t^*$ remains bounded. Since $t^*$ is random, this introduces some 
subtleties which, however, can be handled (see [BG4]). But if  
$\lim_{N\uparrow\infty} \frac MN=\a>0$, we do not get a lower bound for 
balls of radius smaller than $O(\sqrt\a)$ and there is no hope to get 
a large deviation principle in the usual 
sense from Lemma \ver.4. What is more disturbing, 
is the fact that the quantities 
$\PSS$ and $t^*$ are more or less impossible to compute in an explicit form,
and this makes Lemma \ver.4 not a very good starting point for further 
investigations. 
}

\newsubsec{ \ver.2. Transfer principle.}

As we will show now, it is possible to get large deviation {\it estimates} 
that 
do not involve the computation of Legendre transforms. The price to pay will
be that these will not be sharp everywhere. But as we will see,  they are sharp
at the locations of the extrema and thus are 
sufficient 
for many purposes. Let us define the function
$$
\PPP_{N,M}(x)=-\ff(x)+(x,\nabla \ff(x))-\LL_{N,M}(\nabla \ff(x))
\Eq(T.33)
$$

\theo{\ver.5} {\it 
\item{(i)} Let $x^*$ be a point in $\R^M$ such that for some $\rho_0>0$, for 
all $x,x'\in B_{\rho_0}(x^*)$, 
$\|\nabla \ff(x)-\nabla \ff(x')\|_2<c\|x-x'\|_2$. 
Then,
for all $0<\rho<\rho_0$
$$
\frac{1}{N}\log Z_{N,\rho}(x^*)\leq -\PPP_{N,M}(x^*)+\frac 12 c\rho^2
\Eq(T.70)
$$
\item{(ii)} Let $x^*$ be a point such that
$\nabla \LL_{N,M}(\nabla \ff(x^*))=x^*$. Then,
$$
\frac{1}{N}\log Z_{N,\rho}(x^*)\geq
-\PPP_{N,M}(x^*) 
+\sfrac{1}{N}\log(1-\sfrac{1}{\rho^2N}\Delta\LL_{N,M}(\nabla \ff(x^*)))
\Eq(T.71)
$$
}

\remark  {The condition 
$\nabla \LL_{N,M}(\nabla \ff(x^*))=x^*$ is equivalent to 
the condition $\nabla \PSS_{N,M}(x^*)=0$, if $\LL^*$ is essentially smooth. 
This means that the lower bound 
holds at all critical points of the ``true'' rate function. It is easy to see
that $\nabla \PSS_{N,M}(x)=0$ implies $\nabla \PPP_{N,M}(x)=0$, while the 
converse  is not generally true. Fortunately, however, this {\it is }
true for critical points of $\PPP_{N,M}$ that are minima. This fact will 
be established in the remainder of this section. } 

\remark{ It is clear that we could get an upper bound with error term 
$C \rho$ without the hypothesis that $\nabla\ff$ is Lipshitz. However, when 
we apply Theorem \ver.5, a good estimate on the error will be important\note{
The point is that the number of balls of radius $\scriptstyle\rho$ to cover, 
say, the unit ball is of the order $\scriptstyle\rho^{-\a N}$, 
that is exponentially large.
Therefore we want to use as large a $\scriptstyle\rho$ 
as possible with as small an error as possible. Such problems
do not occur when the dimension of the space is independent of 
$\scriptstyle N$.}, while
local Lipshitz bounds on $\nabla\ff$ are readily available.}

\proof {With the definition of $\tilde\nu_{N, t}$ from
Lemma \ver.4, we have,
$$
\eqalign{
 Z_{N,\rho}(x^*)
&=\wt\E_{t}\left(e^{N\{\ff(X)-(t,X)\}}\1_{\{B_{\rho}(x^*)\}}\right)
\wt\E_{0}\left(e^{N(t,X)}\right)\cr
&=e^{N\{\LL_{N,M}(t)+\ff(x^*)-(t,x^*)\}}
\cr
&\times\wt\E_{t}
\left(e^{N\{\ff(X)-\ff(x^*)-(t,(X-x^*))\}}\1_{\{B_{\rho}(x^*)\}}\right)\cr
}
\Eq(T.72)
$$
The strategy is now to chose $t$ in such a way as to get optimal control over 
the last exponent in \eqv(T.72). By the fundamental theorem of calculus,
$$
\eqalign{
&|\ff(X)-\ff(x^*)-(t,(X-x^*))|\cr
&=\left|\int_0^1 ds
\left((\nabla \ff(sX+(1-s)x^*)-t),(X-x^*)\right)\right|\cr
&\leq \sup_{s\in [0,1]} \|(\nabla \ff(sX+(1-s)x^*)-t\|_2\|X-x^*\|_2
}
\Eq(T.73)
$$
Of course we want a bound that is uniform in the set 
of $X$ we consider, so that 
the best choice is of course $t\equiv\nabla \ff(x^*)$.  
Since $\nabla \ff(x)$ was assumed to be Lipshitz 
in $B_\rho(x^*)$ we get
$$
\eqalign{ 
Z_{N,\rho}(x^*)
&\leq e^{N\{\LL_{N,M}(\nabla \ff(x^*))+\ff(x^*)-(\nabla \ff(x^*),x^*)\}}
e^{\frac 12 Nc\rho^2}
\cr
&=e^{-N\PPP_{N,M}(x^*)}e^{\frac 12 Nc\rho^2}
}\Eq(T.74)
$$
where the last equality follows from the definition \eqv(T.33).
This proves the upper bound \eqv(T.70). To prove the lower bound, note that 
since $\ff$ is convex,
$$
\ff(X)-\ff(x^*)-\left(\nabla \ff(x^*), (X-x^*)\right)\geq 0
\Eq(T.75)
$$
Using this in the last factor of \eqv(T.72),
we get
$$
Z_{N,\rho}(x^*)
\geq e^{-N\{\PPP_{N,M}(x^*)\}}\tilde\nu_{N,t}(B_{\rho}(x^*))
\Eq(T.76)
$$
Now, just as in \eqv(T.16), 
$$
1-\tilde\nu_{N,t}(B_{\rho}(x^*))
\leq\sfrac{1}{N}\wt\E_{t}\|X-x^*\|_2^2
\Eq(T.77)
$$
and a simple calculation as in Section \ver.1 shows that 
$$
\wt\E_{t}\|X-x^*\|_2^2=
\sfrac{1}{\rho^2N}\Delta\LL_{N,M}(t)+\|\nabla\LL_{N,M}(t)-x^*\|_2^2
\Eq(T.78)
$$
Here we see that the optimal choice for $t$ would be the solution of
$\nabla\LL_{N,M}(t)=x^*$, an equation we did not like before. However, we now 
have by assumption,
$\nabla\LL_{N,M}(\nabla \ff(x^*))=x^*$.
This concludes the proof of 
Theorem \ver.14.
\endproof}

Sometimes the estimates on the probabilities of $\ell_2$-balls  may not be the 
most suitable ones. 
A charming feature of the upper bound is that it can also be
 extended to sets that are adapted to the function $\ff$. Namely, if we define
$$
\wt Z_{N,\rho}(x^*)\equiv \int e^{N\ff(x)}\1_{\{\|\nabla \ff(x)-\nabla \ff(x^*)
\|_2\leq \rho\}} d \nu_N(x)
\Eq(T.78bis)
$$
we get 

\theo{\ver.6} {\it Assume that for some $q\leq 1$ for all 
$y,y'\in B_{\rho_0}(\nabla \ff(x^*))$, $\|(\nabla\ff)^{-1}(y)-
(\nabla\ff)^{-1}(y')\|_2\leq c \|y-y'\|_2^q$, then  
for all $0<\rho<\rho_0$
$$
\frac{1}{N}\log \wt Z_{N,\rho}(x^*)\leq -\PPP_{N,M}(x^*)+\frac 12 c\rho^{1+q}
\Eq(T.79)
$$
}

The proof of this Theorem is a simple rerun of that of the upper bound 
in  Theorem \ver.5
and is left to the reader.

We now want to make the remark following Theorem \ver.5 precise.

\proposition {\ver.7} {\it Assume that $\ff$ is strictly convex,
and essentially smooth. If $\PPP_{N.M}$ has a local extremum at a point 
$x^*$ in the interior of its domain, then
$\nabla \LL(\nabla \ff(x^*))=x^*$. 
}

\proof {To prove this proposition, we recall a fundamental Theorem on functions
of Legendre type from [Ro].
\definition{\ver.8} {\it Let $h$ be a differentiable real-valued function on a open
subset $C$ of $\R^M$. The {\it Legendre conjugate} of the pair $(C,h)$ is
defined to be the pair $(D,g)$ where $D=\nabla h(C)$ and $g$ is the function
on $D$ given by the formula
$$
g(x^*)=((\nabla h)^{-1}(x^*),x^*)-h((\nabla h)^{-1}(x^*))
\Eq(T.40)
$$
Passing from  $(C,h)$ to  $(D,g)$, if the latter is well defined, is called
the {\it Legendre transformation}.}
\definition{\ver.9} {\it Let $C$ be an open convex set and $h$ an essentially smooth
and strictly convex function on $C$. The pair $(C,h)$ will be called a 
{\it convex function of Legendre type}.}
The Legendre conjugate of a convex  function of Legendre type is related to 
the 
ordinary conjugate as follows:
\theo{\ver.10}{\it {\rm([Ro], Theorem 26.5)} 
 Let $h$ be a closed convex function. Let 
$C=\hbox{int}(\hbox{dom} h)${\it and }$C^*=\hbox{int}(\hbox{dom} h^*)$.  
Then $(C,h)$ is a convex function of Legendre type if and only if $(C^*,h^*)$ 
is a convex function of Legendre type. When these conditions hold, 
$(C^*,h^*)$ is the Legendre conjugate of $(C,h)$, and $(C,h)$ is in turn the 
Legendre conjugate of $(C^*,h^*)$.  The gradient mapping is then one-to-one 
from the open convex set $C$ onto the open convex set $C^*$, continuous in 
both directions and 
$$
\nabla h^*=(\nabla h)^{-1}
\Eq(T.41)
$$
}
With this tool at our hands, let us define the function 
$\psi_{N,M}(x)\equiv \ff^*(x)-\LL_{N,M}(x)$. The crucial point is that since  
$\ff$ is of Legendre type, by Definition \ver.8 and Theorem \ver.10, 
we get
$$
\PPP_{N,M}(x)=\psi_{N,M}(\nabla \ff(x))
\Eq(T.42)
$$
Moreover, since $\nabla\ff$ is one-to-one and continuous, $\PPP_{N,M} $
 has a local 
extremum at $x^*$ if and only if $\psi_{N,M}$ has a local extremum at the 
point $y^*=\nabla \ff(x^*)$. In particular, $\nabla \psi_{N,M}(y^*)=0$.
Thus, $\nabla \ff^*(y^*)=\nabla \LL_{N,M} (y^*)$, and by \eqv(T.41),
$
(\nabla\ff)^{-1}(y^*)=\nabla \LL_{N,M} (y^*)
$, or $x^*=\nabla \LL_{N,M}(\nabla \ff(x^*)$, which was to be proven.
\endproof}


The proposition asserts that at the minima of $\PPP$, the condition of part
(ii) of Theorem \ver.5 is satisfied. Therefore, if we are interested in 
establishing localization properties of our measures, we  only need to compute 
$\PPP$ and work with it {\it as if it was the true rate function}. 
This will greatly
simplify the analysis in the models we are interested in.

\remark { {\it If} $\LL$ is of Legendre type, it follows by the same type of
argument that $x^*$ is a critical point of $\Psi$ if and only if
$\nabla \ff(x^*) $ is a critical point of $\psi$. Moreover,
at such critical points, $\Phi(x^*)=\psi(\nabla \ff(x^*))$. Thus  in this 
situation, if $x^*$ is a critical point of $\Psi$, than $x^*$ is 
a critical point of $\Phi$, and $\Psi(x^*)=\Phi(x^*)$. Conversely, by 
Proposition \ver.7, if $\Phi$ has a local extremum at $x^*$, then $x^*$ is a 
critical point of $\Psi$ and $\Phi(x^*)=\Psi(x^*)$.  
Since generally $\Psi(x)\geq \Phi(x)$, this implies also that if
$\Phi$ has a minimum at $x^*$, then $\Psi$ has a minimum at $x^*$.
One can build on the above observations and establish a more complete 
``duality principle'' between the functions $\PPP$ and $\PSS$ in great
generality, but we will not make use of these observations. The interested 
reader will find details in [G2].
}

\newpage

\chap{4. Bounds on the norm of random matrices}4

One of the crucial observations that triggered the recent progress 
in the Hopfield model was the observation that the properties of the 
random matrix $A(N)\equiv \frac {\xi^T\xi}N$ play a crucial r\^ole in this 
model, and that their main feature is that 
as long as $M/N$ is small, $A(N)$ is close to the identity matrix. This 
observation in a sense provided the proper notion for the intuitive 
feeling that in this case, ``all patterns are almost orthogonal to 
each other''. Credit must go to both Koch [K] and Shcherbina and Tirozzi [TS]
for making this observation, although the properties of the matrices $A(N)$
had been known a long time before. In fact it  is known
that under the hypothesis that $\xi_i^\mu$ are independent, identically 
distributed random variables with 
$\E \xi_i^\mu=0$, $\E\left[ \xi_i^\mu\right]^2=1$ and 
$\E\left[ \xi_i^\mu\right]^4<\infty$, the maximal and minimal eigenvalues 
of $A(N)$ satisfy
$$
\eqalign{
&\lim_{N\uparrow\infty}\l_{max}(A(N))=(1+\sqrt\a)^2, \text{a.s.}\cr
}
\Eq(M.1)
$$
This  statement was proven in [YBK] under the above (optimal) hypotheses.
For prior results under stronger assumptions, see [Ge,Si,Gi]. 
Such results are generally proven by tedious combinatorial methods,
combined with truncation techniques. Estimates for deviations
that were available from such methods give only subexponential estimates;
the best bounds known until recently, to our knowledge,  were due
to Shcherbina and Tirozzi [ST] and gave, in the case where $\xi_i^\mu$ 
are symmetric Bernoulli random variables
$$
\P\left[ \|A(N)-\1\|>[(1+\sqrt\a)^2-1](1+\e)\right]\leq \exp\left(-\frac
{\e^{4/3}M^{2/3}}{K}\right)
\Eq(M.2)
$$
with $K$ a numerical constant and valid for small $\e$. 
More recently, a bound of the form $\exp\left(-\frac {\e^2N}{K}\right)$
was proven by the authors in [BG5], using a concentration estimate 
due to Talagrand. In [T4] a simplified version of that proof is given.
We will now give the simplest proof of such a result we can think of.

Let us define for a $M\times M$-matrix $A$ the norm 
$$
\|A\|\equiv \sup_{{x\in \R^M}\atop{\|x\|_2=1}}(x,Ax)
\Eq(M.3)
$$
For positive symmetric matrices it is clear that $\|A\|  $ is the maximal
  eigenvalue of $A$. 
We shall also use the notation $\|A\|_2\equiv 
\sqrt{\sum_{\mu,\nu} A_{\mu\nu}^2}$. 

\theo {\ver.1} 
{\it Assume that  $\E \xi_i^\mu=0$, $\E\left[ \xi_i^\mu\right]^2=1$
and $|\xi_i^\mu|\leq 1$. Then there exists a numerical constant $K$ such that 
for large enough $N$, the following holds for all $\e\geq 0$ and all
 $\a\geq 0$ 
$$
\eqalign{
\P&\left[|\|A(N)\|-(1+\sqrt\a)^2|\geq \e\right]
\cr
&\leq K\exp\left(-N\frac {(1+\sqrt\a)^2}{K}
\left(\sqrt{\frac {\e}{1+\sqrt\a}+1}-1
\right)^2\right)
}
\Eq(M.4)
$$
}

\proof {Let us define for the  rectangular matrix $\xi$ 
$$
\|\xi\|_{+}\equiv \sup_{{x\in \R^M}\atop{\|x\|_2=1}}
\|\xi x\|_2
\Eq(M.5)
$$
Clearly
$$
\|A(N)\| = \|\xi/\sqrt N\|_{+}^2
\Eq(M.6)
$$
Motivated by this remark we show first that $\|\xi/\sqrt N\|_+$ has
nice concentration properties.  
For this we will use the following theorem due to Talagrand:
\theo {\ver.2} {{\rm (Theorem 6.6 in [T2])}  Let $f$ be a real valued 
function defined on $
[-1,1]^N$. Assume that for each real number $a$, the set $\{f\leq a\}$ is
convex. Suppose that on a convex set $B\subset [-1,1]^N$ the restriction
of $f$ to $B$ satisfies for all $x,y\in B$
$$
|f(x)-f(y)|\leq l_B\|x-y\|_2
\Eq(02.22)
$$
for some constant $l_B>0$.
Let $h$ denote the random variable $h=f(X_1,\dots, X_N)$.Then,
if $M_f$ is a median of $h$,
for all $t>0$,
$$
\P\left[|h-M_f|\geq t\right]
\leq 4b+\frac 4{1-2b} \exp\left(-\frac {t^2}{16 l_B^2}\right)
\Eq(02.23)
$$
where $b$ denotes the probability of the complement of the set $B$.}
To make use of this theorem, we show first that  $\|\xi/\sqrt N\|_+$ is a
Lipshitz function of the i.i.d.  variables $\xi_i^\mu$:
\lemma {\ver.3} {\it For any two matrices $\xi$, $\xi'$, we have that
$$
\left|\|\xi\|_+-\|\xi'\|_+\right|\leq \|\xi-\xi'\|_2
\Eq(M.7)
$$
}
\proof {We have
$$
\eqalign{
\left|\|\xi\|_+-\|\xi'\|_+\right|&\leq 
\sup_{{x\in \R^M}\atop{\|x\|_2=1}}
\left| \|\xi x\|_2 -\|\xi'x\|_2\right|\cr
&\leq 
\sup_{{x\in \R^M}\atop{\|x\|_2=1}}\|\xi x -\xi'x\|_2\cr
&\leq \sup_{{x\in \R^M}\atop{\|x\|_2=1}}
\sqrt{\sum_{i=1}^N x_i^2  \sum_{i=1}^N
\sum_{\mu=1}^M (\xi_i^\mu-{\xi'}_i^\mu)^2}=\|\xi-\xi'\|_2
}
\Eq(M.8)
$$
where in the first inequality we used that the modulus of the difference of
suprema is bounded by the supremum of the modulus of the 
differences, the second follows from the triangle inequality and the third
from the Schwarz inequality.
\endproof}
Next, note  that as a function of the variables $\xi\in [-1,1]^{MN}$,
$\|\xi\|_+$ is convex. Thus, 
by Theorem \ver.2, it follows that for all $t>0$,
$$
\P\left[|\|\xi/\sqrt N\|_+-\M_{\|\xi/\sqrt N\|_+}|\geq t\right]
\leq 4 e^{-N \frac {t^2}{16}}
\Eq(M.9)
$$
where $\M_{\|\xi/\sqrt N\|_+}$ is a median of $\|\xi/\sqrt N\|_+$.
Knowing that $\|A(N)\| $ converges almost surely to the values given 
in \eqv(M.1)
we may without harm replace  the median by this value.  Thus
$$
\eqalign{
&\P\left[ \|A(N)\|_+-(1+\sqrt\a)^2\geq \e\right]\cr
&=
\P\left[ \|\xi/\sqrt N\|_+-(1+\sqrt\a)\geq (1+\sqrt\a)\left(
\sqrt{1+\frac {\e}{(1+\sqrt\a)^2}}-1\right)\right]\cr
&\leq 4 \exp\left(-N (1+\sqrt\a)^2\left(
\sqrt{1+\frac {\e}{(1+\sqrt\a)^2}}-1\right)^2/16
\right)
}
\Eq(M.10)
$$
and similarly, for $0\leq \e\leq (1+\sqrt \a)^2$ 
$$
\eqalign{
&\P\left[ \|A(N)\|_+-(1+\sqrt\a)^2\leq  -\e\right]\cr
&=
\P\left[ \|\xi/\sqrt N\|_+-(1+\sqrt\a)\leq (1+\sqrt\a)\left(
\sqrt{1-\frac {\e}{(1+\sqrt\a)^2}}-1\right)\right]\cr
&\leq 4 \exp\left(-N (1+\sqrt\a)^2\left(
\sqrt{1-\frac {\e}{(1+\sqrt\a)^2}}-1\right)^2/16
\right)
}
\Eq(M.11)
$$
while trivially $\P\left[ \|A(N)\|_+-(1+\sqrt\a)^2\leq  -\e\right]=0$
for $ \e> (1+\sqrt \a)^2$. 
Using that for $0\leq x\leq 1$, $(\sqrt{1-x}-1)^2\geq (\sqrt {1+x}-1)^2$, 
we get Theorem 4.1. \endproof}

\remark{ Instead of using the almost sure results \eqv(M.1), it would also be 
enough to use estimates on the expectation of $\|A(N)\| $ to prove
Theorem 4.1. We see that the proof  required no computation 
whatsoever; it uses however that we know the medians or expectations. 
The boundedness condition on $\xi_i^\mu$ arises from the conditions 
in Talagrand's Theorem. It is likely that these could be relaxed. }

\remark {In the sequel of the paper we will always assume that our general 
assumptions on $\xi$ are such that Theorem \ver.1 holds. Of course, since 
exponential bounds are mostly not really necessary, one may also get away 
in more general situations. On the other hand, we shall see in Section 
6 that unbounded $\xi_i^\mu$ cause other problems as well.}

\newpage

\chap{5. Properties of the induced measures}5

In this section we collect the general results on the localization
(or concentration) of the induced measures in dependence on properties 
of the function $\Phi_{\b,N,M}$ introduced in  the previous section. There are 
two parts to this. Our first theorem will be a rather simple 
generalization to what could be called the ``Laplace method''. 
It states, roughly, 
the (hardly surprising) fact that the Gibbs measures are concentrated
``near'' the absolute minima of $\Phi$.  A second, and less trivial remark 
states that quite generally, the Gibbs measures  ``respect the symmetry
of the law of the disorder''. We will make precise what that means.

\newsubsec{\ver.1 Localization of the induced measures.}

The following Theorem will tell us what we need to know about 
the function $\Phi$ in order to locate the support of the limiting 
measures $\QQ$. 

\theo {\ver.1}{\it Let $\AA\subset \R^\infty$ be a set such that 
for all $N$ sufficiently large
the following holds:
\item{(i)} There is $n\in \AA$ such that for all $m\in \AA^c$,
$$
\Phi_{\b,N,M(N)}[\o](m)-\Phi_{\b,N,M(N)}[\o](n)\geq C \a
\Eq(C.2)
$$
for $C>c$ sufficiently large, with $c$ the constant from (i) of Theorem 3.5. 
\item{(ii)}  $\D\LL_{N,M}(\nabla E_M(n))\leq K M$ for some $K<\infty$, and
$B_{K\sqrt\a}(n)\subset \AA$.
%
Assume further that  $\Phi$ satisfies a tightness condition, i.e. 
there exists a constant, $a$, sufficiently small (depending on $C$),
 such that for all $r>C\a$ 
$$
\ell\left(\{m\,|\Phi_{\b,M,N}[\o](m)-\Phi_{\b,M,N}[\o](n)\leq r\}\right)
\leq r^{M/2} a^M M^{-M/2}
\Eq(C.1)
$$
where $\ell(\cdot)$ denotes the Lebesgue measure. 
Then there is $L>0$ such that 
$$
\QQ_{\b,N,M(N)}[\o]\left(\AA^c\right)\leq e^{-L\b M}
\Eq(C.3bis)
$$
and in particular 
$$
\lim_{N\uparrow\infty}\QQ_{\b,N,M(N)}[\o]\left(\AA\right)=1
\Eq(C.3)
$$
}

\remark {Condition \eqv(C.1) is verified, e.g. if $\Phi$ is bounded from below
by a quadratic function.}

\proof { To simplify notation, we put w.r.g. $ \Phi_{\b,N,M}[\o](n)=0$. 
Note first that by (ii) and \eqv(T.71)  we have that (for suitably chosen 
$\rho$) 
$$
\QQ_{\b,N,M(N)}[\o]\left(\AA\right)\geq \frac 1{Z_{\b,N,M(N)}[\o] }
\frac 12 e^{-\b N \Phi_{\b,N,M}[\o](n)}= \frac1{2Z_{\b,N,M(N)}[\o] }
\Eq(C.4)
$$
It remains to show that the remainder has much smaller mass. Note that
obviously, by (i),
$$
\eqalign{
\QQ_{\b,N,M(N)}[\o]\left(\AA^c\right)
&\leq \int_{C\a}^\infty dr
\QQ_{\b,N,M(N)}[\o]\left(\AA^c\cap\{m\,|\Phi_{\b,N,M}(m)= r\}\right)
\cr
&\leq  \int_{C\a}^\infty dr
\QQ_{\b,N,M(N)}[\o]\left(\{m\,|\Phi_{\b,N,M}(m)= r\}\right)
}
\Eq(C.5)
$$
Now we introduce a lattice $\WW_{M,\a}$ of spacing $1/\sqrt N$ in $\R^M$. 
The point here is that any domain $D\subset \R^M$ is covered by the union of
balls of radius $\sqrt \a$ centered at the lattice points in $D$, while the
number of lattice points in  any reasonably regular set 
$D$ is smaller than $\ell(D)N^{M/2}$ (see e.g. [BG5] for more details).
Combining this observation with the upper bound \eqv(T.70), we get from 
\eqv(C.5) that
$$
\eqalign{
&Z_{\b,N,M(N)}[\o]
\QQ_{\b,N,M(N)}[\o]\left(\AA^c\right)\cr
&\leq 
 \int_{C\a}^\infty dr 
e^{-\b N r } \ell\left(\{m\,|\Phi_{\b,N,M}(m)\leq r\}\right) N^{M/2}
e^{\b M c/2}\cr
&\leq \int_{C\a}^\infty dr 
e^{-\b N r } r^{M/2}  a^M \a^{-M/2}e^{\b \a c/2} \cr
&\leq a^Me^{\b M c/2} \a\int_C^\infty dr e^{-\b Mr}r^{M/2}\cr
&\leq a^Me^{\b \a c/2}e^{-\b M C/2}\a\int_C^\infty dr e^{-\b Mr/2}r^{M/2}\cr
& e^{-\b M [C/2-c/2- \ln a/\b]} N^{-1}\left[\frac {2}{e\b}\right]^{M}
}
\Eq(C.6)
$$
which clearly for $\b\geq 1$ can be made exponentially small in $M$ 
for $C$ sufficiently large. Combined with \eqv(C.4) this proves \eqv(C.3bis).
\eqv(C.3) follows by a standard Borel-Cantelli argument. \endproof}

\remark{ We see at this point why it was important to get the error terms
of order $\rho^2$ in the upper bound of Theorem 3.5; this allows us to choose
$\rho\sim \sqrt\a$. otherwise, e.g. when we are in a situation where we want 
use  Theorem 5.6, we could of course choose $\rho$ to be some higher power of
$\a$, e.g. $\rho=\a$. This then introduces an extra factor $e^{M|\ln \a|}$, 
which can be offset only by choosing $C\sim |\ln \a|$, which of course implies 
slightly worse estimates on the sets where $\QQ$ is localized. }

\newsubsec{\ver.2 Symmetry and concentration of measure.}

Theorem \ver.1 allows us to localize the measure near the 
``reasonable candidates'' for the absolute minima of $\Phi$. As we will see,
frequently, and in particular in the most interesting situation 
where we expect a {\it phase transition}, the smallest set
$\AA$ satisfying the hypothesis of Theorem \ver.1 we can find will still be 
a union of disjoint sets. The components of this set are typically linked by 
``symmetry''.  In such a situation we would like to be able to compare the 
exact mass of the individual components, a task that goes beyond the 
possibilities of the explicit large deviation estimates. It is the idea of 
concentration of measure that allows us to make use of the symmetry of the 
distribution $\P$ here. This fact was first noted in [BGP3], and a more 
elegant proof in the Hopfield model that made use of the Hubbard-Stratonovich 
transformation was given first in [BG5] and independently in [T4]. 

Here we give a very simple proof that works in more general situations.
The basic problem we are facing is the following. Suppose we are in a situation where the set $\AA$ from Theorem \ver.1 can be decomposed as
$\A=\cup_k\AA_k$ for some collection of disjoint sets $\AA_k$.
 Define 
$$
f_N[\o](k)\equiv -\frac 1{\b N}\ln \E_\s e^{-\b  H_{N,M}[\o](\s)}
\1_{\{m_{N,M}[\o](\s)\in \AA_k\}}
\Eq(C.7)
$$
Assume that by for all $k$
$$
\E f_N[\o](k)=\E f_N[\o](1)
\Eq(C.8)
$$
(Think of $\AA_k= B_\rho(m^* e^k)$ in  the standard Hopfield model).
We want so show that this implies that for all $k$,
 $|f_N[\o](k)- f_N[\o](1)|$ is ``small'' with large probability. Of course we 
should show this by proving that each $ f_N[\o](k)$ is close to its mean, 
and such a result is typically given by concentration 
estimates. To prove this would be easy, if it were not for the
indicator function in \eqv(C.7), whose argument depends on the random 
parameter $\o$ as well as the Hamiltonian. 
Our strategy will be to introduce quantities  $f^\e_N(k)$ that are close to 
$f_N(k)$, and for which it is easy to prove the concentration 
estimates. We will then control the difference between $f^\e_N(k)$ and
$f_N(k)$. 
We set
$$
f^\e_N(k)\equiv  -\frac 1{\b N}\ln \E_\s 
\left(\sfrac{\b N}{2\pi\e}\right)^{M/2} \int_{\AA_k}\, dm
\, e^{-\frac {\b N}{2\e} \|m_{N,M}(\s)-m\|_2^2} e^{\b N E_M(m)}
\Eq(C.8bis)
$$
Note that the idea is that 
$\left(\sfrac{\b N}{2\pi\e}\right)^{M/2} e^{-\frac {\b N}{2\e} \|m_{N,M}(\s)-m\|_2^2}$
converges to the Dirac distribution concentrated on $m_{N,M}(\s)$, so that 
 $f^\e_N(k)$ converges to $f_N(k)$ as $\e\downarrow 0$. Of course we will have
to be a bit more careful than just that. However, Talagrand's 
Theorem 6.6 of [T2] 
gives readily 

\proposition {\ver.2} {\it Assume that $\xi$ verifies the assumptions of 
Theorem 4.1 and $\SS$ is compact. Then 
there is a finite universal constant $C$ such that 
for all $\e>0$, 
$$
\P\left[\left|f^\e_N(k)-\E f^\e_N(k)\right|>x\right]\leq
Ce^{-M}+C e^{-\frac {x^2\e^2 N}C}
\Eq(C.9)
$$
}

\proof {We must establish a Lipshitz bound for $f^\e_N[\o](k)$. For notational 
simplicity we drop the superfluous indices $N$ and $k$ and set
$f^\e[\o]\equiv f^\e_N[\o](k)$. Now 
$$
\eqalign{
&\left|f^\e[\o]-f^\e[\o']\right|\cr
&=\frac 1{\b N}\left|\ln \frac
{\E_\s\int_{\AA_k}\, dm
\, e^{-\frac {\b N}{2\e} \|m_{N,M}[\o](\s)-m\|_2^2} e^{\b N E_M(m)}}
{\E_\s\int_{\AA_k}\, dm
\, e^{-\frac {\b N}{2\e} \|m_{N,M}[\o'](\s)-m\|_2^2} e^{\b N E_M(m)}}
\right|\cr
&=
\frac 1{\b N}\Biggl|\ln \frac
{\E_\s\int_{\AA_k}\, dm
\, e^{-\frac {\b N}{2\e} \|m_{N,M}[\o'](\s)-m\|_2^2} e^{\b N E_M(m)}}
{\E_\s\int_{\AA_k}\, dm
\, e^{-\frac {\b N}{2\e} \|m_{N,M}[\o'](\s)-m\|_2^2} e^{\b N E_M(m)}}\cr
&\quad\quad\quad \frac{
 e^{-\frac {\b N}{2\e}\left( \|m_{N,M}[\o](\s)-m\|_2^2-
 \|m_{N,M}[\o'](\s)-m\|_2^2\right)}
}{ \text{}}
\Biggr|\cr
&\leq \frac 1{\e} \sup_{\s\in \SS^N, m\in \AA_k}
\left| \|m_{N,M}[\o](\s)-m\|_2^2-
 \|m_{N,M}[\o'](\s)-m\|_2^2\right|
}
\Eq(C.10)
$$
But 
$$
\eqalign{
&\left| \|m_{N,M}[\o](\s)-m\|_2^2-
 \|m_{N,M}[\o'](\s)-m\|_2^2\right|\cr
&\leq 
\|m_{N,M}[\o'](\s)-m_{N,M}[\o](\s)\|_2
\|2m -m_{N,M}[\o](\s)-m_{N,M}[\o'](\s)\|_2
\cr
&\leq \frac 1{\sqrt N} \|\xi[\o']-\xi[\o]\|_2 \left[R+ c(\sqrt{\|A[\o]\|}+
\sqrt{\|A[\o']\|})\right]
}
\Eq(C.11)
$$
where $R$ is a bound for $m$ on $\AA_k$. We wrote $A[\o]\equiv 
\frac {\xi^T[\o]\xi[\o]}N$ to make the dependence of
the random matrices on the random parameter explicit.
 Note that this estimate is uniform in
$\s$ and $m$. It is easy to see that $f^\e[\o]$ has convex level sets so
 that the 
assumptions of Theorem 6.6 of [T1] are verified.
 Proposition \ver.2 follows from here and the bounds on 
$\|A[\o]\|$ given by Theorem 4.1. 
\endproof}

We see from Proposition \ver.2 that we can choose an $\e= N^{-\d_1}$, and an 
$x=N^{-\d_2}$ with $\d_1,\d_2>0$ and still get a 
probability that decays faster 
than any power with $N$. 

Let us now see more precisely 
 how $f^\e[\o]$ and $f^{0}[\o] $ are related. Let us introduce as an 
intermediate step the $\e$-smoothed measures
$$
\tilde \QQ^\e_{\b,N,M}[\o]\equiv  \QQ_{\b,N,M}[\o]\star 
\NN\left(0,\frac \e{\b N}\right)
\Eq(C.12)
$$
where $\NN\left(0,\frac \e{\b N}\right)$ is a $M$-dimensional normal 
distribution with mean $0$ and variance $\frac \e{\b N}\1$. 
We mention that in the case $E_M(m)=\frac 12 \|m\|_2^2$, 
the choice $\e=1$ is particularly convenient. This convolution is then known 
as the ``Hubbard-Stratonovich transformation'' [HS]. Its use simplifies to
 some extent that particular case and has been used frequently, by us as well
 as other authors. It allows to avoid the complications of Section 3 
altogether. 

We set $\tilde f^\e[\o]\equiv -\frac 1{\b N} \ln\left(Z_{\b,N,M}\tilde 
\QQ^\e_{\b,N,M}(\AA_k)\right)$. 
But
$$
\eqalign{
&Z_{\b,N,M}\tilde 
\QQ^\e_{\b,N,M}(\AA_k)\cr
&= \E_\s 
\left(\sfrac{\b N}{2\pi\e}\right)^{M/2} \int_{\AA_k}\, dm
\, e^{-\frac {\b N}{2\e} \|m_{N,M}(\s)-m\|_2^2} e^{\b N E_M(m_{N,M}(\s))}\cr
&=\E_\s 
\left(\sfrac{\b N}{2\pi\e}\right)^{M/2} \int_{\AA_k}\, dm \1_{\{\|m_{N,M}(\s)-m\|_2\leq \d\}}
\, e^{-\frac {\b N}{2\e} \|m_{N,M}(\s)-m\|_2^2} \cr
&\quad\quad\times e^{\b N E_M(m_{N,M}(\s))}\cr
&+\E_\s 
\left(\sfrac{\b N}{2\pi\e}\right)^{M/2} \int_{\AA_k}\, 
dm \1_{\{\|m_{N,M}(\s)-m\|_2> \d\}}
\, e^{-\frac {\b N}{2\e} \|m_{N,M}(\s)-m\|_2^2}\cr
&\quad\quad\times  e^{\b N E_M(m_{N,M}(\s))}
\cr&\equiv (I)+(II)
}
\Eq(C.13)
$$
for $\d>0$ to be chosen. We will assume that on $\AA_k$, $E_M$ is uniformly 
Lipshitz for some constant $C_L$. Then 
$$
\eqalign{
(I)&\leq e^{+\b NC_L\d}
\E_\s 
\left(\sfrac{\b N}{2\pi\e}\right)^{M/2} \int_{\AA_k}\, dm 
\, e^{-\frac {\b N}{2\e} \|m_{N,M}(\s)-m\|_2^2} e^{\b N E_M(m)}\cr
\cr 
&=e^{\b NC_L\d}e^{-\b N f^\e[\o]}
}
\Eq(C.14)
$$
and 
$$
\eqalign{
(II)&\leq \E_\s 
2^{M/2} \left(\sfrac {\b N}{4\pi\e}\right)^{M/2} e^{-\frac {\b N}{4\e}\d^2}\cr
&\times
 \int_{\AA_k}\, dm 
\, e^{-\frac {\b N}{4\e} \|m_{N,M}(\s)-m\|_2^2} e^{\b N E_M(m_{N,M}(\s))}\cr
&\leq 2^{M/2} e^{-\frac {\b N}{4\e}\d^2} \E_\s  e^{\b N E_M(m_{N,M}(\s))} 
\left(\sfrac {\b N}{4\pi\e}\right)^{M/2}\cr
&\times
 \int \, dm e^{-\frac {\b N}{4\e} \|m_{N,M}(\s)-m\|_2^2}
= 2^{M/2} e^{-\frac {\b N}{4\e}\d^2} Z_{\b,N,m}
}
\Eq(C.15)
$$
In quite a similar way we can also get a lower bound on $(I)$, namely
$$
\eqalign{
(I)&\geq e^{-\b NC_L\d} 
\E_\s 
\left(\sfrac{\b N}{2\pi\e}\right)^{M/2} \int_{\AA_k}\, dm 
\, e^{-\frac {\b N}{2\e} \|m_{N,M}(\s)-m\|_2^2} e^{\b N E_M(m)}\cr
&- e^{-\b NC_L\d} 2^{M/2}  e^{-\frac {\b N}{4\e}\d^2} e^{\b N\sup_{m\in \AA_k}
E_M(m)}\cr
&= e^{-\b NC_L\d}e^{-\b N f^\e[\o]} - e^{-\b NC_L\d} 2^{M/2}  
e^{-\frac {\b N}{4\e}\d^2} e^{\b N\sup_{m\in \AA_k}
E_M(m)}
}
\Eq(C.16)
$$
Since we anticipate that $\e=N^{-\d_1}$, the second term in \eqv(C.16)
is negligible 
compared to the first, and (II) is negligible compared to (I), with room even 
to choose $\d$ tending to zero with $N$; e.g., if we choose $\d=\e^{1/4}$,
we get that
$$
|\tilde f^\e[\o]-f^\e[\o]|\leq const. \e^{1/4}
\Eq(C.17)
$$
for sufficiently small $\e$. (We assume that $|f^\e[\o]|\leq C $).

Finally we must argue that $\tilde f^\e[\o]$ and $f^0[\o]$ differ only by 
little. This follows since $\NN(0,\frac {\b N}\e)$ is sharply concentrated
on a sphere of radius $\frac {\e\a}\b$ (although this remark alone would be 
misleading). In fact, arguments quite similar to those that yield
\eqv(C.17)(and that we will not reproduce here)
give also
$$
|\tilde f^\e[\o]-f^0[\o]|\leq const. \e^{1/4}
\Eq(C.18)
$$
Combining these observations with Proposition \ver.2
gives 

\theo {\ver.3} {\it  Assume that $\xi$ verifies the assumptions of 
Theorem 4.1 and $\SS$ is compact. Assume that $\AA_k\subset \R^M$  verifies 
$$
\QQ_{\b,N,M}[\o](\AA_k)\geq e^{-\b N c }
\Eq(C.19) 
$$
for some finite constant $c$, with probability greater than $1-e^{-M}$. 
Then there is a finite  constant $C$ such that for $\e>0$ small enough 
for any $k,l$,
$$
\P\left[|f_N(k)-f_N(l)|\geq C\e^{1/4} +x\right]\leq Ce^{-M}
+C e^{-\frac {x^2\e^2 N}C}
\Eq(C.20)
$$
}

\newpage
\def\vo{\hat v}
\chap{6. Global estimates on the free energy function}6

After the rather general discussion in the last three sections, we see
that all results on a specific model depend on the analysis of the
(effective) rate function $\Phi_{\b,N}[\o](x)$.  The main idea we
want to follow here is to divide this analysis in two steps:

\item{(i)} Study the average $\E \Phi_{\b,N}[\o](x)$ and obtain explicit 
bounds from which the locations of the global  minima can be read off. 
This part is typically identical to what we would have to do 
in the case of finitely many patterns.

\item{(ii)} Prove that with large probability, $|\Phi_{\b,N}[\o](x)-\E
\Phi_{\b,N}[\o](x)|$ is so small that the deterministic result from 
(i)  holds essentially outside small balls around the locations of the 
minima  for $\Phi_{\b,N}[\o](x)$ itself. 

These results then suffice to use Theorems 5.1 and 5.3 in order to construct 
the limiting induced measures. 
The more precise analysis of $\Phi$ close to the minima is of interest in its 
own right and will be discussed in the next section. 

We mention that this strict separation into two steps was not followed in 
[BG5]. However, it appears to be the most natural and reasonable 
procedure. Gentz [G1] used this strategy in her proof of the central limit
theorem,  but only in the regime 
$M^2/N\downarrow 0$. To get sufficiently good estimates when $\a>0$, 
a sharper analysis is required in part (ii). 

To get explicit  results, we will from now on work in a more restricted class
of examples that includes the Hopfield model. We will take $\SS=\{-1,1\}$, 
with $q(\pm 1)=1/2$ and 
$E_M(m)$ of the form
$$
E_M(m)\equiv\frac{1}{p}\|m\|_p^p
\Eq(X.1)
$$    
with $p\geq 2$ and we will only require of the variables $\xi_i^\mu$
to have mean zero, variance one and to be bounded. To simplify notation,
we assume $|\xi_i^\mu|\leq 1$. 
We do not strive to get optimal estimates on constants in this generality,
but provide all the tools necessary do so in any specific situation, if 
desired\note{A word of warning is due at this point. We will treat these
 generalized models assuming always  $\scriptstyle M=\a N$. 
But from the memory point 
of view, these models should and do work with 
$\scriptstyle M=\a N^{p-1}$ (see e.g. 
[Ne] for a proof in the context of storage capacity). For $\scriptstyle 
p>2$ our approach 
appears perfectly inadequate to deal with so many patterns, as the 
description of system in terms of so many variables (far more than the 
original spins!) seems quite absurd. Anyhow, there is some fun in these models
even in this more restricted setting, and since this requires only a little 
more work, we decided to present those results.}.

A simple calculation shows that the function of Theorem 3.5 defined in \eqv(T.33)
in this case is
given by (we make explicit reference to $p$ and $\b$, but drop the $M$)
$$
\Phi_{p,\b, N}[\o](m)=\frac{1}{q}\|m\|_p^p
-\frac{1}{\b N}\sum_{i=1}^N\ln\cosh\left(
\b\sum_{\mu=1}^M \xi_i^{\mu}[\o]|m_\mu|^{p-1}\hbox{sign}(m_\mu)
\right)
\Eq(X.2)
$$
where $\frac 1p+\frac 1q=1$.\note{Throughout this section, $\scriptstyle q$ 
will stand for 
the conjugate of $\scriptstyle p$.}
Moreover 
$$
\Phi_{p,\b, N}[\o](m)=\left(
\psi_{q,\b, N}[\o]\circ\nabla E_M
\right)(m)
\Eq(X.3)
$$
where $\psi_{q,\b, N}[\o]$ : $\R^M\rightarrow \R$ is given by
$$
\psi_{q,\b, N}[\o](x)=\frac{1}{q}\|x\|_q^q
-\frac{1}{\b N}\sum_{i=1}^N\ln\cosh\left(
\b(\xi_i[\o],x)
\right)
\Eq(X.4)
$$
and $\nabla E_M$ :  $\R^M\rightarrow \R^M$, by
$$
\nabla E_M(m)=\left(\nabla_1 E_M(m_1),\dots,\nabla_{\mu} E_M(m_{\mu}),\dots,
\nabla_M E_M(m_M)\right)
\Eq(X.4bis)
$$ 
where
$$
\nabla_{\mu} E_M(m_{\mu})=\sign(m_{\mu})|m_\mu|^{p-1}
\Eq(X.4ter)
$$
Since $\nabla_{\mu}E_M$ is a continuous and strictly increasing function
going to $+\infty$, resp. $-\infty$, as $m_{\mu}$ goes to $+\infty$, resp. 
$-\infty$, (and being zero at $m_{\mu}=0$) its inverse 
$\nabla_{\mu}E_M^{-1}$ exists and has the same properties as  
$\nabla_{\mu} E_M$. It is thus enough, in order to study the structure of 
the minima of $\Phi_{p,\b, N}[\o]$, to study that of $\psi_{q,\b, N}[\o]$.

Before stating our main theorem we need to make some comments on the 
generalized Curie-Weiss functions
$$
\phi_{q,\b}(z)= \frac 1q |z|^q-\frac 1\b \ln\cosh(\b z)
\Eq(X.7a)
$$
The  standard Curie-Weiss case $q=2$ is well documented (see e.g. 
[El]), but the general situation can be analyzed in the same way.
In a quite general setting, this can be found in [EE]. 
A new feature for $q<2$ is that now zero is always a {\it local} 
minimum and that there is a range of temperatures where three local
minima exist while the absolute minimum is the one at zero. For sufficiently
low temperatures, however, the two minima away from zero are always the 
lowest ones. The critical temperature $\b_c$ is defined as the one where 
$\phi_{q,\b}$ takes the same value at all
three local minima. Thus a particular feature for all $q<2$ is that   
for $\b\geq \b_c$, the position of the 
deepest minimum, $x^*(\b)$, satisfies $x^*(\b)\geq x_q^*(\b_c)>0$. 
Of course $x^*_q(\b_c)$ tends to 
$0$ as $q$ tends to $2$. For integer $p\geq 3$ we have thus the situation
that $x^*(\b)= O(1)$, and only in the case $p=2$ do we have to
take the possible smallness of $x^*(\b)$ near the critical point into account.


\proposition {\ver.1} {\it Assume that $\xi_i^\mu$ are i.i.d., symmetric 
bounded random variables with variance $1$. 
Let either $p=2$ or $p\geq 3$. Then for all $\b>\b_c(p)$ there exists a 
strictly positive constant $C_p(\b)$ and a subset $\O_1\subset\O$ with 
$\P[\O_1]\geq 1-O(e^{-\a N})$ such that for all
$\o\in \O_1$ the following holds for 
all $x$ for which $x_\mu=sign(m_\mu)|m_\mu|^{p-1}$ with $\|m\|_2\leq 2$:
\break
 There is  
$\g_a>0$ and a finite 
numerical constant
 $c_1$    such that for all $\g\leq\g_a$ if $\inf_{s,\mu}\|x-se^\mu x^*_q\|_2
\geq \g c_1 x^*_q$,
$$ 
\psi_{p,\b,N}[\o](x)-\frac 1q 
(x^*_q)^q + \frac{1}{\b}\ln\cosh(\b x^*_q)
\geq C_p(\b)\inf_{s,\mu}\|x-se^\mu x_q^*\|_2^2
\Eq(X.t1)
$$
where $C_2(\b)\sim (m^*(\b))^2$ as $\b\downarrow 1$, and $C_p(\b)\geq C_p>0$
for  $p\geq 3$. The infima are over $s\in \{-,1,+1\}$ and $\mu=1,\dots, M$.
}

\remark{ Estimates on the various constants can be collected from the proofs.
In case (i), $C_2(\b)$ goes like $10^{-5}$, and  $\g_a\sim 10^{-8}$ and
$c_1\sim 10^{-7}$. These numbers are of course embarrassing. 
}

 From Proposition \ver.1 one can immediately deduce localization 
properties of
the Gibbs measure with the help of the theorems in Section 5. 
In fact one obtains

\theo {\ver.2} {\it  Assume that $\xi_i^\mu$ are i.i.d.
Bernoulli random variables taking the values $\pm 1$ with equal probability. 
Let either $p=2$ or $p\geq 3$. Then there exists a 
finite constant $c_p$  such that for all $\b>\b_c(p)$   
there is subset $\O_1\subset\O$ with 
$\P[\O_1]\geq 1-O(e^{-\a N})$ such that for all
$\o\in \O_1$ the following holds:
\item{(i)} In the case $p=2$,
$$
\QQ_{\b,N,M(N)}[\o]\left(\cup_{s,\mu} B_{c_2 \g m^*}(se^\mu m^*)\right)
\geq 1-\exp\left(-KM(N)\right)
\Eq(X.t2)
$$
\item{(ii)} In the case $p\geq 3$,
$$
\eqalign{
&\QQ_{\b,N,M(N)}[\o]\left(\cup_{s,\mu} \left\{ m\in \R^M\,| x(m)\in  
B_{c_p \a|\ln \a| }(se^\mu x^*_q)\right\}\right)\cr
&\quad\geq  
1-\exp\left(-KM(N)\right)
}
\Eq(X.t3)
$$
Moreover, for $h=  \e s e^\mu$, and any $\e>0$, for $p=2$
$$
\QQ^h_{\b,N,M(N)}[\o]\left( B_{c_2 \g m^*}(se^\mu m^*)\right)
\geq 1-\exp\left(-K(\e)M(N)\right)
\Eq(t4)
$$
and for $p\geq 3$,
$$
\eqalign{
&\QQ^h_{\b,N,M(N)}[\o]\left( \left\{ m\in \R^M\,| x(m)\in  
B_{c_p \a|\ln \a| }(se^\mu x^*_q)\right\}\right)\cr
&\quad\geq  
1-\exp\left(-K(\e)M(N)\right)
}
\Eq(X.t5)
$$
with $K(\e)\geq const. \e>0$. 
}

\remark{ Theorem \ver.2 was first proven, for the case $p=2$,
 with imprecise estimates
on the radii of the balls in [BGP1,BGP3]. The correct asymptotic behaviour 
(up to constants) given here was proven first in [BG5]. A somewhat different 
proof was given recently in [T4], after being announced in [T3] (with 
additional restrictions on $\b$).
The case $p\geq 3$ is new. It may be that the $|\ln \a|$ in the estimates 
there can be avoided. We leave it to the reader to deduce Theorem \ver.2 from 
Proposition \ver.1 and Theorems 5.1 and 5.3. In the case $p\geq 3$, Theorem 
3.6 and  the remark following the proof of Theorem 5.1 should be kept in mind.
}
 
\proofof{Proposition \ver.1}{
We follow our basic strategy to show first that the 
mean of $\psi_{q,\b,N}[\o]$ 
has the desired properties and to control the fluctuations via concentration 
estimates. 
We rewrite $\psi_{q,\b, N}[\o](x)$ as
$$
\eqalign{
\psi_{q,\b, N}[\o](x)=&
+\E\left\{\frac{1}{q}|(\xi_1,x)|^q-\frac{1}{\b}\ln\cosh(\b(\xi_1,x))
\right\}\cr
&+\frac{1}{q}\|x\|_q^q-\frac{1}{q}\E|(\xi_1,x)|^q
\cr
&+\frac{1}{\b N}\sum_{i=1}^N\left\{
\E\ln\cosh(\b(\xi_i,x))-\ln\cosh(\b(\xi_i,x))
\right\}\cr
}
\Eq(X.5)
$$
We will study the first, and main, term in a moment. The middle term 
``happens'' to 
be positive:
\lemma{\ver.3}{\it Let $\{X_j, j=1,\dots,n\}$ be i.i.d. random variables with
$\E X_i=0$, $\E X_i^2=1$, and let $x=(x_1,\dots,x_n)$ be a vector in $\R^n$. 
Then, for $1<q\leq 2$,
$$
\|x\|_q^q-\E|\sum_{j=1}^n x_jX_j|^q\geq 0
\Eq(X.6)
$$
Equality holds if all but one component of the $x_j$ are zero.
}
\proof{ A straightforward application of the H\"older inequality yields
$$
\E|\sum_{j=1}^n x_jX_j|^q\leq (\E|\sum_{j=1}^n x_jX_j|^2)^\frac{q}{2}
=\|x\|_q^q
\Eq(X.7)
$$
\endproof}
Let us now consider the first term in \eqv(X.5). 
For $q=2$ we have from [BG5] the following bound:
Let 
$$
\hat c(\b)\equiv \frac {\ln\cosh(\b x^*)}{\b (x^*)^2}-\frac 12
\Eq(3.4)
$$
Then for all  $\b>1$ and for all $z$
$$
\phi_{2,\b}(z)-\phi_{2,\b}(x^*)\geq \hat c(\b) (|z|-x^*)^2
\Eq(3.05)
$$
Moreover 
$\hat c(\b) $ tends to $\frac 12$ as $\b\uparrow 
\infty$, and behaves like $\frac 1{12} (x^*(\b))^2$, as $\b\downarrow 1$.
\proposition {\ver.4} {\it Assume 
that $\xi_1^\mu$ are i.i.d., symmetric and 
$\E(\xi_1^\mu)^2=1$ and $|\xi_1^\mu|\leq 1$. Let either 
$p=2$ or $p\geq 3$. Then for all $\b>\b_c$ (of $p$)  
there exists a positive constant $C_q(\b)$ such that 
for all $x$ such that $x_\mu=sign(m_\mu)|m_\mu|^{p-1}$ with
$\|m\|_2\leq 2$,
$$
\eqalign{
& \E\left\{\frac{1}{q}|(\xi_1,x )|^q-\frac{1}{\b}\ln\cosh(\b(\xi_1,x))
\right\}-\frac 1q (x^*)^q + \frac{1}{\b}\ln\cosh(\b x^*)
\cr
&
\quad\quad\geq C_q(\b) 
\inf_{\mu,s}\|x-se^\mu x^*\|_2^2
}\Eq(X.7b)
$$
where $x^*$ is the largest solution of the equation 
$x^{q-1}=\tanh \b x$.
In the case $q=2$ $C_2(\b)=\frac 1{5000
}\left(\frac {\ln\cosh(\b x^*)}
{\b (x^*)^2}-\frac 12\right)
\approx \frac 1{600000  
} (x^*)^2$ for $\b\downarrow 1$. 
}
\remark {Note that nothing depends on $\a$ in this proposition.
The constants appearing here are quite poor, but the proof is fairly
 nice and universal. In a very recent paper [T4] has a similar result where
the constant seems to be $1/256 L$, but so far we have not been able to 
figure out what his estimate for $L$ would be. Anyway, there are other options
if the proof below is not to your taste! }
\proof {It is not difficult to convince oneself of the fact that there exist
positive constants $\tilde C_q(\b) $ such that for all $Z=(\xi_1,x)$ satisfying the assumption of the proposition 
$$
\frac{1}{q}|Z|^q-\frac{1}{\b}\ln\cosh(\b Z)
-\frac 1q (x^*)^q + \frac{1}{\b}\ln\cosh(\b x^*)
\geq \tilde C_q(\b)\left(|Z|-x^*\right)^2
\Eq(X.7c)
$$
For $q=2$ this follows from Lemma \eqv(3.05).  For $q\geq 3$, note first
that the allowed $|Z|$ are bounded. Namely,
$$
|(\xi_1,x)|\leq \left|\sum_{\mu=1}^M |\xi_1^\mu|| m_\mu^{p-1}|\right|
\leq \sqrt{\sum_\mu m_\mu^2 
}\sqrt{\sum_{\mu}|m_\mu|^{2(p-2)}}
\leq 
 \|m\|_2^2
\Eq(X.7d)
$$
using that $\|m\|_\infty \leq 
1
$ and the H\"older inequality in the case 
$p>3$. 
Moreover 
since by definition $\pm x^*$ are the only points where the function
 $\phi_{q,\b}(z)$ takes its absolute minimum, and $x^*$ is uniformly bounded
 away from
$0$, it is clear that a lower bound of the form \eqv(X.7c) can be constructed 
on the  bounded interval $[-2,2]$.
\hfill\break
We have to bound the expectation of the right hand side of  \eqv(X.7c).
\lemma{\ver.5} {\it Let $Z=X+Y$ where $X,Y$ are independent
real valued random variables.
  Then for any $\e>0$
$$
\eqalign{
\E (|Z|-x^*)^2 
&\geq \frac 12\left(\sqrt{\E Z^2}-x^*\right)^2 +
\frac 12\e^2 \P[|X|>\e]\cr
&\times\min\left(
\P[Y>\e],\P[Y<-\e]\right)
}
\Eq(X.Y1)
$$
}
\proof { First observe that, since  $\E|Z|\leq\sqrt{\E Z^2}$,
$$
\eqalign{
\E (|Z|-x^*)^2 &= \left(\sqrt{\E Z^2}-x^*\right)^2+2x^*\E\left(\sqrt{\E Z^2}
-|Z|\right) \cr
&\geq  \left(\sqrt{\E Z^2}-x^*\right)^2
}
\Eq(X.Y2)
$$
On the other hand, Tchebychev's 
inequality gives that for any positive $\e$,
$$
\E\left(|Z|-x^*\right)^2 \geq \e^2
\P\left[\left||Z|-x^*\right|>\e\right]
 \Eq(X.Y3)
$$
Now it is clear that if $|X|>\e$, then   $||X+Y|-x^*|>\e$ either if $Y>\e$ or
if $Y<-\e$ (or in both cases). This gives the desired estimate.
Thus \eqv(X.Y3) implies that
$$
\E\left(|Z|-x^*\right)^2\geq\e^2 \P[|X|>\e]\min\left(
\P[Y>\e],\P[Y<-\e]\right)
\Eq(X.Y4)
$$
 \eqv(X.Y2) and \eqv(X.Y4) together imply \eqv(X.Y1).\endproof}
%
%
%
In the case of symmetric random variables, the estimate simplifies to
$$
\E (|Z|-x^*)^2 
\geq \frac 12\left(\sqrt{\E Z^2}-x^*\right)^2 +\frac 14 \e^2 \P[|X|>\e]
\P[|Y|>\e] 
\Eq(X.Y1a)
$$
which as we will see is more easy to apply in our situations.
In particular, we have the following estimates.
\lemma {\ver.6}{\it Assume that $X=(x,\xi)$ where $|\xi^\mu|\leq 1$,
$\E\xi^\mu=0$
and $\E(\xi^\mu)^2=1$.
Then for any $1>g>0$, 
$$
\P\left[|X|>g\|x\|_2\right]\geq \frac 14\left(1-g^2\right)^2 
\Eq(X.Y6a)
$$
}
\proof{ 
A trivial generalization of the  Paley-Zygmund inequality [Ta1]
implies that for any $1>g>0$
$$
\P\left[|X|^2\geq  g^2 \E|X|^2\right]\geq  (1-g^2)^2
\frac{(\E|X|^2)^2}{\E X^4}
\Eq(X.Y7) 
$$
On the other hand,  the Marcinkiewicz-Zygmund inequality (see [CT], page 
367) yields that
$$
\E|(x,\xi)|^4\leq 4\E\left(\sum_{\mu}x_\mu^2(\xi^\mu)^2\right)^2
\leq 4 
 \|x\|_2^4
\Eq(X.Y7a)
$$
while $\E X^2=\|x\|_2^2$. This gives \eqv(X.Y6a).
\endproof}
Combining these two results we arrive at 
\lemma {\ver.7} {\it Assume that $Z=(x,\xi)$ with $\xi^\mu$ as in 
Lemma \ver.6 and 
symmetric. 
Let $I\subset \{1,\dots M\}$ and set $\tilde x_\mu =x_\mu$, if $\mu\in I$,
$\tilde x_\mu=0$ if $\mu\not\in I$. Put $\hat x=m-\tilde x$. Assume 
$\|\tilde x\|_2\geq \|\hat x\|_2$.    
Then 
$$
\E (|Z|-x^*)^2 
\geq \frac 12\left(\|x\|_2-x^*\right)^2 + \frac 1{500} 
\|\hat x\|_2^2 
\Eq(X.Y10)
$$
}
\proof {We put $\e=g\|\hat x\|_2$ in \eqv(X.Y1a) and set $g^2=\frac 15$. Then 
Lemma \ver.6 gives the desired bound.
\endproof }
\lemma {\ver.8} {\it Let $Z$ be as in Lemma \ver.7. Then there is a finite 
positive constant $c$ 
such that 
$$
\E (|Z|-x^*)^2\geq c \inf_{\mu,s}\|x-se^\mu x^*\|_2^2
\Eq(X.Y12)
$$
where $c\geq \frac 1{4000 
}$.
}
\proof{  We assume w.r.g. that $x_\geq |x_2|\geq |x_3|\geq\dots\geq |x_M|$
and distinguish  three cases.
\noindent {\bf Case 1:} $x_1^2\geq \frac 12 \|x\|_2^2$. Here we 
set $\hat x\equiv (0,x_2,\dots,x_M)$. We have that 
$$
\eqalign{
\|x-e^1 x^*\|_2^2&=\|\hat x\|_2^2 +(x_1-x^*)^2\cr
&\leq \|\hat x\|_2^2
+2(x_1-\|x\|_2)^2+2(\|x\|_2-x^*)^2\cr
&\leq 3\|\hat x\|_2^2 +2(\|x\|_2-x^*)^2\cr
}
\Eq(X.Y14)
$$
Therefore \eqv(X.Y10) yields
$$
\eqalign{
&\frac 12\left(\|x\|_2-x^*\right)^2 + \frac 1{500} 
\|\hat x\|_2^2 
\geq 
\frac 1{3\cdot 500 
}\left(3 \|\hat x\|_2^2 +1500 
/2 (\|x\|_2-x^*)^2  
\right)\cr
&\geq \frac 1{3\cdot 500 
}\|x-e^1 x^*\|_2^2
}
\Eq(X.Y15)
$$
which is the desired estimate in this case.
\hfill\break
\noindent {\bf Case 2:} $x_1^2< \frac 12 \|x\|_2^2$, $x_2^2\geq \frac 14
 \|x\|_2^2$. Here we may choose $\hat x=(0,x_2,0,\dots,0)$. We set 
$\tilde x=(0,0,x_3,\dots,x_M)$. Then
$$
\|x-e^1 x^*\|_2^2\leq (x_1-x^*)^2+\|\hat x\|_2^2+\|\tilde x\|_2^2
\Eq(X.Y17)
$$
But 
$\|\tilde x\|_2^2\leq \|x\|_2^2-\frac 12\|x\|_2^2\leq 2\|\hat x\|_2$ and
$$
\eqalign{
(x_1-x^*)^2&\leq (\sfrac 12 \|x\|_2-x^*)^2\leq 
2( \|x\|_2-x^*)^2+ \frac 12 \|x\|_2^2
\frac 1{2(1-\e)}x^*\|\hat x\|_2\cr
&\leq 2( \|x\|_2-x^*)^2+2\|\hat x\|_2^2
}
\Eq(X.Y16)
$$
Thus $\|x-e^1 x^*\|_2^2\leq 4\|\hat x\|_2^2 +2(\|x\|_2-x^*)^2$, 
from which follows as above that
$$
\frac 12\left(\|x\|_2-x^*\right)^2 + \frac 1{500} 
\|\hat x\|_2^2 
\geq \frac 1{4\cdot 500 
}\|x-e^1 x^*\|_2^2
\Eq(X.Y17bis)
$$
\hfill\break
\noindent {\bf Case 3:} $x_1^2< \frac 12 \|x\|_2^2$, $x_2< \frac 14
 \|x\|_2^2$. In this case it is possible to find $1\leq t<M$ such that
 $\tilde x=(x_1,x_2,\dots,x_t,0,\dots,0)$ and
$\hat x=(0,\dots,0,x_{t+1},\dots,x_M)$
satisfy $|\|\tilde x\|_2^2-\|\hat x\|_2^2|\leq \frac 14\|x\|_2^2$. 
In particular, $\|\tilde x\|_2^2\leq \frac 53 \|\hat x\|_2^2$, and
$(x^*)^2\leq 2(\|x\|_2-x^*)^2+2\|x\|_2^2\leq  2(\|x\|_2-x^*)^2+\frac {16}3
\|\hat x\|_2^2$.
Thus
$$
\|x-e^1 x^*\|_2^2\leq (x^*)^2+\|\tilde x\|_2^2+\|\hat x\|_2^2\leq 
 2(\|x\|_2-x^*)^2+8 \|\hat x\|_2^2
\Eq(X.Y18)
$$
and thus 
$$
\frac 12\left(\|x\|_2-x^*\right)^2 + \frac 1{500} 
\|\hat x\|_2^2 
\geq \frac 1{8\cdot 500 
}\|x-e^1 x^*\|_2^2
\Eq(X.Y17ter)
$$
Choosing the worst estimate for the constants of all three cases proves the 
lemma. Proposition \ver.4 follows by putting al together.\endproof}}
We thus want an estimate on the fluctuations of the last term in the r.h.s. 
of \eqv(X.5). 
We will do this uniformly inside balls
$B_{R}(x)\equiv\left\{x'\in\R^M \mid \|x-x'\|_2\leq R\right\}$ of radius 
$R$ centered at the point $x\in\R^M$.
\proposition{\ver.9} {\it Assume $\a\leq 1$. 
Let $\{\xi^{\mu}_i\}_{i=1,\dots,N; \mu=1,\dots, M}$ be 
i.i.d. random variables taking values in $[-1,1]$ and satisfying 
$\E\xi^{\mu}_i=0$, $\E(\xi^{\mu}_i)^2=1$. 
For any $R<\infty$ and $x_0\in\{s m^* e^\mu, s=\pm1, \mu=1,\dots,M\}$ we have:
\item{ i)} For $p=2$ and  $\b<11/10$, there exist finite numerical constants
$C$, $K$ such that
\note{The absurd number $\scriptstyle 11/10$ is of course an 
arbitrary choice.
It so happens that, numerically, $\scriptstyle m^*(1.1)\approx 0.5$ 
which seemed like a 
good place to separate cases.}
$$
\eqalign{
&\P\Biggl[
\sup_{x\in B_{R}(x_0)}\left|\frac{1}{\b N}\sum_{i=1}^N\left\{
\E\ln\cosh(\b(\xi_i,x))-\ln\cosh(\b(\xi_i,x))\right\}\right|\cr
&\quad
\geq C\sqrt\a R(m^*+R)+C\a m^* 
+4\a^3 (m^*+R)\Biggr]\cr
&\leq  \ln \left(\sfrac {R}{\a^3}\right)
e^{-\a N} +e^{-\a^2 N}
}
\Eq(X.8)
$$
\item{ii)} For $p\geq 3$ and $\b>\b_c$,  and for $p=2$ and  $\b\geq11/10$,
$$
\eqalign{
&\P\Biggl[
\sup_{x\in B_{R}(x_0)}\left|\frac{1}{\b N}\sum_{i=1}^N\left\{
\E\ln\cosh(\b(\xi_i,x))-\ln\cosh(\b(\xi_i,x))\right\}\right|\cr
&\quad>C\sqrt\a R(R+\|x_0\|_2)+C\a+4\a^3
\Biggr]\leq 
 \ln \left(\sfrac {R}{\a^3}\right)
e^{-\a N}+e^{-\a^2 N}
}
\Eq(X.8bis)
$$
} \advance\foot by 1
\proof{ 
We will treat the case (i) first, as it is the more difficult one.
To prove Proposition \ver.9 we will have to employ some quite 
heavy machinery, known as ``chaining''  in the probabilistic 
literature\note{Physicists would more likely 
call this ``coarse graining'' of even ``renormalization''.} 
(see [LT]; we follow closely the
strategy outlined in Section 11.1 of that book). Our problem is
to estimate the probability of a supremum over an 
$M$-dimensional set, and the 
purpose of chaining is to reduce this to an estimate of suprema over countable
(in fact finite) sets. Let us use in the following the abbreviations
$f(z)\equiv \b^{-1}\ln\cosh(\b z)$ and $F(\xi,x)\equiv \frac 1N\sum_{i=1}^N
f((\xi_i,x))$. 
We  us denote by $\WW_{M,r}$ 
the lattice in $\R^M$ with spacing $r/\sqrt{M}$.
Then, for any $x\in\R^M$ there exists a lattice point $y\in\WW_{M,r}$ such 
that $\|x-y\|_2\leq r$. Moreover, the cardinality of the set of lattice 
points inside the ball $B_R(x_0)$ is bounded by\note{For the (simple) 
proof see [BG5].}
$$
\left|\WW_{M,r}\bigcap B_R(x_0)\right|\leq e^{\a N[\ln(R/r)+2]}
\Eq(X.52)
$$
We introduce a set of exponentially decreasing numbers $r_n=e^{-n}R$
(this choice is somewhat arbitrary and maybe not optimal) and 
set
$\WW(n)\equiv \WW_{M,r_n}\cap B_{r_{n-1}}(0)$.
The point is that if $r_0=R$,  any point $x\in B_R(x_0)$ can be subsequently
 approximated arbitrarily well by a sequence of points $k_n(x)$ with the 
property that 
$$
k_n(x)-k_{n-1}(x)\in \WW(n)
\Eq(XC.1)
$$
 As a consequence, we may write, for any $n^*$ conveniently chosen,
$$
\eqalign{
&|F(\xi,x)-\E F(\xi,x)|\leq 
|F(\xi,k_0(x))-\E F(\xi,k_0(x))|\cr
&\quad+
\sum_{n=1}^{n^*}
|F(\xi,k_n(x))- F(\xi,k_{n-1}(x))-\E(F(\xi,k_n(x))- F(\xi,k_{n-1}(x)))|\cr
&\quad
+ |F(\xi,x)- F(\xi,k_{n^*}(x))-\E(F(\xi,x)- F(\xi,k_{n^*}(x)))| 
}
\Eq(XC.2)
$$
At this point it is useful to observe that the functions $F(\xi,x)$ have some
good regularity properties as functions of $x$.
\lemma{\ver.10}{\it For any $x\in\R^M$ and $y\in\R^M$,
$$
\eqalign{
&\frac{1}{\b N}\left|\sum_{i=1}^N
\left\{\ln\cosh(\b(\xi_i,x))-\ln\cosh(\b(\xi_i,y))\right\}\right|\cr
&\leq
\cases{
\|x-y\|_2\max(\|x\|_2,\|y\|_2)\|A\| & if $\b<11/10$\cr
&\cr
\|x-y\|_2\|A\|^{1/2} & if $\b\geq 11/10$\cr
}
\cr
}
\Eq(X.9)
$$
}
\proofof{Lemma \ver.10}{  Defining $F$ as before, we 
use  the mean value 
theorem to write 
that, for some $0<\theta<1$, 
$$
\eqalign{
&|F(\xi,x)-F(\xi,y)|
=\frac{1}{N}\sum_{i=1}^N(x-y,\xi_i)f'((\xi_i,x+\theta(y-x)))\cr
&\leq\left[\frac{1}{N}\sum_{i=1}^N(x-y,\xi_i)^2\right]^{\frac{1}{2}}
\left[\frac{1}{N}\sum_{i=1}^N\Bigl(f'\bigl((\xi_i,x+\theta(y-x))\bigr)
\Bigr)^2\right]^{\frac{1}{2}}\cr
}
\Eq(X.47)
$$
By the Schwarz inequality we have.
$$
\frac{1}{N}\sum_{i=1}^N(x-y,\xi_i)^2
\leq\|x-y\|_2^2\|A\|
\Eq(X.48)
$$
To treat the last term in the r.h.s. of \eqv(X.47) we will distinguish the 
 two cases $\b\leq \frac {11}{10} $ and $\b\geq \frac {11}{10} $. 
\item{1)} If $\b\leq \frac {11}{10} $ we use that $|f'(x)|=|\tanh(\b x)|
\leq \b |x|$
 to write
$$
\eqalign{
\frac{1}{N}\sum_{i=1}^N&\Bigl(f'\bigl((\xi_i,x+\theta(y-x))\bigr)
\Bigr)^2
\leq 
\left(\sfrac{11}{10}\right)^2
\frac{1}{N}\sum_{i=1}^N(\xi_i,x+\theta(y-x))^2\cr
\leq &
\left(\sfrac{11}{10}\right)^2
\frac{1}{N}\sum_{i=1}^N(\theta(x,\xi_i)^2+(1-\theta)(y,\xi_i)^2)\cr
= &
\left(\sfrac{11}{10}\right)^2
(\theta\|x\|_2^2+(1-\theta)\|y\|_2^2)\|A\|\cr
\leq &
\left(\sfrac{11}{10}\right)^2
\max(\|x\|_2^2,\|y\|_2^2)\|A\|
}
\Eq(X.49)
$$
which, together with \eqv(X.47) and  \eqv(X.48), yields
$$
|F(\xi,x)-F(\xi,y)|\leq
\|x-y\|_2\max(\|x\|_2,\|y\|_2)\|A\|
\Eq(X.50)
$$
\item{2)} If A $\b\geq \frac {11}{10} $ we use that $|f'(x)|\leq 1$ to get
$$
|F(\xi,x)-F(\xi,y)|\leq
\|x-y\|_2\|A\|^{1/2}
\Eq(X.51)
$$
This concludes the proof of Lemma \ver.10.
\endproof}
Lemma \ver.10 implies that 
the last term in \eqv(XC.2) satisfies 
$$
 |F(\xi,x)- F(\xi,k_{n^*}(x))-\E(F(\xi,x)- F(\xi,k_{n^*}(x)))| 
\leq const. r_{n^*} 
\Eq(XC.3quater)
$$
which can be made irrelevantly small by choosing, e.g., $r_{n^*}=\a^3$.
\hfill\break
From this it follows that for any sequence of positive real numbers $t_k$
such that $\sum_{n=0}^\infty t_n\leq t$, we have the estimate
$$
\eqalign{
&\P\left[ \sup_{x\in B_R(x_0)}|F(\xi,x)-\E F(\xi,x)|
\geq t+ \bar t+r_{n^*} \|x\|_2(\|A\|+\E\|A\|)
\right]\cr
&\leq\P\left[|F(\xi,x_0)-\E F(\xi,x_0)|
\geq \bar t\right]\cr
&+\P\Biggl[  \sup_{x\in B_R(x_0)}\bigl|F(\xi,k_0(x))-F(\xi,x_0)\cr
&\quad\quad\quad-\E\left( F(\xi,k_0(x))-F(\xi,x_0)\right)\bigr|
\geq t_0\Biggr]\cr
&+\sum_{n=1}^{n^*}  \P\Biggl[  \sup_{x\in B_R(x_0)}
\bigl|F(\xi,k_n(x))- F(\xi,k_{n-1}(x))\cr
&\quad\quad\quad-\E(F(\xi,k_n(x))- F(\xi,k_{n-1}(x)))\bigr|\geq t_n
\Biggr]
\cr
&\leq \P\left[|F(\xi,x_0)-\E F(\xi,x_0)|
\geq \bar t\right]\cr
&+e^{M[\ln \frac R{r_0}+2]} \P\left[|
|F(\xi,x)-\E F(\xi,x)|
\geq t_0\right]\cr
&+\sum_{n=1}^{n^*} e^{M[\ln \frac R{r_n}+2]}
 \P\Bigl[| 
F(\xi,k_n(x))- F(\xi,k_{n-1}(x))\cr
&\quad\quad\quad-\E(F(\xi,k_n(x))- F(\xi,k_{n-1}(x)))|\geq t_n
\Bigr]
}
\Eq(XC.3)
$$
where we used that the cardinality of the set 
$$
\eqalign{
&\hbox{Card}\biggl\{ 
|F(\xi,k_n(x))- F(\xi,k_{n-1}(x))\cr
&\quad\quad-\E(F(\xi,k_n(x))- F(\xi,k_{n-1}(x)))|\,
; x\in  B_R(x_0)\biggr\}\cr
&\leq \hbox{Card}\{ \WW_{M,r_{n-1}}\cap B_R(x_0)\}
\leq \exp\left(M[\ln\sfrac {R}{r_n}+2]\right)
}
\Eq(XC.3bis)
$$
We must now estimate the probabilities occurring in \eqv(XC.3); the first one
is simple and could be bounded by using Talagrand' s Theorem 6.6 
cited in Section 4. Unfortunately, for the other terms this does not seem 
possible since the functions involved there do not satisfy the hypothesis 
of convex level sets. We thus proceed by elementary methods, exploiting the 
particularly simple structure of the functions $F$ as sums over independent 
terms. Thus we get from the exponential Tchebychev inequality that
$$
\eqalign{
&\P\left[F(\xi,x)-F(\xi,y)-\E[F(\xi,x)-F(\xi,y)]\geq \d\right]\cr
&\leq \inf_{s\geq 0}e^{-\d s}\prod_{i=1}^N\E e^{+\frac sN \left(f((\xi_i,x))
-f((\xi_i,y)) -\E[f((\xi_i,x))
-f((\xi_i,y))]\right)}\cr
&\leq  \inf_{s\geq 0}e^{-\d s} \prod_{i=1}^N\Biggl[1+\frac {s^2}{2N^2}
\E \Bigl(f((\xi_i,x))
-f((\xi_i,y)) \cr
&-\E[f((\xi_i,x))
-f((\xi_i,y))]\Bigr)^2 
e^{\frac sN\left|f((\xi_i,x))
-f((\xi_i,y)) -\E[f((\xi_i,x))
-f((\xi_i,y))]\right|}\Biggr]
}
\Eq(X.CC1)
$$
We now use that both $|\tanh(\b x)|\leq 1$ and $ |\tanh(\b x)|\leq \b|x|$
to get that 
$$
|f((\xi_i,x))
-f((\xi_i,y))|\leq |(\xi_i,(x-y))| \max_{z}|f'((\xi_i,z))|\leq 
 |(\xi_i,(x-y))|
\Eq(X.CC2)
$$
and
$$
\eqalign{
|f((\xi_i,x))
-f((\xi_i,y))&\leq |(\xi_i,(x-y))|\cr
&\leq \b |(\xi_i,(x-y))|\max\left(|(\xi_i,x)|,
|(\xi_i,y)|\right)
}
\Eq(X.CC2bis)
$$
The second inequality will only be used in the case $p\geq 3$ and 
if $\b\leq 1.1$
Using the Schwarz inequality together with \eqv(X.CC2) we get 
$$
\eqalign{
&\E \left(f((\xi_i,x))
-f((\xi_i,y)) -\E[f((\xi_i,x))
-f((\xi_i,y))]\right)^2\cr
&\quad\times e^{\frac sN\left|f((\xi_i,x))
-f((\xi_i,y)) -\E[f((\xi_i,x))
-f((\xi_i,y))]\right|}\cr
&\leq 
\left[8\E\left(f((\xi_i,x))
-f((\xi_i,y))\right)^4\right]^{1/2}\cr
&\quad\times
\left[\E e^{\frac {2s}N |f((\xi_i,x))
-f((\xi_i,y))-\E (f((\xi_i,x))
-f((\xi_i,y)))|}\right]^{1/2}\cr
&\leq \sqrt 8 \left[\E(\xi_i,x-y)^4\right]^{1/2}
\left[ \E e^{\frac {2s}N |(\xi_i,(x-y))|}\right]^{1/2}
e^{\frac sN \E|(\xi_i,(x-y))|}\cr
 }
\Eq(X.CC4)
$$
Using \eqv(X.CC2bis) and once more the Schwarz inequality
 we get an alternative bound for this quantity by
$$
\eqalign{
 &\sqrt 8 \b^2  \left[\E(\xi_i,x-y)^8\right]^{1/4}
\max\left(  \left[\E(\xi_i,x)^8\right]^{1/4} 
\left[\E(\xi_i,y)^8\right]^{1/4}\right)\cr
&\times
\left[ \E e^{\frac {2s}N |(\xi_i,(x-y))|}\right]^{1/2}
e^{\frac sN \E|(\xi_i,(x-y))|}
}
\Eq(X.CC4bis)
$$
The last line is easily bounded using essentially Khintchine resp. 
Marcinkiewicz-Zygmund  
inequalities (see [CT], pp. 366 ff.), in particular
$$
\eqalign{
\E|(\xi_i,x)|&\leq \sqrt 2\|x\|_2  \quad\text{reps.}\|x\|_2/\sqrt 2
\text {if $\xi_i^\mu$ are Bernoulli}\cr 
\E e^{s|(\xi_i,x)|}&\leq 2e^{\frac {s^2}2 c}  \quad\text{with $c=1$ 
if $\xi_i^\mu$ are Bernoulli}\cr
\E(\xi_i,(x-y))^{2k}&\leq  2^{2k}k^{k} \|(x-y)\|_2^{2k}
\quad\text{no $2^{2k}$ if $\xi_i^\mu$ are Bernoulli} 
}
\Eq(X.CC5)
$$ 
Thus
$$
\eqalign{
&\E \left(f((\xi_i,x))
-f((\xi_i,y)) -\E[f((\xi_i,x))
-f((\xi_i,y))]\right)^2\cr
&\quad\times e^{\frac sN\left|f((\xi_i,x))
-f((\xi_i,y)) -\E[f((\xi_i,x))
-f((\xi_i,y))]\right|}\cr
&\leq
\sqrt{8}\sqrt{32} \|(x-y)\|_2^2 e^{\frac sN \sqrt 2\|x-y\|_2+
c\frac {2s^2}{N^2}\|x-y\|_2^2}   \cr
&\text{respectively}\cr
&\leq \b^2\sqrt{8} 2^4 4^2\|x-y\|_2^2 \left(\|x\|_2+\|y\|_2\right)^2
 e^{\frac sN \sqrt 2\|x-y\|_2+
c\frac {2s^2}{N^2}\|x-y\|_2^2} 
} 
\Eq(X.CC6)
$$
In the Bernoulli case the constants can be improved to $2\sqrt 8$ and 
$\sqrt 8 4^2$, resp., and $c=1$. 
\hfill\break
Inserting \eqv(X.CC6) into \eqv(X.CC1), using that $1+x\leq e^x$ 
and choosing $s$ gives the desired bound on the probabilities. The trick here 
is not to be tempted to choose 
$s$ depending on $\d$. Rather, depending on which 
bound we use, we choose $\,\,\,s=\frac{N\sqrt\a}{\|x-y\|_2}$ 
or  $\,\,\,s=\frac{N\sqrt\a}{\|x-y\|_2\|\left(\|x\|_2+\|y\|_2\right)}$.
This gives
$$
\eqalign{
&\P\left[F(\xi,x)-F(\xi,y)-\E[F(\xi,x)-F(\xi,y)]\geq \d\right]\cr
&\leq  \exp\left(-N\frac{\sqrt\a \d}{\|x-y\|_2} +8 \a N e^{\sqrt {2\a}+2c \a}
\right)
}
\Eq(X.CC7)
$$ 
respectively
$$
\eqalign{
&\P\left[F(\xi,x)-F(\xi,y)-\E[F(\xi,x)-F(\xi,y)]\geq \d\right]\cr
&\leq \exp\left(-N\sfrac{\sqrt\a \d}{\|x-y\|_2(\|x\|_2+\|y\|_2)}
+\a N \b^2 \sqrt 8 2^4 4^2 e^{\frac {2\sqrt\a}{\|x\|_2+\|y\|_2}
+\frac {c\a}{(\|x\|_2+\|y\|_2)^2}}\right)
}
\Eq(X.CC7bis)
$$
In particular
$$
\eqalign{
&\P\biggl[\bigl|F(\xi,k_n(x))-F(\xi,k_{n-1}(x))\cr
&\quad-
\E[|F(\xi,k_n(x))-F(\xi,k_{n-1}(x))]\bigr|\geq t_n\biggr]
\cr
&
\leq 
2\exp\left(-N\frac {t_n\sqrt\a }{r_{n-1}}+N 8 \a N 
e^{\sqrt {2\a}+2c \a}
\right)
}
\Eq(XC.9)
$$
and
$$
\eqalign{
&\P\biggl[\bigl|F(\xi,k_n(x))-F(\xi,k_{n-1}(x))\cr
&\quad-
\E[|F(\xi,k_n(x))-F(\xi,k_{n-1}(x))]\bigr|\geq t_n\biggr]\cr
&\leq 2\exp\left(-N\frac {t_n\sqrt\a }{r_{n-1}(R+\|x_0\|_2)}
+\a N \b^2 \sqrt 8 2^4 4^2 e^{\frac {2\sqrt\a}{R+\|x_0\|_2}
+\frac {c\a}{(\|x_0\|_2+R)^2}}
\right)
}
\Eq(XC.9bis)
$$
We also have 
$$
\eqalign{
&\P\left[\left|F(\xi,k_0(x))- F(\xi,x_0)-
\E\left[ F(\xi,k_0(x))-F(\xi,x_0)\right]
\right|\geq t_0\right]\cr
&\leq 
2\exp\left(-N\frac {t_0\sqrt\a }{ R}+ 8\a N  
e^{\sqrt {2\a}+2c \a}
\right)
}
\Eq(XC.10)
$$
and
$$
\eqalign{
&\P\left[\left|F(\xi,k_0(x))- F(\xi,x_0)-
\E\left[ F(\xi,k_0(x))-F(\xi,x_0)\right]
\right|\geq t_0\right]
\cr
&\leq 
2\exp\left(-N\frac {t_0\sqrt\a }{ R(\|x_0\|_2+R)}
+\a N \b^2 \sqrt 8 2^4 4^2 e^{\frac {2\sqrt\a}{R+\|x_0\|_2}
+\frac {c\a}{(\|x_0\|_2+R)^2}}
\right)
}
\Eq(XC.10bis)
$$
Since $\|x_0\|_2+R\geq 
x^*$ so that 
$\frac {2\sqrt\a}{R+\|x_0\|_2}
+\frac {c\a}{(\|x_0\|_2+R)^2}\leq \sqrt 2 \g x^* + c \g^2 (x^*)2\leq c'\g$. 
Thus in the case $\b\leq 1.1$ we may choose
$ t_0$ and $t_n$ as
$$
t_0 =\sqrt\a R(\|x_0\|_2+R) \left[1
+2+ \b^2 2^4 4^2\sqrt 8 e^{c'\g}+1\right]
\Eq(XC.11)
$$
and
$$
t_n=\sqrt\a e^{-(n-1)}R(\|x_0\|_2+R) \left[ n +2+\b^2 2^44^2e^{c'\g}+1\right]
\Eq(XC.12)
$$
Finally a simple estimate gives that (for $x_0=\pm m^* e^\mu$)
$$
\P\left[|F(\xi,x_0)-\E F(\xi,x_0)|\geq \bar t\right]\leq 2e^{-\frac{\bar t^2}
{20 (m^*)^2} N}
\Eq(XC.12bis)
$$
Choosing $\bar t=m^*\a$, setting $n^*=\ln \left(\sfrac
{\a^3}{R}\right)$, and putting all this into \eqv(XC.3) we get that
$$
\eqalign{
&\P\Biggl[\sup_{x\in B_R(x_0)}\left|F(\xi,x)-\E F(\xi,x)\right|
\geq\cr
&\quad  \sqrt \a R(\|x_0\|_2+R) \biggl(  
4+ \b^2 2^44^2\sqrt 8 e^{c'\g}\cr
&\quad
+ 3+\sfrac e{e-1} 
+\b^2 2^44^2e^{c'\g}\biggr)+m^*\a+\a^3 (\|x_0\|_2+R)
\Biggr]\cr
&\leq  \ln \left(\sfrac
{\a^3}{R}\right)e^{-\a N}+2e^{-\a^2 N/20}
}
\Eq(XC.14bis)
$$ 
This proves part (i) of Proposition \ver.9 and allows us to estimate 
the  constant $C$ in \eqv(X.8). 
In the same way, but using \eqv(XC.9) and \eqv(XC.10), 
we get the analogous bound in case (ii), namely
$$
\eqalign{
&\P\Biggl[\sup_{x\in B_R(x_0)}\left|F(\xi,x)-\E F(\xi,x)\right|
\geq \cr
&\quad \sqrt \a R (\|x_0\|_2+R)\left( 4+  8 e^{c'\g}
+ 3+\sfrac e{e-1}+ 8e^{c'\g}\right) +4\a^3 
\Biggr]\cr
&\leq  \ln \left(\sfrac
{\a^3}{R}\right)e^{-\a N}
}
\Eq(XC.14)
$$ 
This concludes the proof of Proposition \ver.9.\endproof}

\remark { The reader might wonder whether this heavy looking chaining 
machinery used in the proof of Proposition \ver.9 is really 
necessary. Alternatively, one might use just a single lattice approximation
and use Lemma \ver.10 to estimate how far the function can be from the 
lattice values. But for this we need at least a lattice with $r=\sqrt \a$,
and this would force us to replace the $\sqrt\a$ terms in
\eqv(X.8) and \eqv(X.8bis) by $\sqrt{\a|\ln \a|}$. While this may not look 
too serious, it would certainly spoil the correct scaling between
the critical $\a$ and $\b-1$ in the case $p=2$. }

We are now ready to conclude the proof of Proposition {\ver.1}. 
To do this, we consider the $2M$ sectors in which $s x^* e^\mu$ is the 
closest of all the points $s x ^* e^\nu$ and use Proposition \ver.9
with $x_0= s e^\mu x^*$ and $R$ the distance from that point. One sees easily 
that if that distance is sufficiently large (as 
stated in the theorem), then with probability exponentially close to 
one, the modulus of the last term in \eqv(X.5) is bounded by one half
of the lower bound on the first term given by Proposition \ver.4.
Since it is certainly enough to consider a discrete set of radii
(e.g. take $R\in \Z/N$), and the individual estimates 
fail only with a  probability of order $\exp(-\a N)$, it is clear that the 
estimates on $\psi$ hold indeed uniformly in $x$ with probability 
exponentially close to one. This concludes the proof of Proposition \ver.1. 
\endproof}

\newpage

\def\vo{\hat v}

\chap {7. Local analysis of $\Phi$ }7

To obtain more detailed information on the Gibbs measures requires 
to look more precisely at the behaviour of the functions $\Phi_{p,\b,N}(m)$
in the vicinities of points $\pm m^*(\b)e^\mu$. Such an analysis has first 
been performed in the case of the standard Hopfield model in [BG5]. 
The basic idea was simply to use second order Taylor expansions combined with 
careful probabilistic error estimates. One can certainly do the same
in the general case with sufficiently smooth energy function $E_M(m)$, but 
since results  (and to some extent techniques) depend 
on specific properties of these functions, we restrict our attention again to 
the cases where $E_M(m)=\frac 1p\|m\|_p^p$, with $p\geq 2$ integer, as in the previous section. 
For reasons that will 
become clear in a moment, the (most interesting) case $p=2$ is special, and we 
consider first the case $p\geq 3$. Also throughout this section, the 
$\xi_i^\mu$ take the values $\pm 1$.

\newsubsec{ \ver.1. The case $p\geq 3$.}

As a matter of fact, this case is ``misleadingly simple''\note{But 
note that we consider only the case $\scriptstyle M\sim \a N$ rather than 
$\scriptstyle M\sim \a N^{p-1}$}. Recall that we deal 
with the function $\Phi_{p,\b,N}(m)$ given by \eqv(X.2). Let us consider 
without restriction of generality the vicinity of
$m^* e^1$. Write $m=m^* e^1 +v$ where $v$ is assumed ``small'', e.g. 
$\|v\|_2\leq \e<m^*$. We have to consider mainly the regions over which 
Proposition 6.1 does not give control, i.e. where
$\|\sign(m)|m|^{p-1}-e^1 (m^*)^{p-1}\|_2 \leq c_1 \sqrt {\a}$
(recall \eqv(X.4ter)).
In terms of the variable $v$ this condition implies that
both
$|v_1|^2 \leq C  \sqrt {\a}$ and 
$\|\vo \|_{2p-2}^{2p-2} \leq C  \sqrt {\a}$ for some constant 
$C$ (depending on $p$), where we have set $\vo=(0,v_2,v_3,\dots,v_M)$.
Under these conditions we want to study
$$
\eqalign{
&\Phi_{p,\b,N}(m^*e^1+v)-\Phi_{p,\b,N}(m^* e^1)=
\frac 1q\left( (m^*+v_1)^p -(m^*)^p+\|\vo\|_p^p\right)\cr
&\quad-\frac 1{\b N}\sum_{i=1}^N \Biggl[\ln\cosh\left(\b ((m^*+v^1)^{p-1}
+\sum_{\mu\geq 2} \hat\xi_i^\mu v^{p-1}_\mu)\right)\cr
&\quad-\ln\cosh(\b (m^*)^{p-1})
\Biggr]
}
\Eq(L.1)
$$
where we have set  $\hat\xi_i^\mu\equiv
\xi_i^1\xi_i^\mu$. The crucial point is now that we can expand 
each of the terms in the sum over $i$ without any difficulty: for 
$|(m^*+v_1)^{p-1}-(m^*)^{p-1}|\leq |v_1| (p-1)(m^*+|v_1|)^{p-2}\leq C|v_1|$, 
and, more importantly,
the  H\"older inequality gives
$$
\left|\sum_{\mu\geq 2} \hat\xi_i^\mu \hbox{sign}(v_\mu)|v_\mu|^{p-1}\right|
\leq \|\vo\|_2^2 \|\vo\|_\infty^{p-3}
\Eq(L.2)
$$
As explained earlier, we need to consider only $v$ for which  $\|v\|_2\leq 2$, 
and 
$\|\vo\|_\infty\leq \|\vo\|_{2p-2}\leq 
\left( C  \sqrt {\a}\right)^{1/(2p-2)}$ is small on the set 
we consider. 
Such a result does not hold if $p=2$, and this makes the whole analysis 
much more cumbersome in that case --- as we shall see. 

What we can already read off from \eqv(L.1) otherwise is that $v_1$ and $\vo$ 
enter in a rather asymmetric way. We are thus well-advised to treat 
$|v_1|$ and $\|\vo\|_2$ as independent  small parameters.  Expanding, 
and using 
that $m^*=\tanh( \b(m^*)^{p-1})$ gives therefore
$$
\eqalign{
&\Phi_{p,\b,N}(m^*e^1+v)-\Phi_{p,\b,N}(m^* e^1)\cr
&=v_1^2\frac {p-1}2
(m^*)^{p-2}\left[1-\b (1-(m^*)^2) (m^*)^{p-2}(p-1)\right]\cr
&+\frac 1 q\|\vo\|_p^p-\frac \b 2(1-(m^*)^2)\left(\sign(\vo)
|\vo|^{p-1},\frac{\xi^T\xi}N 
\sign(\vo)|\vo|^{p-1}\right)\cr
&-\frac 1N\sum_{i=1}^N\left(\hat\xi_i,\sign(\vo)|\vo|^{p-1}\right)\left[
m^*+\frac \b 2(1-(m^*)^2) (m^*)^{p-2} v_1\right]\cr
&+R(v)
}
\Eq(L.3)
$$
where
$$
\eqalign{
&|R(v)|\leq |v_1|^3\frac{(p-1)(p-2)(p-3)}6 (m^*+|v_1|)^{p-3}\cr
&+\frac {2^{9/4}}6\left[|v_1|^3    (p-1)^3(m^*+|\e|)^{3(p-2)}+
 \frac 1N\sum_{i=1}^N|(\hat\xi_i, \sign(\vo)|\vo|^{p-1})|^3\right]\cr
&\times
\frac {2\b^2 \tanh\b\left((m^*+|v_1|)^{p-1}+\|\vo\|_2^2 \|\vo\|_\infty^{p-3}
\right)}
{\cosh^2\b\left((m^*-|v_1|)^{p-1}-\|\vo\|_2^2 \|\vo\|_\infty^{p-3}
\right)}
}
\Eq(L.4)
$$
where the last factor is easily seen to be bounded uniformly by some 
constant, provided $|v_1|$ and $\|\vo\|_2$ are small compared to 
$m^*(\b)$.  Recall that the latter is, for $\b\geq \b_c$, bounded away from 
zero if $p\geq 3$. (Note that we have used that for positive $a$ and $b$,
$(a+b)^3\leq 2^{9/4}(a^3+b^3)$).  
Note further that 
$$
\eqalign{
&\left(\sign(\vo)|\vo|^{p-1},\frac{\xi^T\xi}N 
\sign(\vo)|\vo|^{p-1}\right)\leq \|A(N)\| \sum_{\mu\geq 2} v_\mu^{2p-2}
\cr
&\leq  \|A(N)\| \|\vo\|_p^p \|\vo\|_\infty^{p-2}
}\Eq(L.5)
$$
and 
$$
\eqalign{
 &\frac 1N\sum_{i=1}^N|(\hat\xi_i, \sign(\vo)|\vo|^{p-1})|^3
\leq  \frac 1N\sum_{i=1}^N|(\hat\xi_i, \sign(\vo)|\vo|^{p-1})|^2 \|\vo\|_2^2\|\vo\|_\infty^{p-3}
\cr
&\leq \|A(N)\|  \|\vo\|_p^p \|\vo\|_\infty^{2p-5} \|\vo\|_2^2
}
\Eq(L.6)
$$
so that in fact 
$$
\eqalign{
&\Phi_{p,\b,N}(m^*e^1+v)-\Phi_{p,\b,N}(m^* e^1)\cr
&=v_1^2\frac {p-1}2
(m^*)^{p-2}\left[1-\b (1-(m^*)^2) (m^*)^{p-2}(p-1)\right]
+\frac 1 q\|\vo\|_p^p\cr
&-\frac 1N\sum_{i=1}^N\left(\hat\xi_i,\sign(\vo)|\vo|^{p-1}\right)\left[
m^*+\frac \b 2(1-(m^*)^2) (m^*)^{p-2} v_1\right]
+R(v)
}
\Eq(L.7)
$$
where
$|\tilde R(v)|\leq c\left (|v_1|^3+ \|\vo\|_p^p\|\vo\|_\infty^{p-2}
\right)$.

These bounds give control  over the local minima near the Mattis states. 
In fact, we can compute easily the first corrections to their precise 
(random) positions. The approximate equations for them have the form
$$
\eqalign{
 v_1&=c_1(\b) \frac 1{\sqrt N} (z, \sign(\vo)|\vo|^{p-1})    \cr
v_\mu&=\frac 1{\sqrt N}z_\mu (m^*+c_2(\b)v_1), \text{for $\mu\neq 1$}
}
\Eq(L.8)
$$
where $z_\mu=\frac 1{\sqrt N}\sum_i \hat\xi_i^\mu$ 
and $c_1(\b), c_2(\b)$ are constants that can be read off \eqv(L.7).
These equations are readily solved and give
$$
\eqalign{
v_1&=\frac 1{\sqrt N} c_1 X\cr
v_\mu&=\frac 1{\sqrt N}z_\mu \left(m^*+\frac {c_1c_2}{\sqrt N} X\right)
}
\Eq(L.9)
$$
where $X$ is the solution of the equation
$$
X=\frac 1{N^{(p-1)/2}}\|z\|_p^p\left(m^*+\frac {c_1c_2}{\sqrt N} X\right)^p
\Eq(L.10)
$$
Note that for $N$ large, 
$$
\E \|z\|_p^p \approx \frac {M}{N^{(p-1)/2}} 
\frac{(1-(-1)^p)2^{p/2}\G\left(\sfrac {1+p}2\right)}
     {2\sqrt{\pi}}
\Eq(L.11)
$$
Moreover, an estimate of Newman ([N], Proposition 3.2)  shows 
that 
$$
\P\left[\left|\|z\|_p^p-\E \|z\|_p^p\right| >\g M^{\frac{p-2}{2p-2}}\right]
\leq 2 e^{-c_p(\g) M^{1/(p-1)}}
\Eq(L.12)
$$
for some function $c_p(\g)>0$ for $\g>0$. This implies in particular that 
$\frac X{\sqrt N}$ is sharply concentrated around the value
$\frac M{N^{p/2}}$ (which tends to zero rapidly for our choices of $M$). 
Thus under our assumptions on $M$, the location of the minimum 
 in the limit as $N$ tends to infinity is
$v_1=0$ and $v_\mu=\frac{m^*}{\sqrt N}z_\mu$, and
at this point  $\Phi_{p,\b,N}(m^*e^1+v)-\Phi_{p,\b,N}(m^* e^1)=O\left(
M/N^{p/2}\right)$.

On the other hand,  for $\|\vo\|_p\geq 2 \sqrt\a (m^*+c)$,
$$
\Phi_{p,\b,N}(m^*e^1+v)-\Phi_{p,\b,N}(m^* e^1)\geq 
c_1  v_1^2 + c_3 \|\vo\|_p^p>0
\Eq(L.13)
$$
which completes the problem of localizing the minima of $\Phi$ in the 
case $p\geq 3$. Note the very asymmetric shape of the function in their 
vicinity.

\vskip 0.5cm
\line{\bf \ver.2 The case $p=2$.\hfill}

The case of the standard Hopfield model turns out to be the 
more difficult, but also the most interesting one. The major source of 
this is the fact that an inequality like 
\eqv(L.2) does {\it not} hold here. Indeed, it is easy to see that there 
exist $v$ such that 
$\left|\sum_\mu\hat\xi_i^\mu v_\mu\right|=\sqrt M\|\vo\|_2$.
The idea, however, is that this requires that $v$ be adapted to the particular
$\hat\xi_i$, and that it will be impossible, typically, to find a $v$ such that
for {\it many} indices, $i$,  $\left|\sum_\mu\hat\xi_i^\mu v_\mu\right|$ 
would be much bigger 
than $\|v\|_2$  and to take advantage of that fact. The corresponding analysis
has been carried out in [BG5] and we will not repeat all the intermediate 
technical steps here. We will however present the main arguments in a
streamlined form. The key idea is to perform a Taylor expansion like in 
the previous case only for those indices $i$ for which $(\xi_i,v)$ is small,
and to use a uniform bound for the others. The upper and lower bounds must be 
treated slightly differently, so let us look first at the lower bound.

The uniform bound we have here at our disposal is that
$$
-\frac 1\b \ln\cosh\b x\geq \frac {(m^*)^2}2 -\frac 1\b \ln\cosh \b m^*-\frac
{x^2}2
\Eq(L.14)
$$
Using this we get, for suitably chosen parameter $\t>0$, 
by a simple computation that for some $0\leq \theta\leq 1$, 
$$
\eqalign{
&\Phi_{2,\b,N}(m^*e^1+v)-\Phi_{2,\b,N}(m^* e^1)\cr
&\geq
\frac 12 \|v\|_2^2-\frac 12 \b(1-(m^*)^2)\frac 1N \sum_{i=1}^N
(\xi_i,v)^2-\frac {m^*}N \sum_{i=1}^N (\hat\xi_i,\vo)\cr
&-\frac 16\frac 1N \sum_{i=1}^N \1_{\{|(\xi_i,v)|\leq \t m^*\}}
|(\xi_i,v)|^3 2\b^2\frac {\tanh\b(m^*+\th (\xi_i,v))}
                         {\cosh^2\b(m^*+\th (\xi_i,v))}
\cr
&-\frac 12(1-\b(1-(m^*)^2))\frac 1N\sum_{i=1}^N \1_{\{|(\xi_i,v)|>\t m^*\}}
(\xi_i,v)^2
}
\Eq(L.15)
$$
The first two lines are the main second order contributions. The third line 
is the standard third order remainder, but improved by the characteristic
 function that forces $(\xi_i,v)$ to be small. The last line is the price 
we have to pay for that, and we will have to show that with large probability 
this is also very small.
This is the main ``difficulty''; for the third order remainder 
one may use simply that
$$
\eqalign{
&\frac 16\frac 1N \sum_{i=1}^N \1_{\{|(\xi_i,v)|\leq \t m^*\}}
|(\xi_i,v)|^3 2\b^2\frac {\tanh\b(m^*+\th (\xi_i,v))}
                         {\cosh^2\b(m^*+\th (\xi_i,v))}\cr
&\leq \frac 1{2N} \sum_{i=1}^N(\xi_i,v)^2 \t m^* \frac 13
\b^2\frac {\tanh\b(m^*(1+\t))} {\cosh^2\b(m^*(1-\t))}\cr
&\leq  \frac 1{2N} \sum_{i=1}^N(\xi_i,v)^2
\t (1+\t)(m^*)^2 \frac {\b^3}3 \cosh^{-2}\b(m^*(1-\t))
}
\Eq(L.16)
$$
For $\t$ somewhat small, say $\t\leq 0.1$, it is not difficult to
see that $\frac {\b^3}3 \cosh^{-2}\b(m^*(1-\t))$ 
is bounded uniformly in $\b$ by a constant of order 
$1$. Thus we can for our purposes use 
$$ 
\eqalign{
\frac 1{6N}& \sum_{i=1}^N \1_{\{|(\xi_i,v)|\leq \t m^*\}}
|(\xi_i,v)|^3 2\b^2\frac {\tanh\b(m^*+\th (\xi_i,v))}
                         {\cosh^2\b(m^*+\th (\xi_i,v))}
\cr
&\leq \t(1+\t) (m^*)^2 \frac 1{2N} \sum_{i=1}^N(\xi_i,v)^2
}\Eq(L.17)
$$ 
which produces  just a small perturbation of the quadratic term.
Setting 
$$
X_a(v)\equiv \frac 1N\sum_{i=1}^N \1_{\{|(\xi_i,v)|>a\}}
(\xi_i,v)^2
\Eq(L.18)
$$
we summarize our finding so far as 

\lemma {\ver.1} {\it There exists $\t_c>0$ ($\approx 0.1$) such that 
for all $\b$, for $\t\leq \t_c$,
$$
\eqalign{
&\Phi_{2,\b,N}(m^*e^1+v)-\Phi_{2,\b,N}(m^* e^1)\cr
&\geq 
\frac 12\left(v,\left[\1-(\b(1-(m^*)^2)+\t(1+\t)(m^*)^2)
\frac {\xi^T\xi}N
\right]v\right)-\frac {m^*}N\sum_{i=1}^N (\hat\xi_i,\vo)\cr
&-\frac 12 (1-\b(1-(m^*)^2))X_{\t m^*}(v)   
}
\Eq(L.19)
$$
}

Before turning to the study of $X_a(v)$, we derive corresponding lower bounds.
For this we need a complement to \eqv(L.14). Using the  
Taylor formula with second order remainder we have that for some
$\tilde x$
$$
\eqalign{
-\frac 1\b \ln\cosh\b x&\leq \frac {(m^*)^2}2 -\frac 1\b \ln\cosh \b m^*-\frac
{x^2}2\cr
&
+\frac {(x-m^*)^2}{2}\left[1-\b\left(1-\tanh^2\b(\tilde x)
\right)\right]\cr
&\leq \frac {(m^*)^2}2 -\frac 1\b \ln\cosh \b m^*-\frac
{x^2}2
+\frac {(x-m^*)^2}{2}
}
\Eq(L.20)
$$
By a similar computation as before this gives 

\lemma {\ver.2} {\it There exists $\t_c>0$ ($\approx 0.1$) such that 
for all $\b$, for $\t\leq \t_c$,
$$
\eqalign{
&\Phi_{2,\b,N}(m^*e^1+v)-\Phi_{2,\b,N}(m^* e^1)\cr
&\leq 
\frac 12\left(v,\left[\1-(\b(1-(m^*)^2)-\t(1+\t)(m^*)^2)
\frac {\xi^T\xi}N
\right]v\right)\cr
&-\frac {m^*}N\sum_{i=1}^N (\hat\xi_i,\vo)
+\frac 12 \b(1-(m^*)^2))X_{\t m^*}(v)   
}
\Eq(L.21)
$$
}

To make use of these bounds, we need to have uniform control over 
the $X_a(v)$. In [BG5] we have proven for this the following 

\proposition {\veroparagrafo.3} {\it Define
 $$
\eqalign{
\G(\a,a/\rho) &=
\left(2\sqrt{
2\sqrt 2 e^{-\frac {(1-3\sqrt\a)^2}{(1-\sqrt\a)^2}\frac {a^2}{4\rho^2}}
+\a(|\ln\a| +2)}+   \a\sqrt {1+r(\a)}\right)^2\cr
&+2\a^2(1+ r(\a)) +  \sfrac 12\a
  \left(2e^{- \frac {a^2}{\a\rho^2}} 
+2  \sqrt{3\a(|\ln \a|+2)}\right)
}
\Eq(L.22)
$$
Then 
$$
\eqalign{
\P\left[\sup_{v\in B_\rho} X_a(v)\geq \rho^2 \G(\a,a/\rho)\right]
\leq e^{-\a N} +\P[\|A-\1\|\geq r(\a)]
}
\Eq(L.23)
$$
}

We see that $\G(\a,a,\rho)$ is small if $\a$ is small and $\rho^2$
is small compared to $a$ which for us is fine: we need the proposition
with $a=\t m^*$ and with $\rho\leq \g m^*c_1$, where $\g$ is our small
parameter.
The proof of this proposition can be found in [BG5]. 
It is quite technical and uses a chaining procedure quite similar 
to the one used in Section 6 in the proof of Proposition 6.9. Since we
have not found a way to simplify or improve it, we will not reproduce it here. 
Although in [BG5]
only the Bernoulli case was considered, but the extension to centered bounded
$\xi_i^\mu$ poses no particular problems and can be left to the reader; 
of course constants will 
change, in particular if the variables are asymmetric.  

The expression for $\G(\a,a,\rho) $ looks quite awful. However, for $\a $ small
(which is all we care for here), it is in fact bounded by
$$
\G(\a,a/\rho)\leq C 
\left[ e^{- (1-2\sqrt\a)^2\frac {a^2}{4\rho^2}}
+\a(|\ln\a| +2)\right]
\Eq(L.24)
$$
with $C\approx 25$.
We should now choose $\t$ in an optimal way. It is easy to see that 
in \eqv(L.19), for $\rho\leq c\g m^*$, this leads to 
$\t\sim \g \sqrt{|\ln \g|}$, uniformly in $\b>1$. This uses that 
 the coefficient
of $X_{\t m^*}(v)$ is proportional 
to $(m^*)^2$. Unfortunately, that is not the 
case in the upper bound of Lemma \ver.2, so that it turns out that while this
estimate is fine for $\b $ away from $1$ 
(e.g. $\b>1.1$, which means $m^*>0.5$), for $\b$ near one we have been too
careless! This is only just: replacing $\b(1-\tanh^2\b\tilde x)$ by zero
and hoping to get away with it was overly optimistic. This is, however, 
easily remedied by dealing more carefully with that term. We will not give the
(again somewhat tedious) details here; they can be found in [BG5].
We just quote from [BG5] (Theorem 4.9) 

\lemma {\ver.4} {\it Assume that $\b\leq 1.1$. Then 
 there exists $\t_c>0$ ($\approx 0.1$) such that 
for $\t\leq \t_c$,
$$
\eqalign{
&\Phi_{2,\b,N}(m^*e^1+v)-\Phi_{2,\b,N}(m^* e^1)\cr
&\leq 
\frac 12\left(v,\left[\1-(\b(1-(m^*)^2)-\t(1+\t)(m^*)^2)
\frac {\xi^T\xi}N
\right]v\right)-\frac {m^*}N\sum_{i=1}^N (\hat\xi_i,\vo)\cr
&+\frac 12 (m^*)^2\|v\|_2^2\left(\g +240 e^{-(1-2\sqrt\a)^2\frac {(m^*)^2}{4
\|v\|_2^2}}\right)   
}
\Eq(L.25)
$$
}

For the range of $v$ we are interested in, all these bounds combine to 

\theo {\ver.5}{\it For all $\b>1$ and for all $\|v\|_2\leq c\g m^*$, there 
exists  a finite numerical constant $0<C<\infty$ such that
$$
\eqalign{
&\Biggl|\Phi_{2,\b,N}(m^*e^1+v)-\Phi_{2,\b,N}(m^* e^1)\cr
&-\frac 12\left[1-\b(1-(m^*)^2)
\right]\|v\|_2^2-\frac {m^*}N\sum_{i=1}^N (\hat\xi_i,\vo)\Biggr|
\leq \g\sqrt {|\ln \g|} C (m^*)^2 \|v\|_2^2
}
\Eq(L.26)
$$
with probability greater that $1-e^{-\a N}$. 
}

As an immediate consequence of this bound we can localize the position of the
minima of $\Phi$ near $m^*e^\mu$ rather precisely.

\corollary {\ver.6}{\it   Let $v^*$ denote the position of the lowest minimum 
of the function $\Phi_{2,\b,N}(m^*e^1+v)$
in 
the ball $\|v\|_2\leq c\g m^*$. 
Define the vector $z^{(\nu)} \in \R^M$ with components
$$
z^{(\nu)}_\mu\equiv\cases{ 
\frac 1N\sum_{i}\xi_i^\nu\xi_i^\mu,& for $\mu\neq\mu$\cr 0,&for $\mu=\nu$}
\Eq(L.27)
$$
There exists a finite
constant $C$ such that 
$$
\left\|v^*-\frac {m^*} {1-\b(1-(m^*)^2)} z^{(1)}\right\|_2
\leq C \ \g \sqrt {|\ln \g|}\frac{\|z^{(1)}\|_2(m^*)^3}{(1-\b(1-(m^*)^2))^2}
\Eq(L.28)
$$
with the same probability as in Theorem \ver.5. Moreover, with 
probability greater than $1-e^{-4 M/5}$,
$$
\|z^{(1)}\|_2\leq 2\sqrt\a
\Eq(L.29)
$$
so that in fact
$$
\left\|v^*-\frac {m^*} {1-\b(1-(m^*)^2)} z^{(1)}\right\|_2
\leq C \g^2\sqrt{|\ln \g|} m^*
\Eq(L.30)
$$
}

\proof{ \eqv(L.28) is straightforward from Theorem \ver.5. The bound on 
$\|z^{(1)}\|_2$ was given in [BG5], Lemma 4.11 and follows from 
quite straightforward exponential estimates. \endproof}

\remark {We will see in the next section that for $\b$ not too large 
(depending on $\a$), there is actually a unique minimum for 
$\|v\|_2\leq c\g m^*$.}

\newpage

\def\m{\hat\mu}

\chap{8. Convexity, the replica symmetric solution,  convergence}8

In this final section we restrict our attention to the standard
Hopfield model.  Most of the results presented here were inspired by a
recent paper of Talagrand [T4].

In the last section we have seen that the function $\Phi$ is locally bounded 
from above and below by quadratic functions. A natural question 
is to ask whether this function may even be locally covex.
The following theorem 
(first proven in [BG5]) shows that this is true under some further 
restrictions on the range of the parameters.

\theo{\ver.1} {\it Assume that $1<\b<\infty$. 
If the parameters $\a,\b,\rho$ are such that for $\e>0$, 
$$
\eqalign{
\inf_{\t}&\Bigl(\b(1-\tanh^2(\b m^*(1-\t)))(1+3\sqrt\a)\cr
&+
2\b \tanh^2(\b m^*(1-\t))\G(\a,\t m^*/\rho)\Bigr)\leq 1-\e
}\Eq(LL.19zero)
$$
Then with probability one for all 
but a finite number of indices $N$, $\Phi_{N,\b}[\o](m^* e^1+v)$ is a twice  
differentiable and strictly convex function of $v$ on the set $\{v:\|v\|_2\leq
 \rho\}$, and  
$\l_{min}\left(\nabla^2 \Phi_{N,\b}[\o](m^* e^1+v)\right)>\e$
on this set.}

\remark {The theorem should of course be used for $\rho=c \g m^*$. 
One checks easily that with such $\rho$, the conditions mean: 
(i) For $\b$ close to $1$: $\g$ small and,
(ii) For $\b$ large:  $\a\leq c\b^{-1}$.} 

\remark {In deviation from our general policy not to speak about the 
high-temperature regime, we note that it is of course trivial to show that
$\l_{min}\left(\nabla^2 \Phi_{N,\b}[\o](m)\right)\geq \e$ for all $m$ if
$\b\leq \frac {1-2\e}{(1+\sqrt\a)^2}$. Therefore all the results below can be 
easily extended into that part of the high-temperature regime. Note that this 
does {\it not} cover {\it all} of the high temperature phase, 
which starts already 
at $\b^{-1}=1+\sqrt \a$.}

\proof{  The differentiability for fixed $N$ is no problem. The non-trivial
assertion of the theorem is the local convexity. Since 
$\frac {d^2}{dx^2}\ln\cosh (\b x)=\b\left(1-\tanh^2(\b x)\right)$
we get 
$$
\eqalign{
\nabla^2\Phi&(m^*e^1 +v)= \1-\frac 1N\sum_{i=1}^N
f_\b''(m^*\xi_i^1+(\xi_i,v))\xi_i^T\xi_i\cr
&= \1-\frac \b N\sum_{i=1}^N\xi_i^T\xi_i+
\frac\b N\sum_i \xi_i^T\xi_i \tanh^2(\b (m^*\xi_i^1+(\xi_i,v)))\cr
&\geq \1 -\b \frac {\xi^T\xi}N +\frac \b N\sum_i \xi_i^T\xi_i 
 \1_{\{|(\xi_i,v)|\leq \t m^*\}} \tanh^2(\b m^*(1-\t))\cr
&=\1 -\b\left[1- \tanh^2(\b m^*(1-\t))\right] \frac {\xi^T\xi}N 
\cr
&-  \b \tanh^2(\b m^*(1-\t))\frac 1N  \sum_i \xi_i^T\xi_i 
 \1_{\{|(\xi_i,v)|> \t m^*\}}
}
\Eq(LL.19)
$$
Thus
$$
\eqalign{
\l_{min}&\left(\nabla^2\Phi(
m^*e^1 +v)\right)\geq  
1-\b\left[1- \tanh^2(\b m^*(1-\t))\right]\|A(N)\|\cr
&-
 \b \tanh^2(\b m^*(1-\t))
\left\|
\sfrac 1N\sum_{i=1}^N \1_{\{|(\xi_i,v)|>\t m^*\}}
\xi_i^T\xi_i\right\|
}
\Eq(LL.20)
$$
What we need to do is to estimate the
norm of the last term in \eqv(LL.20).
Now,
$$
\eqalign{
&\sup_{v\in B_\rho}\left\|
\sfrac 1N\sum_{i=1}^N \1_{\{|(\xi_i,v)|>\t m^*\}}
\xi_i^T\xi_i\right\|\cr
& = \sup_{v\in B_\rho}\sup_{w:\|w\|_2=\rho}
\sfrac 1{\rho^2}\sfrac 1N\sum_{i=1}^N \1_{\{|(\xi_i,v)|>\t m^*\}}(\xi_i,w)^2\cr
&\leq \sfrac 1{\rho^2}  \sup_{v\in B_\rho}\sup_{w\in B_\rho}
\sfrac 1N\sum_{i=1}^N \1_{\{|(\xi_i,v)|>\t m^*\}}(\xi_i,w)^2
}
\Eq(LL.21)
$$
To deal with this last expression, notice that
$$
\eqalign{
(\xi_i,w)^2\cr
&=\1_{\{|(\xi_i,v)|>\t m^*\}}
(\xi_i,w)^2\left(\1_ {\{|(\xi_i,w)|<|(\xi_i,v) |\}}+
\1_{\{|(\xi_i,w)|\geq|(\xi_i,v)|\}}\right)\cr
&\leq \1_{\{|(\xi_i,v)|>\t m^*\}}
(\xi_i,v)^2 +\1_{\{|(\xi_i,w)|>\t m^*\}}
(\xi_i,w)^2
\cr
}
\Eq(LL.27)
$$
Thus 
$$
\sfrac 1N\sum_{i=1}^N  \1_{\{|(\xi_i,v)|>\t m^*\}}
(\xi_i,w)^2  =X_{\t m^*}(v)+X_{\t m^*}(w)
\Eq(LL.27bis)
$$
and so  we are reduced to estimating the same  quantities 
as in Section 7. 
Thus using Proposition 7.3 and the estimate \eqv(M.10) with $\e=\sqrt\a$, 
we obtain therefore that with probability greater than $1-e^{-const. \a N}$
for all $v$ with norm less than $\rho$,
$$
\eqalign{
\l_{min}\left(\nabla^2\Phi(
m^*e^1 +v)\right)&\geq  
1-\b\left[1- \tanh^2(\b m^*(1-\t))\right](1+3\sqrt\a)\cr
&-
 2\b \tanh^2(\b m^*(1-\t))
\G(\a,\t m^*/\rho)
}
\Eq(LL.29)
$$
Optimizing over $\t$ gives the claim of the theorem.
\endproof}

\remark {Note that the estimates derived from \eqv(LL.29) become
quite bad if $\b$ is large. Thus local convexity appears to break down 
for some critical $\b_{conv}(\a)$ that tends to infinity, as $\a\downarrow 0$.
In the heuristic picture [AGS] such a critical line appears as the 
boundary of the region where the so-called replica symmetry is supposed to 
hold. It is very instructive to read what Amit et al. write on replica symmetry
breaking in the retrieval phases: {\it ``....the very occurrence of RSB\note{
= replica symmetry breaking} implies that the energy landscape of the basin 
of each of the retrieval phases has features that are similar to the SG\note{=
spin glass} phase. In particular, each of the retrieval phases represents 
many degenerate retrieval states. All of them have the same macroscopic 
overlap  $m$, but they differ in the location of the errors. These states are 
organized in an ultrametric structure''} 
\advance\foot by 2([AGS], page 59). Translated to our 
language, this means that replica symmetry breaking is seen as a failure of 
local 
convexity and the appearance of many local minima. 
On this basis we conjectured in [BG5]
that  replica symmetry is closely 
related to the local convexity of the free energy functional 
\note{We should note, however, that our condition for local convexity 
(roughly $\scriptstyle {\b^{-1}>\a}$) does {\smit not}
have the same behaviour as is found for the stability of  
the replica symmetric solution in [AGS] ($\scriptstyle \b^{-1}>\exp(-1/2\a)$).
 It is rather 
clear that our condition for convexity cannot be substantially improved. On 
the other hand, Talagrand has informed us that his method of deriving the 
replica symmetric solution which does not require convexity, can be 
extended to work under essentially the conditions of [AGS].}}

We can now make these observations more precise. 
While we have so far avoided this,
now is the time to make use of the {\it Hubbard-Stratonovich transformation} 
[HS] 
for the case of quadratic $E_M$. That is, we consider the new measures
$\wt\QQ_{\b,N,M}\equiv \wt\QQ_{\b,N,M}^1$ defined in \eqv(C.12).
They have the remarkable property that they are absolutely continuous 
w.r.t. Lebesgue measure with density
$$
\frac 1{Z_{\b,N,M}} \exp\left(-\b N\Phi_{\b,N,M}(z)\right)
\Eq(L.40)
$$
(do the computation or look it up in [BGP1]). Moreover, in many computations 
it can conveniently replace the original measure $\QQ$.  In particular, 
the following identity holds for all $t\in \R^M$.
$$
\int d\QQ_{\b,N,M}(m) e^{(t,m)}= e^{\frac {\|t\|_2^2}{\b N}}
\int d \wt\QQ_{\b,N,M}(z) e^{(t,z)}
\Eq(L.41)
$$ 
Since for $t$ with bounded norm the first factor tends to one rapidly, 
this shows that the exponential moments of $\QQ$ and $\wt\QQ$ are 
asymptotically 
equal. 
We will henceforth assume that we are in a range of 
$\b$ and $\a$ such that the union of the balls $ B_{\rho(\e)}(sm^*e^\mu)$
has essentially full mass under $\wt\QQ$. 

To study one of the balls, we 
define for simplicity the conditional measures
$$
\wt\QQ^{(1,1)}_{\b,N,M}
\left(\cdot\right)\equiv \wt\QQ_{\b,N,M}
\left(\cdot\,\big|z\in B_{\rho(\e)}(m^*e^1)\right)
\Eq(L.42)
$$
with $\rho(\e)$ such  that  Theorem \ver.1 holds.
(Alternatively we could consider tilted measures with $h$ proportional to $e^1$
and arbitrarily small).
For notational convenience we
will introduce the abbreviation $\E_{\wt\QQ}$ for the expectation w.r.t. the measure
$\wt\QQ^{(1,1)}_{\b,N,M}$. 

Now intuitively one would think that since $\wt\QQ^{(1,1)}_{\b,N,M}$
has a density of the form $e^{-N V(z)}$ with a convex $V$ with 
strictly positive second derivative, this measure should
have similar properties as for quadratic $V$. 
It turns out that this is to some extent true. 
For instance, we have:

\theo {\ver.2} {\it Under the hypothesis of Theorem \ver.1, and with the 
same probability as in the conclusion of that theorem, for any 
$t\in \R^M$ with $\|t\|_2\leq C<\infty$,
$$
e^{(t,\E_{\tilde\QQ} z )}  -O(e^{-M})\leq 
\E_{\wt\QQ} e^{(t,z)} \leq
e^{(t,\E_{\tilde\QQ} z)} e^{\|t\|_2^2/\e N} +O(e^{-M})
\Eq(L.43)
$$
In particular, the marginal distributions of $\QQ$ converge to 
Dirac distributions concentrated on the corresponding projections
of $\E_{\wt\QQ} z$.
}

\proof{ The main tool in proving this Theorem are the so-called 
{\it Brascamp-Lieb inequalities}\note{We thank Dima Ioffe 
for having brought these to our attention} [BL]. 
We paraphrase them as follows.
\lemma {\ver.3}{[Brascamp-Lieb[BL]]{\it Let $V:\R^M\rightarrow\R$
 be non-negative and strictly convex
with $\l_{min}(\nabla^2V)\geq \e$.
Denote by $\E_V$ expectation with respect to the 
probability measure 
$$
  \frac{e^{-NV(x)}d^Mx}
{ \int e^{-NV(x)}d^Mx}
\Eq(L.44)
$$
 Let $f:\R^M\rightarrow \R$ be any  continuously differentiable 
function. Then 
$$
\E_V(f-\E_V f)^2\leq \frac 1{\e N}\E_V(\|\nabla f\|_2^2)
\Eq(L.45)
$$
}}
We see that we are essentially in a situation where we can apply Lemma \ver.3.
The only difference is that our measures are supported only on a subset of 
$\R^M$.
This is however no problem: we may either continue the function 
$\Phi(m)$ as a strictly convex function to all $\R^M$ and study the
 corresponding measures noting that all reasonable expectations differ only by exponentially small terms, or one may run through the proof
of Lemma \ver.3 to see that the boundary terms we introduce only lead to 
exponentially small error terms in \eqv(L.45). We will disregard this issue 
in order not to complicate things unnecessarily. 
To see how Lemma \ver.3 works, we deduce the following 
\corollary {\ver.4} {\it Let $\E_V$ be as in Lemma \ver.3. Then
\item {(i)} $\E_V\left\|x-\E_V x\right\|_2^2 \leq  \frac M{\e N}$
\item{(ii)} $\E_V\left\|x-\E_V x\right\|_4^4 \leq 4\frac  M{\e^2 N^2}$
\item{(iii)} For any function $f$ such that $V_t(x)\equiv
 V(x)-tf(x)/N$ for $t\in [0,1]$ is still 
strictly convex and $\l_{min}(\nabla^2 V_t)\geq \e'>0$, then
$$
0\leq\ln \E_V e^f -\E_V f\leq \frac 1{2\e' N}
\sup_{t\in [0,1]} \E_{V_t}\|\nabla f\|_2^2 
\Eq(L.45bis)
$$
In particular,
\item{(iv)}$\ln \E_V e^{(t,(x- \E_V x))}
 \leq   \frac {\|t\|_2^2}{2\e N}$
\item{(v)} $\ln \E_V e^{ \left\|x-\E_V x\right\|_2^2}
- \E_V\left\|x-\E_V x\right\|_2^2 \leq  \frac  M{\e^2 N^2}$
}
\proof{ (i) Choose $f(x)=x_\mu$ in \eqv(L.45). Insert and sum.
(ii) Choose $f(x)=x_\mu^2$ and use (i). (iii) Note that
$$
\eqalign{
\ln \E_V e^{f}&=\E_V f+\int_{0}^1 ds\int_0^s d s'\frac{
\E_V\left[e^{s'f}\left(f-
\frac{\E_Ve^{s'f}f}
{\E_Ve^{s'f}}\right)^2\right]}
{\E_Ve^{s'f}}\cr
&
=\E_V f+\int_{0}^1 ds\int_0^s d s' \E_{V_{s'}} 
\left(f-\E_{V_{s'}}f\right)^2
}
\Eq(L.46)
$$
where by assumption
$V_s(x)$ has the same properties as $V$ itself.
Thus using \eqv(L.45) gives \eqv(L.46) (iv) and (v) follow with 
the corresponding choices for $f$ easily. \endproof}
Theorem \ver.2 is thus an immediate consequence of (iv). \endproof}  

We now come to  the main result of this section. 
We will show that Theorem \ver.1 in fact 
implies that the replica symmetric solution of [AGS] is correct in  
the range of 
parameters where  Theorem \ver.1 holds. Such a result was recently proven by
Talagrand [T4], but we shall see that using Theorem \ver.1 and the 
Brascamp-Lieb inequalities, we can give a greatly simplified proof.  

\theo{\ver.5} {\it Assume that the  parameters $\b,\a$ are such that the 
conditions both of Theorem 6.2 and of  Theorem \ver.1 are satisfied,
with $\e>0$ and $\rho\geq 
c\g m^*$, where $c$ is such that the mass of the complement of the set
$\cup_{s,\mu} B_{c \g m^*}(sm^*e^\mu)$ is negligible. Then, the replica 
symmetric solution of [AGS] holds in the sense that, asymptotically, as
$N\uparrow \infty$, $\E_{\wt\QQ} z_1$, and 
$\E\|\E_{\wt\QQ} \hat z\|_2^2$  (recall that 
$\hat z\equiv(0,z_2,\dots)$ converge almost surely to the 
positive solution $\m$ and $r$ of the system of equations
$$
\m =\int d\NN (g) \tanh(\b( \m +\sqrt{\a r}g))
\Eq(L.RS1)
$$
$$
q=\int d\NN (g)  \tanh^2(\b( \m +\sqrt{\a r}g))
\Eq(L.RS2)
$$
$$
r=\frac q{(1-\b+\b q)^2}
\Eq(L.RS3)
$$
(note that $q$ is an auxiliary variable that could be eliminated).
}

\remark {As far as Theorem \ver.5 is considered as a result on {\it conditional
measures} only, it is possible to extend its validity beyond the regime of 
Theorem 6.2. In that case, what is needed is only Theorem \ver.1 {\it and} 
the control of the location of the local minima given by Theorem 7.5. 
One may also, in this spirit, consider the extension of this result to other 
local minima (corresponding to the so-called ``mixed patterns''), which would, 
of course, require to prove the analogues of Theorem 7.5, \ver.1 in this case, 
as well as carrying out the stability analysis of a certain dynamical system
(see below). We do not doubt that this can be done.  }

\remark {We will not enter into the discussion on how these equations were 
originally derived with the help of the replica trick. This is well 
explained in [AGS]. In [T4] it is also shown how one can derive on 
this basis the formula for the free energy as a function of 
$\m,r$, and $q$ that is given in [AGS] and for which the above 
equations are the saddle point equations. 
We will not repeat these arguments here.}

\remark {In [PST] it was shown that the replica symmetric solution 
holds if the so-called Edwards-Anderson parameter, $\frac 1N\sum_i
[\mu_{\b,N,M}(\s_i)]^2$ is self-averaging. 
Some of the basic ideas in that paper are used both in 
Talagrand's and in our proof below. In fact we follow the strategy of 
[PST] more closely than Talagrand, and we will see that this leads 
immediately to the possibility of studying the limiting Gibbs measures. }

\medskip
\noindent{\bf Proof.}  
It may be well worthwhile to outline the strategy  of the proof in a slightly 
informal way before we
go into the details. This may also give a new explanation to the 
mysterious looking equations above. It turns out that in a very specific sense,
the idea of these equations and their derivation is closely related to the 
original idea of  ``mean field theory''. 
Let us briefly recall what this means. 
The standard derivation of ``mean field'' equations for 
homogeneous magnets in most textbooks 
on statistical mechanics does not start from the Curie-Weiss model
but from (i) the hypothesis that in the infinite volume limit, the 
spins are independent and identically distributed 
under the limiting (extremal) Gibbs measure and that (ii) their
distribution is of the form $e^{\b \s_i m}$ where $m$ is the mean value of the 
spin under this same measure, and that is assumed to be an almost sure
 constant with 
respect to the Gibbs measure. The resulting consistency equation is then 
$m=\tanh\b m$. This derivation breaks down in random systems, since it would be
unreasonable to think that  the spins are identically distributed. Of course one may
keep the assumption of independence, and write down a set of consistency equations 
(in the spin-glass case, these are know as TAP-equations [TAP]).
Let us try the idea in Hopfield model. The spin $\s_i$ here couples to a 
``mean field'' $ h_i(\s)=(\xi_i,m(\s))$, which is a function of the entire 
vector of magnetizations. To obtain a self-consistent set of equations 
we would have to compute all of these, leading to the system
$$
m_\mu = \frac 1N\sum_i\xi_i^\mu \tanh(\b(\xi_i,m))
\Eq(H.1)
$$ 
Solving this  is a hopelessly difficult task when $M$ is growing somewhat 
fast with $N$, 
and it is not clear why one should expect these quantities to be constants 
when $M=\a N$. 
\hfill\break
But now suppose it were true that we could somehow compute the {\it distribution} 
of $h_i(\s)$ {\it a priori}  as a function of a small number of parameters,
not depending on $i$. Assume 
further that these parameters are again functions of the distribution 
of the mean field. Then we could write down consistency conditions for 
them and (hopefully) solve them. In this way  the expectation of $\s_i$ could 
be computed.  
The tricky part is thus to find the {\it distribution} of the mean field
\note{This idea seems related to statements of physicists one finds sometimes
in the literature that in spin glasses, that 
the relevant ``order parameter'' is a 
actually a probability distribution.}. 
Miraculously, this can be done, and the relevant parameters turn out to 
be the quantities $\m$ and $r$, with \eqv(L.RS1)-\eqv(L.RS3) the 
corresponding consistency equations\note{In fact, we will see that 
the situation is just a bit more complicated. For finite $\scriptstyle N$, the 
distribution of the mean field will be seen to depend essentially on 
three $\scriptstyle N$-dependent, non-random quantities whose limits, 
{\smit should they exist}, are related to   $\scriptstyle \m$, 
$\scriptstyle r$ and $\scriptstyle q$. Unfortunately, 
one of the notorious problems in disordered mean field type models is that one 
cannot prove a priori such intuitively obvious facts like that the 
mean values  of  thermodynamic quantities (such as the free energy, etc.) 
converge, even when it is possible to show that their fluctuations converge 
to zero (this sad fact is  sometimes overlooked). We shall see that 
convergence of the quantities involved here
can be proven in the process, using properties of the recurrence equations for 
which the equations above are the fixed point equations, and a priori control 
on the overlap distribution as results from Theorem 6.2 (or 7.5).}

We will now follow these ideas and give the individual steps a precise 
meaning. 
In fact, the first step in our proof corresponds to proving  a 
version of Lemma 2.2 of [PST], or
if one prefers, a sharpened version of Lemma 4.1 of [T4]. Note that 
we will never introduce any auxiliary Gaussian fields in the Hamiltonian,
as is done systematically in [PST] and sometimes in [T4]; all 
comparison to quantities in these papers is thus understood modulo removal 
of such terms.  
Let us begin by mentioning that the crucial quantity $u(\t)$ defined in 
Definition 5 of [PST] has the following nice representation\note{Actually,
our definition differs by an irrelevant constant from that of 
[PST].}
$$
u(\t)= \ln \int d\wt\QQ_{\b,N,M}^{(1,1)}(z) e^{\t\b (\eta,z)} 
\Eq(L.52)
$$
where, like  Talagrand in [T4], we singled out the site $N+1$
(instead of $1$ as in [PST]) and set $\xi_{N+1}=\eta$. 
For notational simplicity we will
denote the expectation w.r.t. the measure $\wt\QQ_{\b,N,M}^{(1,1)}$ by
$\E_{\wt\QQ}$ and we will set $\bar z = z-\E_{\wt\QQ} z$.

\lemma{\ver.6} {\it Under the hypotheses of Theorem \ver.5 we have that
\item{(i)} With probability exp. close to 1,
$$
\E_\eta  \E_{\wt\QQ} e^{\t\b (\eta,\bar z)} 
=e^{ \frac {\t^2\b^2}{2}\E_{\wt\QQ} \| \bar z \|_2^2+R}
\Eq(L.53)
$$
where $|R|\leq \frac CN$.
\item{(ii)} Moreover,
$$
\E_\eta\left(  \E_{\wt\QQ} e^{\t\b (\eta,\bar z)}-
\E_\eta\E_{\wt\QQ} e^{\t\b (\eta,\bar z)}\right)^2\leq \frac CN
\Eq(L.54)
$$
}

\proof{   Note first that
$$
\E_\eta  \E_{\wt\QQ} e^{\t\b (\eta,\bar z)} 
\leq \E_{\wt\QQ} e^{\frac {\t^2\b^2}2 \|\bar z\|_2^2}
\Eq(L.55)
$$
and also 
$$
\E_\eta  \E_{\wt\QQ} e^{\t\b (\eta,\bar z)} 
\geq \E_{\wt\QQ} e^{\frac {\t^2\b^2}2 \|\bar z\|_2^2-
\frac {\t^4\b^4}{4} \|\bar z\|_4^4 }
\Eq(L.56)
$$
\eqv(L.55) looks most encouraging 
and (ii) of Corollary \ver.4 leaves hope for the $\|\bar z\|_4^4$ to be 
irrelevant. Of course for this we want the expectation to move up into the 
exponent. To do this, we use (iii) of Corollary \ver.4 with $f$ 
chosen as $\frac {\t^2\b^2}2 \|\bar z\|_2^2$ and
$\frac {\t^2\b^2}2 \|\bar z\|_2^2-\frac {\t^4\b^4}{12} \|\bar z\|_4^4$,
respectively. For this we have to check the strict convexity of 
$\Phi +\frac sN f$ in these cases. But a simple 
computation shows that in both cases 
$\l_{\min}\left(\nabla^2(\Phi +\frac sN f)\right)
\geq \e  -\frac {\t \b}N$, so that for any $\t,\b$ there is no problem if $N$
is large enough (Note that the quartic term has the good sign!). 
A straightforward  calculation shows 
that this gives \eqv(L.53). 
\hfill\break
To prove (ii), it is enough to compute 
$$
\E_\eta \left(\E_{\wt\QQ} e^{\t\b (\eta,\bar z)}\right)^2=
\E_\eta \E_{\wt\QQ} e^{\t\b (\eta,\bar z+\bar z')}
\Eq(L.59)
$$
where we (at last!) introduced the ``replica'' $z'$ that is an independent copy
of the random variable $z$. By some abuse of notation $\E_{\wt\QQ}$ also denotes 
the product measure for these two copies. By the same token as in the proof
of (i), we see that,
$$
\E_\eta \E_{\wt\QQ} e^{\t\b (\eta,\bar z+\bar z')}=
e^{\frac {\t^2\b^2}2 \E_{\wt\QQ}\|\bar z+\bar z'\|_2^2 +O(1/N)}
\Eq(L.60)
$$
Finally,
$$
\E_{\wt\QQ}\|\bar z+\bar z'\|_2^2 =2\E_{\wt\QQ}\|\bar z\|_2^2
+2\E_{\wt\QQ}(\bar z,\bar z')=2\E_{\wt\QQ}\|\bar z\|_2^2
\Eq(L.61)
$$
Inserting this and \eqv(L.53) into the left hand side of \eqv(L.54) 
establishes that bound. 
This concludes the proof of Lemma \ver.6. \endproof}

An easy corollary gives what Talagrand's Lemma 4.1 should be:

\corollary {\ver.7} {\it Under the hypotheses of Lemma \ver.6, 
there exists a finite numerical constant $c$ such that 
$$
u(\t)= \b\t (\eta,\E_{\wt\QQ} z)+\frac{\t^2\b^2}2 \E_{\wt\QQ}\|\bar z\|_2^2 + R_N
\Eq(L.63)
$$ 
where 
$$
\E| R_N|^2\leq \frac c{ N}
\Eq(L.64)
$$
}

\proof{  Obviously
$$
\E_{\wt\QQ} e^{\t\b (\eta,z)}=
 e^{\t\b (\eta,\E_{\wt\QQ} z)} \E_\eta \E_{\wt\QQ} e^{\t\b (\eta,\bar z)}
\frac{ \E_{\wt\QQ} e^{\t\b (\eta,\bar z)}}
{ \E_\eta \E_{\wt\QQ} e^{\t\b (\eta,\bar z)}}
\Eq(L.65)
$$
Taking logarithms, the first two factors in \eqv(L.65) together with
\eqv(L.53) give  the two first terms in \eqv(L.63) plus a remainder 
of order $\frac 1N$. For the last factor, we notice first that by Corollary
\ver.4, (iii),
$$
e^{-\t^2\b\frac {M}{\e N}}\leq  \E_{\wt\QQ} e^{\t\b (\eta,\bar z)}
\leq e^{\t^2\b\frac {M}{\e N}}
\Eq(L.66)
$$
so that for $\a$ small, $\t$ and $\b\a$ bounded, 
$\E_{\wt\QQ} e^{\t\b (\eta,\bar z)}$
is bounded away from $0$ and infinity; we might for instance 
think that $ \frac 12\leq \E_{\wt\QQ} e^{\t\b (\eta,\bar z)}\leq 2$.
But for $A,B$ in a compact interval of the positive half line not containing
zero, there is a finite constant $C$ such that
$|\ln \frac AB|=|\ln A-\ln B|\leq C |A-B|$. Using this gives
$$
\E_\eta \left[\ln\frac { \E_{\wt\QQ} e^{\t\b (\eta,\bar z)}}
{ \E_\eta \E_{\wt\QQ} e^{\t\b (\eta,\bar z)}}\right]^2 \leq C^2\E_\eta
\left( \E_{\wt\QQ} e^{\t\b (\eta,\bar z)}- \E_\eta \E_{\wt\QQ} e^{\t\b (\eta,\bar z)}
\right)^2
\Eq(L.67)
$$
From this and \eqv(L.54)
follows the estimate \eqv(L.64). \endproof}

 We have almost proven the equivalent of Lemma 2.2 in [PST]. What remains to 
be shown is

\noindent {\bf Lemma {\ver.8}:} {\it 
Under the assumptions of Theorem \ver.1 
$(\eta, \E_{\wt\QQ} z)$ converges in law to $\eta_1\m +\sqrt{\a r} g$
where $\m=\lim_{N\uparrow \infty}\E_{\wt\QQ} z_1$ and 
$r\equiv \a^{-1}\lim_{N\uparrow \infty} \left\|\E_{\wt\QQ} \hat z\right\|_2^2$, 
where
$\hat z\equiv (0,z_2,z_3,\dots,)$ and $g$ is a standard normal random variable.
}
\medskip
\noindent {\bf Quasiproof:}[PST] The basic idea behind this 
lemma is that for all $\mu>1$, 
$\E_{\wt\QQ} z_\mu$ tends to zero, the $\eta_\mu$ are independent 
amongst each other and 
of the $\E_{\wt\QQ} z_\mu$ and that therefore 
 $\sum_{\mu>1} \eta_\mu \E_{\wt\QQ} z_\mu$  converge to a Gaussians with 
variance $\lim_{N\uparrow \infty} \left\|\E_{\wt\QQ} \hat z\right\|_2^2$. 
\endproof\medskip

To make this idea precise is somewhat subtle.
First, to prove a central limit theorem, 
one has to show that some version of the Lindeberg condition [CT] is
 satisfied in an 
appropriate sense.  To do this  
we need some more facts about self-averaging.
Moreover, one has to make precise to what extent the quantities 
$\E_{\wt\QQ} z_1$ and 
$ \left\|\E_{\wt\QQ} \hat z\right\|_2^2$ converge, as $N$ tends to infinity.
There is no way to prove this a priori, and only at the end of the proof of Theorem 
\ver.5 will it be clear that this is the case. Thus we cannot and will not use 
Lemma \ver.8 in the proof of the Theorem, but a weaker statement formulated as Lemma
\ver.13 below.

The following lemma follows easily from the proof of Talagrand's Proposition 
4.3 in [T5].

\lemma {\ver.9} {\it Assume that $f(x)$ is a convex random 
function defined on some open neighborhood $U\subset \R$. 
Assume that $f$ verifies for all $x\in U$ that 
$|(\E f)''(x)|\leq C<\infty$ and  
$\E(f(x)-\E f(x))^2\leq S^2$. Then, if $x\pm S/C\in U$
$$
\E\left( f'(x)-\E f'(x)\right)^2\leq 12 C S
\Eq(L.68) 
$$
}
 
But as so often in this problem, variance estimates are not quite sufficient. 
We will need the following, sharper estimate (which may be well known):

\lemma {\ver.10} {\it Assume that $f(x)$ is a  random 
function defined on some open neighborhood $U\subset \R$. 
Assume that $f$ verifies for all $x\in U$
that for all $0\leq r\leq 1$, 
$$
\P\left[|f(x)-\E f(x)|>r\right]\leq c \exp\left(-\frac {Nr^2}c\right)
\Eq(L.69)
$$
and that, at least with probability $1-p$,
 $|f'(x)|\leq C$, $|f''(x)|\leq C<\infty$ both hold uniformly in $U$. 
Then, for any $0<\zeta\leq 1/2$, and for any $0<\d<N^{\zeta/2}$, 
$$
\P\left[|f'(x)-\E f'(x)|> \d N^{-\zeta/2}\right]
\leq \frac {32 C^2}{\d^2}N^\zeta\exp\left(- \frac {\d^4N^{1-2\zeta}}{256c}\right)+p
\Eq(L.69bis)
$$
}

\proof{ Let us assume that $|U|\leq 1$. We may  first assume 
that the boundedness conditions for the derivatives of $f$ 
hold uniformly; by standard arguments one 
shows that if they only hold with probability $1-p$, the effect is nothing more
than the final summand $p$ in \eqv(L.69bis).
The first step in the proof consists in showing that \eqv(L.69)
together with the boundedness of the derivative of $f$ 
implies that $f(x)-\E f(x)$ is uniformly small.
To see this introduce a grid of spacing $\e$, i.e. let
 $U_\e =U\cap \e\Z$. Clearly
$$
\eqalign{
&\P\left[\sup_{x\in U}|f(x)-\E f(x)|>r\right]\cr
&\leq 
\P\Biggl[\sup_{x\in U_\e}|f(x)-\E f(x)|\cr
&\quad\quad+\sup_{x,y: |x-y|\leq \e}
|f(x)-f(y)| +|\E f (x)-\E f(y)|>r\Biggr] \cr
&\leq \P\left[\sup_{x\in U_\e}|f(x)-\E f(x)|>r- 2C\e\right]
\cr
&\leq \e^{-1}  \P\left[|f(x)-\E f(x)|>r- 2C\e\right]
}
\Eq(L.71)
$$
If we choose $\e=\frac r{4C}$, this yields 
$$
\P\left[\sup_{x\in U}|f(x)-\E f(x)|>r\right]\leq 
\frac {4C}r \exp\left(-\frac {N r^2}{4 c}\right)
\Eq(L.72)
$$
Next we show that {\it if } $\sup_{x\in U}|f(x)-g(x)|\leq r$ for two functions 
$f$, $g$ with bounded second derivative, then
$$
|f'(x)-g'(x)|\leq \sqrt {8Cr}
\Eq(L.73)
$$
For notice that
$$
\left|\frac 1\e [f(x+\e)-f(x)]- f'(x)\right|\leq \frac \e 2 
\sup_{x\leq y\leq x+\e} f''(y) \leq C\frac \e 2
\Eq(L.74)
$$
so that
$$
\eqalign{
|f'(x)-g'(x)|&\leq \frac 1\e|f(x+\e) -g(x+\e)-f(x)+g(x)|+ C \e
\cr 
&\leq \frac {2r}\e +C\e
}
\Eq(L.75)
$$
Choosing the optimal $\e=\sqrt{2r/C}$ gives \eqv(L.73). It suffices to combine
\eqv(L.73) with \eqv(L.72) to get
$$
\P\left[|f'(x)-\E f'(x)|>\sqrt {8 rC}\right]\leq \frac {4C}r 
\exp\left(-\frac {N r^2}{4 c}\right)
\Eq(L.76)
$$
Setting $r=\frac {\d^2}{ C N^{\zeta}}$, we arrive at \eqv(L.69bis). \endproof}

We will now use Lemma \ver.10 to control $\E_{\wt\QQ} z_\mu$. We define
$$
f(x)=\frac 1{\b N} \ln \int_{B_\rho(m^*e^1)} d^Mz e^{\b N x z_\mu}
e^{-\b N\Phi_{\b,N,M}(z)}
\Eq(L.77)
$$
and denote by $\E_{\wt\QQ^x}$ the corresponding modified expectation.
As has by now been shown many times [T2,BG5,T4], $f(x)$ verifies 
\eqv(L.69). Moreover,
$f'(x)= \E_{\wt\QQ^x}z_\mu$ and
$$
f''(x)=\b N \E_{\wt\QQ^x}\left(z_\mu- \E_{\wt\QQ^x}z_\mu\right)^2
\Eq(L.78)
$$
Of course the addition of the linear term to $\Phi$ does not change its 
second derivative, so that we can apply the Brascamp-Lieb inequalities
also to the measure $\E_{\wt\QQ^x}$. This shows that 
$$
 \E_{\wt\QQ^x}\left(z_\mu- \E_{\wt\QQ^x}z_\mu\right)^2\leq \frac 1{\e N\b}
\Eq(L.79)
$$ 
which means that $f(x)$ has a second derivative bounded by $c=\frac 1\e$. 

\remark {In the sequel we will use Lemma \ver.10 only in situations where
$p$ is irrelevantly small compared to the main term in \eqv(L.69bis). We 
will
thus ignore its existence for simplicity.}

This gives the 

\corollary {\ver.11} {\it Under the assumptions of Theorem \ver.1, there are 
finite positive constants $c, C$ such that, for 
any $\zeta\leq \frac 12$ and $\d\leq N^{\zeta/2}$, for any $\mu$, 
$$
\P\left[|\E_{\wt\QQ} z_\mu-\E\E_{\wt\QQ} z_\mu|\geq \d N^{-\zeta/2}\right]\leq
 \frac { C}{\d^2}N^\zeta\exp\left(- \frac {\d^4N^{1-2\zeta}}{c}\right)
\Eq(L.80)
$$
}

This leaves us only with the control of $\E\E_{\wt\QQ} z_\mu$. But by symmetry,
for all $\mu>1$, 
$\E\E_{\wt\QQ} z_\mu=\E\E_{\wt\QQ} z_2$ while on the other hand
$$
\sum_{\mu=2}^M (\E\E_{\wt\QQ} z_\mu)^2\leq c^2 \g^2 (m^*)^2 
\Eq(L.81)
$$
so that $|\E\E_{\wt\QQ} z_\mu|\leq \frac{c}{m^*} N^{-1/2}$. Therefore,
with probability of order, say $1-\exp(-N^{1-2\zeta})$ it is true that 
for all $\mu>2$,
$|\E_{\wt\QQ} z_\mu|\leq \d N^{-\zeta/2}$. 

Finally we must control the behaviour of the prospective variance of 
our gaussian. We set
$T_N \equiv \sum_{\mu=2}^{M (N)} (\E_{\wt\QQ} z_\mu)^2$. 
Let us introduce
$$
g(x)\equiv \frac 1{\b N}\ln 
 \E_{\wt\QQ} e^{\b N x (\hat z, \hat z')}
\Eq(L.82)
$$
where $\E_{\wt\QQ}$ is understood as the product measure for the two independent
copies $z$ and $z'$.
The point is that $T_N= g'(0)$. On the other hand, $g$ satisfies the same 
self-averaging conditions as the function $f$ before, and its second 
derivative is bounded (for $x\leq \e/2$), since
$$
\eqalign{
g''(x)&=\b N
\E_{\wt\QQ^x}\left((\hat z,\hat z')-\E_{\wt\QQ^x}(\hat z,\hat z')\right)^2\cr
&\leq \frac {2\b}{\e} 2 \E_{\wt\QQ^x}\|\hat z\|_2^2\leq 2\rho \frac \b\e
}
\Eq(L.83)
$$
where here $\E_{\wt\QQ}^x$ stands for the coupled measure corresponding 
to \eqv(L.82) (and is not the same as the the measure with the same name in 
\eqv(L.78)).
Thus we get our second corollary:

\corollary {\ver.12} {\it Under the assumptions of Theorem \ver.1, there are 
finite positive constants $c, C$ such that,
 for 
any $\zeta\leq \frac 12$ and $\d\leq N^{\zeta/2}$, 
$$
\P\left[|T_N-\E T_N|\geq \d N^{-\zeta/2}\right]\leq
 \frac { C}{\d^2}N^\zeta\exp\left(- \frac {\d^4N^{1-2\zeta}}{c}\right)
\Eq(L.84)
$$
}

Thus $T_N$ converges almost surely to a constant if $\E  T_N$ converges.
We are now in a position to prove

\lemma {\ver.13} {\it Consider the 
random variables
$
X_N\equiv \frac 1{\sqrt{\E T_N}}\sum_{\mu=2}^{M(N)} \eta_\mu \E_{\wt\QQ} z_\mu
$. Then, if the hypotheses of Theorem \ver.5 are satisfied, 
$X_N$ converges weakly to a gaussian random variable of mean zero and variance
one.}

\proof{ Let us show that $\E e^{itX_N}$ 
converges to $e^{-t^2/2}$. To see this, 
let $\O_N$ denote the subset of $\O$ on which  the various nice things 
we want to impose on $ \E_{\wt\QQ} z_\mu$ hold; we know that the complement of that 
set has measure smaller than $O(e^{-N^{1-2\zeta}})$.
We write
$$
\eqalign{
\E e^{itX_N}&=
\E_\xi\left[\1_{\O_N} \E_\eta e^{itX_N}+\1_{\O_N^c} \E_\eta e^{itX_N}\right]\cr
&=\E_\xi\left[\1_{\O_N} \prod_{\mu}\cos\left(\frac t{\sqrt {\E T_N}}
\E_{\wt\QQ} z_\mu\right)\right]+O\left(e^{-N^{1-2\zeta}}\right)
}
\Eq(L.85)
$$
Thus the second term tends to zero rapidly and can  be forgotten. On the 
other hand, 
on $\O_N$,  
$$
\sum_{\mu=2}^M (\E_{\wt\QQ} z_\mu)^4\leq \d^2 N^{-\zeta}\sum_{\mu=2}^M (\E_{\wt\QQ} z_\mu)^2
\leq \d^2 N^{-\zeta} \frac {c\a}{(m^*)^2}
\Eq(L.86)
$$ 
tends to zero,
so that using for instance $|\ln \cos x -x^2/2|\leq c x^4$ for $|x|\leq 1$,
$$
\eqalign{
&\E_\xi\1_{\O_N} \E_\eta e^{itX_N}\cr
&\leq e^{-t^2/2} 
\sup_{\O_N}\left[\exp\left(-\frac {T_N-\E T_N}{2\E T_N}+c 
\frac{t^4 \d^2 N^{-\zeta}}{(\E T_N)^2}\right) \right] \P_\xi(\O_N)
}
\Eq(L.87)
$$
Clearly, since also $|T_N-\E T_N|\leq \d N^{-\zeta/2}$, the right hand side 
converges to $e^{-t^2/2}$ and this proves the lemma. \endproof}

Corollary \ver.7 together with Lemma \ver.13  represent the  complete 
analogue  of Lemma 2.2 of 
[PST]. To derive from here the equations
\eqv(L.RS1)-\eqv(L.RS3) requires actually a little more, namely a 
corresponding statement on the convergence of the derivative of $u(\t)$.
Fortunately, this is not very hard to show.

\lemma {\ver.14} {\it Set $u(\t)=u_1(\t)+u_2(\t)$, where
$u_1(\t)=\t\b (\eta,\E_{\wt\QQ} z)$ and $u_2(\t)=\ln \E_{\wt\QQ} e^{\b\t(\eta,\bar z)}$.
Then under the assumption of Corollary \ver.13,
\item{(i)}
$\frac 1{\b\sqrt {\E T_N}}\frac d{d\t} u_1(\t)$ converges weakly to 
a standard gaussian random variable. 
\item{(ii)} 
$\left|\frac d{d\t} u_2(\t)-\t \b^2\E\E_{\wt\QQ}\|\bar z\|_2^2\right|$
converges to zero in probability.
}

\proof{ (i) is obvious from Corollary \ver.13. To prove (ii), note that
$u_2(\t)$ is convex and $\frac {d^2}{d\t^2} u_2(\t)\leq \frac {\b\a}{\e}$. 
Thus, {\it if} $\hbox{var}\left(u_2(\t)\right)\leq \frac C{\sqrt {N}}$, then 
 $\hbox{var}\left(\frac {d}{d\t} u_2(\t)\right)\leq \frac {C'}{ N^{1/4}}$
by Lemma \ver.9. On the other hand, 
$|\E u_2(\t)-\frac {\t^2\b^2}2\E\E_{\wt\QQ}\|\bar z\|_2^2|\leq \frac K{\sqrt N}$,
by Corollary \ver.7, which, together with the boundedness of the second 
derivative of $u_2(\t)$ implies that 
$|\frac d{d\t} \E u_2(\t)-\t\b^2
\E\E_{\wt\QQ}\|\bar z\|_2^2|\downarrow 0$. This 
means that  $\hbox{var}\left(u_2(\t)\right)\leq \frac C{\sqrt {N}}$
implies the Lemma. Since we already know that $\E R_N^2\leq \frac KN$, it is
enough to prove $\hbox{var}\left(\E_{\wt\QQ}\|\bar z\|_2^2\right)\leq 
\frac C{\sqrt {N}}$. But this is a, by now, familiar exercise. The point is 
to use that 
$\E_{\wt\QQ}\|\bar z\|_2^2=\frac d{dx}\tilde g(x)$,
where
$$
\tilde g(x)
\equiv \frac 1{\b N}\ln 
 \E_{\wt\QQ} e^{\b N x \|\bar z\|_2^2}
\Eq(L.88)
$$
and to prove that 
$\hbox{var}\left(\tilde g(x)\right)\leq \frac KN$. using what we know about
$\|\E_{\wt\QQ} z\|_2$ this follows as in the case of the function $g(x)$. 
The proof is finished.\endproof}

From here we can
follow [PST]. Let us denote by $\E_\QQ $ the expectation with respect to the 
(conditional) induced measures $\QQ_{\b,N,M}^{(1,1)}$. 
Note first that \eqv(L.41) implies that\note{This relation is exact, if 
the tilted measures
are considered, and it is true up to irrelevant error terms if one 
considers the conditioned measures.}{  }
$\E_{\QQ}m_\mu=\E_{\wt\QQ}z_\mu$. 
On the other hand,
$$
\E_\QQ m_\mu=\frac 1N\sum_{i=1}^N\xi_i^\mu \mu^{(1,1)}_{\b,N,M}(\s_i)
\Eq(L.89)
$$
and so, by symmetry
$$
\E\E_{\wt\QQ_{N+1}}(z_\mu)= \eta^\mu  \E\mu_{\b,N+1,M}(\s_{N+1})
\Eq(L.90)
$$
Note that from here on we will make the $N$-dependence of our mesures explicit,
as we are going to derive recursion relations.
Now, $u(\t)$ was defined such that 
$$
\eqalign{ 
\E\mu_{\b,N+1,M}(\s_{N+1}) &=\E \frac{e^{u(1)}-e^{u(-1)}}{e^{u(1)}+e^{u(-1)}}
\cr &= \E \tanh(\b(\eta_1 \E_{\wt\QQ_N} z_1 + \sqrt{\E T_N} X_N)) +o(1)
}
\Eq(L.91)
$$
Thus, if $ \E_{\wt\QQ_N} z_1$ and $\E T_N$ converge, by Lemma \ver.13, the limit 
must satisfy \eqv(L.RS1).
Of course we still need an equation for $\E T_N$ which is somewhat  
tricky. 
Let us first {\it define} a quantity $\E Q_N$ by 
$$
\E Q_{N}\equiv  \E \tanh^2(\b(\eta_1 \E_{\wt\QQ_{N}} z_1 + \sqrt{\E T_{N}} 
X_{N}))
\Eq(L.911)
$$
This corresponds of course to \eqv(L.RS2). Now note that 
$T_N= \|\E_{\wt\QQ_N} z\|_2^2 -(\E_{\wt\QQ_N} z_1)^2$ and 
$$
\eqalign{ 
\E\|&\E_{\wt\QQ_{N+1}} z\|_2^2
=\sum_{\mu=1}^M\E\left(\frac 1{N+1}\sum_{i=1}^{N+1}\xi_i^\mu
\mu_{\b,{N+1},M}(\s_i)\right)^2\cr
&=\frac {M-1} {N+1} \E  \left(\mu^{(1,1)}_{\b,N+1,M}(\s_{N+1})\right)^2\cr
&+\sum_{\mu=1}^M \E \xi_{N+1}^\mu \mu^{(1,1)}_{\b,N+1,M}(\s_{N+1})
\left(\frac 1{N+1} \sum_{i=1}^N \xi_i^\mu \mu_{\b,{N+1},M}(\s_i)\right)
}
\Eq(L.912)
$$
We see that the first term gives, by definition and \eqv(L.91), $\a \E Q_{N}$.
For the second term, we use the identity form [PST]
$$
\sum_{\mu=1}^M \xi_{N+1}^\mu\left(\frac 1{N}
 \sum_{i=1}^N \xi_i^\mu \mu_{\b,{N+1},M}(\s_i)\right)
=\b^{-1}\frac{\sum_{\t=\pm 1} u'(\t) e^{u(\t)}}{\sum_{\t=\pm 1}e^{u(\t)}}
\Eq(L.913)
$$ 
which it is not too hard to verify.
Together with Lemma \ver.14 one concludes that in law up to small 
errors
$$
\eqalign{
&\sum_{\mu=1}^M \xi_{N+1}^\mu\left(\frac 1{N}
 \sum_{i=1}^N \xi_i^\mu \mu_{\b,{N+1},M}(\s_i)\right)
=\xi_{N+1}^1\E_{\wt\QQ_N} z_1 +\sqrt{\E T_N} X_N \cr
&\quad+\b \E_{\wt\QQ_N}\|\bar z\|_2^2
\tanh\b\left(\xi_{N+1}^1\E_{\wt\QQ_N} z_1 +\sqrt{\E T_N} X_N\right)
}
\Eq(L.914)
$$
and so
$$ 
\eqalign{  
\E\|\E_{\wt\QQ_{N+1}} z\|_2^2
 &=\a \E Q_{N} +\E \Biggl[ \tanh\b\left(\xi_{N+1}^1\E_{\wt\QQ_N} z_1 +
\sqrt{\E T_N} X_N\right) \cr
&\times\left[\xi_{N+1}^1\E_{\wt\QQ_N} z_1 +\sqrt{\E T_N} X_N\right]\Biggr]\cr
&+\b\E \E_{\wt\QQ_N}\|\bar z\|_2^2  
\tanh^2\b\left(\xi_{N+1}^1\E_{\wt\QQ_N} z_1 +
\sqrt{\E T_N} X_N\right)
}
\Eq(L.915)
$$
Using the self-averaging properties of $\E_{\wt\QQ_N}\|\bar z\|_2^2$,
the last term is of course essentially equal to 
$$
\b\E \E_{\wt\QQ_N}\|\bar z\|_2^2 \E Q_{N}
\Eq(L.916)
$$
The appearance of $ \E_{\wt\QQ_N}\|\bar z\|_2^2$ is disturbing, as it
introduces a new quantity into the system. Fortunately, it is the last one.
The point is that proceeding as above, we can show that
$$
\eqalign{
\E \E_{\wt\QQ_{N+1}}\|z\|_2^2=&\a+\E \Biggl[ \tanh\b\left(\xi_{N+1}^1
\E_{\wt\QQ_N} z_1 +
\sqrt{\E T_N} X_N\right) \cr
&\times\left[\xi_{N+1}^1\E_{\wt\QQ_N} z_1 +\sqrt{\E T_N} X_N\right]\Biggr]
+\b\E \E_{\wt\QQ_N}\|\bar z\|_2^2 \E Q_{N}
}
\Eq(L.917)
$$
so that setting $U_{N}\equiv \E_{\wt\QQ_N}\|\bar z\|_2^2$, we get, subtracting 
\eqv(L.915) from \eqv(L.917), the simple 
recursion
$$
\E U_{N+1} =\a(1-\E Q_N)+\b(1-\E Q_N)\E U_N
\Eq(L.918)
$$
From this we get (since all quantities considered are self-averaging, we drop
the $\E $ to simplify the notation), setting  $M_N\equiv \E_{\wt\QQ_N} z_1$,
$$
\eqalign{
 T_{N+1} &=- ( M_{N+1})^2 +\a Q_N  +\b U_N Q_N
\cr
&+\int d\NN(g)[M_N +\sqrt {T_N}g]
\tanh\b(M_N+  \sqrt {T_N}g)
\cr
&= M_{N+1}(M_N-M_{N+1}) +\b U_N Q_N + \b T_N (1-Q_N)+\a Q_N
}
\Eq(L.919)
$$
where we used integration by parts.
The complete system of recursion relations can thus be written as
$$
\eqalign{
 M_{N+1} &=\int d\NN(g) \tanh \b\left(  M_N +\sqrt{ T_N} g\right)\cr
 T_{N+1} & = M_{N-1}(M_N-M_{N+1}) +\b U_N Q_N + \b T_N (1-Q_N)+\a Q_N\cr
 U_{N+1} &=\a(1- Q_N)+\b(1- Q_N) U_N\cr
Q_{N+1} &= \int  d\NN(g) \tanh^2 \b\left(  M_N +\sqrt{ T_N} g\right)\cr
}
\Eq(L.920)
$$
We leave it to the reader to check that the fixed points of this system 
lead to the equations \eqv(L.RS1)-\eqv(L.RS3) with 
$r=\lim_{N\uparrow \infty} T_N/\a$, $q=\lim_{N\uparrow \infty} Q_N$ and
$m_1=\lim_{N\uparrow \infty}M_N$ (where the variable 
$u=\lim_{N\uparrow \infty} U_N$ is eliminated). 

We have dropped both the $o(1)$ errors and the fact that the parameters $\b$ 
and $\a$ are slightly changed on the left by terms of order $1/N$. 
The point is that, as explained in [T4], these things are irrelevant. The point
is that from the localization results of the induced measures we know a priori
that for all $N$, if $\a$ and $\b$ are in the appropriate domain, 
the four quantities are in a well defined domain. Thus, if this 
domain is attracted by the ``pure'' recursion \eqv(L.920), then  we may choose
some function $f(N)$ tending (slowly) to infinity (e.g. $f(N)=\ln N$) 
would be a good choice) and iterate $f(N)$ times; letting $N$ tend to infinity
then gives the desired convergence to the fixed point.

The necessary 
 stability analysis, which is finally an elementary analytical
problem can be found in  
 [T4], Lemma 7.9 where it was apparently carried out for the first time
in rigorous form (a numerical investigation can of course be found in 
[AGS]). 
 It shows that all is well if $\a\b$ and $\g$ are small 
enough. \endproof\medskip

It is a particularly satisfying feature of the proof of Theorem \ver.5 that 
in the process we have obtained via Corollary \ver.7 and Lemma \ver.13
control over the limiting probability distribution of the ``mean field'',
$(\xi_i,m)$, 
felt by an individual spin $\s_i$. In particular, the facts we have gathered 
also prove Lemma \ver.8.  
Indeed, since $u(\t)$ is the logarithm of the 
Laplace transform of that field we can identify it with a gaussian of variance 
$\E\E_{\wt\QQ_N}\|\bar z\|_2^2$ and mean 
$\E_{\wt\QQ_N} z_1+\sqrt{\a r}g_i$, where 
$g_i$ is itself a standard gaussian. Moreover, esssentially the same analysis
allows to control not only the distribution of a single field $(\xi,m)$,
but of any {\it finite } collection,  $(\xi_i,m)_{i\in V}$, of them.  
Form this we are able to 
reconstruct {\it the probability distribution of the Gibbs measures}:  
 
\theo {\ver.15} {\it Under the conditions of Theorem \ver.5,  
for any finite set $V\subset \N$, 
the corresponding marginal distributions of the Gibbs measures
$
\mu_{\b,N,M(N)}^{(1,1)}(\s_i=s_i, \forall i\in V)$ converge in  law to 
$$
\prod_{i\in V} \frac {e^{\b s_i( \hat\mu\xi_i^1+\sqrt{\a r} g_i)}}
{2\cosh(\b ( \hat\mu\xi_i^1+\sqrt {\a r} g_i)}
$$
where $g_i$, $ i\in V$ are independent standard gaussian random variables.
}

\remark {In the language of Newman [NS] the above theorem identifies the 
limiting Aizenman-Wehr metastate\note{It would be interesting to 
study also the ``empirical metastate'.'}  
for our system. Note that there seems 
to be no (reasonable) way to enforce almost sure convergence of Gibbs states
for $\a>0$.  In fact, the $g_i$ are continuous unbounded random variables,
and by chosing suitable random subsequences $N_i$, we can construct 
{\it any} desired product measure as limiting measure!! Thus in the sense of
the definition of limiting Gibbs states in Section II, we must conclude that
for positive $\a$, all product measures are extremal measures for our system,
a statement that may seem surprising and that misses most of the interesting
information contained in Theorem \ver.12. Thus we stress that this provides 
an example where the only way to express the full available information
on  the asymptotics of the Gibbs measures is in terms of their probability
distribution, i.e. through  metastates. Note that in our case, the 
metatstate is concentrated on product mesures which can be seen as a statement
on ``propagation of chaos'' [Sn]. Beyond the ``replica symmetric regime''
this should no longer be true, and the metastate should then live on 
mixtures of product measures.
}

\proof{  We will give a brief sketch of the proof of Theorem
\ver.15. More details are given in [BG6]. 
It is a simple matter to show that 
$$
\eqalign{
& \mu_{\b,N,M}^{(1,1)}(\s_i=s_i, \forall i\in V)\cr
&=\sfrac{\int_{B_{\rho}(m^* e^1)} d^M z e^{- \b N\left[\frac {\|z\|_2^2}2
-\frac 1{\b N}\sum_{i\not\in V}\ln \cosh(\b(\xi_i,z))\right]}
e^{\b\sum_{i\in V}s_i(\xi_i,z)}}
{\int_{B_{\rho}(m^* e^1)} d^M z e^{-\b N\left[\frac {\|z\|_2^2}2
-\frac 1{\b N}\sum_{i\not\in V}\ln \cosh(\b(\xi_i,z))\right]}
\prod_{i\in V}2\cosh(\b(\xi_i,z))}
}
\Eq(L.48)
$$
Note that there is, for $V$ fixed and $N$ tending to infinity, virtually no 
difference between the function $\Phi_{\b,N,M}$ and
$\frac {\|z\|_2^2}2
-\frac 1{\b N}\sum_{i\not\in V}\ln \cosh(\b(\xi_i,z))$
so we will simply pretend they are the same. So we may write in fact
$$
 \mu_{\b,N,M}^{(1,1)}(\s_i=s_i, \forall i\in V)=\frac {\E_{\wt\QQ_{N-|V|}} 
e^{\b\sum_{i\in V}s_i(\xi_i,z)}}{\sum_{\s_V}\E_{\wt\QQ_{N-|V|}} 
e^{\b\sum_{i\in V}\s_i
(\xi_i,z)}}
\Eq(L.49)
$$
Now we proceed as in Lemma \ver.6. 
$$
\E_{\wt\QQ} 
e^{\b\sum_{i\in V}s_i(\xi_i,z)}=e^{\b\sum_{i\in V}s_i(\xi_i,\E_{\wt\QQ} z)}
\E_{\wt\QQ} e^{\b\sum_{i\in V}s_i(\xi_i,\bar z)}
\Eq(L.50)
$$
The second factor is controlled just as in Lemma \ver.6, and up to 
terms that converge to zero in probability is independent of $s_V$. It will 
thus drop out in the ratio in \eqv(L.49). The exponent in the first term is
treated as in Lemma \ver.8; since all the $\xi_i$, $i\in V$ are
independent, we obtain that the $(\xi_i,\E_{\wt\QQ} \hat z)$ converge indeed
to independent gaussian random variables. We omit the details of the
proof of the analogue of Lemma \ver.9; but note that $(\xi_i,\E_{\wt\QQ} \hat z)$
are uncorrelated, and this is enough to get independence in the limit (since 
uncorrelated gaussians are independent). From here the proof of Theorem \ver.15
is obvious.\endproof}

We stress that we have proven that the Gibbs measures converge weakly 
in law (w.r.t. to $\P$) to some random product measure on the spins. Moreover 
it should be noted that the probabilities of local events  (i.e. the 
expressions considered in Theorem \ver.15) in the limit 
 are not measurable with respect 
to a local sigma-algebra, since they involve the gaussians $g_i$. 
These are, as we have seen, obtained in a most complicated way from 
the entire set 
of the $\E_{\wt\QQ} z_\mu$, which depend of course 
on all the $\xi_i$. It is just 
fortunate that the covariance structure of the family of gaussians $g_i$, 
$i\in V$, is actually deterministic. This means in particular that if we take
 a fixed configuration of the $\xi$ and pass to the limit, we cannot expect
to converge. 

Fianlly let us point out that to get propagation of chaos not all what was 
needed to prove Theorem \ver.8 is really necessary. The main fact we used in 
the proof is the self-averaging of the quantity 
$\E_{\wt\QQ} e^{\b\sum_{i\in V}s_i(\xi_i,\bar z)}$, i.e. essentially (ii) of 
Lemma \ver.6, while (i) is not needed. The second property is that
$(\xi_i, \E_{\wt\QQ} z)$ converges in law, while it is irrelevant 
what the limit would 
be (these random variables might well  be dependent). Unfortunately(?), to 
prove (ii) of Lemma \ver.6 requires more or less the same hypotheses as 
everything else (i.e. we need Theorem \ver.1!), so this observation makes 
little difference. Thus ist may be that propagation of chaos and the 
exactness of the replica symmetric solution always go together (as the 
results in [PST]
imply). 

While in our view the results presented here shed some light on the 
``mystery of the replica trick'', we are still far from understanding 
the really interesting phenomenon of ``replica symmetry breaking''. 
This remains a challenge for the decade to come.

\newpage 
\frenchspacing
\chap{References}4
\item{[A]} D.J. Amit, ``Modelling brain function'', 
Cambridge University Press, Cambridge  (1989).
\item{[AGS]} D.J. Amit, H. Gutfreund and H.
Sompolinsky, ``Statistical mechanics of neural networks near saturation'',
Ann. Phys. {\bf 173}, 30-67 (1987).
\item{[AGS2]}  D.J. Amit, H. Gutfreund and H.
Sompolinsky, ``Spin-glass model of neural network'', Phys. Rev. {\bf A 32},
1007-1018  (1985).
\item{[B]} A. Bovier, ``Self-averaging in a class of generalized Hopfield 
models, J. Phys. {\bf A 27}, 7069-7077 (1994).
\item{[BG1]} A. Bovier and V. Gayrard, ``Rigorous bounds on the storage
capacity for the dilute Hopfield  model'',
J. Stat.Phys. {\bf 69}, 597-627 (1992)
\item{[BG2]} A. Bovier and V. Gayrard, ``Rigorous results on the
thermodynamics of the dilute Hopfield model'', J. Stat. Phys. {\bf 72},
643-664 (1993).
\item{[BG3]} A. Bovier and V. Gayrard, ``Rigorous results on the
 Hopfield model of neural networks'', Resenhas do IME-USP {\bf 2}, 161-172
 (1994).
\item{[BG4]} A. Bovier and V. Gayrard, ``An almost sure large deviation 
principle for the Hopfield model'', Ann. Probab {\bf 24}, 1444-1475  (1996).
\item{[BG5]} A. Bovier and V. Gayrard, ``The retrieval phase of the 
Hopfield model, A rigorous analysis of the overlap distribution'',
Prob. Theor. Rel. Fields {\bf 107}, 61-98 (1995).
\item{[BG6]} A. Bovier and V. Gayrard, in preparation.
\item{[BGP1]} A. Bovier, V. Gayrard, and P. Picco, ``Gibbs states
of the Hopfield model in the regime of perfect  memory'',
Prob. Theor. Rel. Fields {\bf 100}, 329-363 (1994).
\item{[BGP2]} A. Bovier, V. Gayrard, and P. Picco,
``Large deviation principles for the Hopfield model and the
Kac-Hopfield model'', Prob. Theor. Rel. Fields {\bf 101}, 511-546 (1995).
\item {[BGP3]} A. Bovier, V. Gayrard, and P. Picco,
``Gibbs states of the Hopfield model with extensively many patterns'',
 J. Stat. Phys. {\bf 79}, 395-414 (1995).
\item{[BGP4]}  A. Bovier, V. Gayrard, and P. Picco,
``Distribution of overlap profiles in the one-dimensional Kac-Hopfield model'',
WIAS-preprint 221, to appear in Commun. Math. Phys. (1997).
\item{[BF]} A. Bovier and J. Fr\"ohlich, ``A heuristic theory of the
spin glass phase'', J. Stat. Phys. {\bf 44}, 347-391 (1986).
\item{[BL]} H.J. Brascamp and E.H. Lieb, ``On extensions of the 
Brunn-Minkowski and P\'ekopa-Leindler theorems, 
including inequalities for log concave functions, and with an application to 
the diffusion equation'', J. Funct. Anal. {\bf 22}, 366-389 (1976).
\item{[Co]} F. Comets, ``Large deviation estimates for a conditional 
probability distribution. Applications to random Gibbs measures'',
Probab. Theor. Rel. Fields {\bf 80}, 407-432 (1989).   
\item{[CT]} Y.S. Chow and H. Teicher, ``Probability theory'', 2nd edition, 
Springer, Berlin-Heidelberg-New York  (1988)
\item{[DHS]} E. Domany, J.L. van Hemmen, and K. Schulten (eds.),
``Models of neural networks'', Springer Verlag, Berlin  (1991).
\item{[DS]} J.-D. Deuschel and D. Stroock, ``Large deviations'',
Academic Press, Boston  (1989).
\item{[DZ]} A. Dembo and O. Zeitouni, Large deviation techniques and
applications, Jones and Bartlett, Boston (1992).
 \item{[El]} R.S. Ellis, ``Entropy, large deviations, and statistical
 mechanics'',  Springer-Verlag, Berlin (1985).
\item {[EA]} S.F. Edwards and P.W. Anderson, ``Theory of spin glasses'', 
J. Phys. {\bf F 5}, 965-974 (1975).
\item{[EE]} T. Eisele and R.S. Ellis, ``Multiple phase transitions in the 
generalized Curie-Weiss model'', J. Stat. Phys. {\bf 52}, 161-202 (1988).
\item{[vE]} A.C.D. van Enter, ``Stiffness exponent, number of pure states,
and Almeida-Thouless line in spin glasses'', J. Stat. Phys. {\bf 60}, 275-279
 (1990).
\item{[vEvHP]} A.C.D. van Enter, J.L. van Hemmen and C. Pospiech, 
``Mean-field theory of random-
site q-state Potts models'', J. Phys. {\bf A 21}, 791-801 (1988).
\item{[FH]}  D.S. Fisher and D.A. Huse, ``Pure phases in spin glasses'', 
J. Phys. {\bf A 20}, L997-L1003 (1987); ``Absence of many states in magnetic 
spin glasses'', J. Phys.  {\bf A 20}, L1005-L1010 (1987).
\item{[FP1]} L.A. Pastur and A.L. Figotin, ``Exactly soluble model
of a spin glass'', Sov. J. Low Temp. Phys. {\bf 3(6)}, 378-383
(1977).
\item{[FP2]} L.A. Pastur and A.L. Figotin, ``On the theory of
disordered spin systems'', Theor. Math. Phys. {\bf 35}, 403-414
(1978).
\item{[FP2]} L.A. Pastur and A.L. Figotin, ``Infinite range limit for a class
 of disordered spin systems'', Theor. Math. Phys. {\bf 51}, 564-569 (1982).
\item{[FZ]} C. Fassnacht and A. Zippelius, ``Recognition and Categorization in 
a structured neural network with attractor dynamics'',
Network {\bf 2}, 63-84 (199?).
\item{[G1]} V. Gayrard, ``The thermodynamic limit of the $q$-state 
Potts-Hopfield model with infinitely many patterns'', 
J. Stat. Phys. {\bf 68}, 977-1011 (1992).
\item{[G2]} V. Gayrard, ``Duality formulas and the transfer principle'',
in preparation. 
\item{[Ge]} S. Geman, ``A limit theorem for the norms of random matrices'',
Ann. Probab. {\bf 8}, 252-261 (1980).
\item{[Geo]} H.-O. Georgii, ``Gibbs measures and phase transitions'',
Walter de Gruyter (de Gruyter Studies in Mathematics, Vol. 19),
Berlin-New York  (1988).
\item{[Gi]}  V.L. Girko,  ``Limit theorems for maximal and minimal 
eigenvalues of
random matrices'', Theor. Prob. Appl. {\bf 35}, 680-695 (1989). 
\item{[GK]} D. Grensing and K. K\"uhn, ``On classical spin-glass
models'', J. Physique {\bf 48}, 713-721 (1987).
\item{[GM]} E. Golez and S. Mart\'\i nez, ``Neural and automata networks'',
Kluwer Academic Publ., Dodrecht  (1990)
\item{[HKP]} J. Hertz, A. Krogh, and R. Palmer, ``Introduction to the 
theory 
of neural computation'', Addison-Wesley, Redwood City  (1991).
\item{[Ho]} J.J. Hopfield, ``Neural networks and physical systems
with emergent collective computational abilities'', Proc. Natl.
Acad. Sci. USA {\bf 79}, 2554-2558 (1982).
\item{[HS]} R.L. Stratonovich, ``On a method of calculating quantum
distribution functions'', Doklady Akad. Nauk S.S.S.R. {\bf 115},  1097
(1957)[translation: Soviet Phys. Doklady {\bf 2}, 416-419 (1958)],
\item{} J. Hubbard, ``Calculation of partition functions'',
Phys. Rev. Lett. {\bf 3}, 77-78 (1959).
\item{[JK]} J. Jedrzejewski and A. Komoda,  ``On equivalent-neighbour, 
random-site models of
disordered systems'', Z. Phys. {\bf B 63}, 247-257 (1986).
\item{[vH1]} J.L. van Hemmen, ``Equilibrium theory of spin-glasses:
mean-field theory and beyond'', 
in ``Heidelberg colloquium on spin glasses'', Eds. J.L. van Hemmen and 
I.Morgenstern, 203-233 (1983),  LNP 192 Springer, Berlin-Heidelberg-New York 
(1983)
\item{[vH2]} J.L. van Hemmen,  ``Spin glass models of a neural network'',
Phys. Rev. A {\bf 34}, 3435-3445 (1986).
\item{[vHGHK]} J.L. van Hemmen, D. Grensing, A. Huber and R. K\"uhn, 
``Elementary solution of classical spin-glass models, Z. Phys. {\bf B 65}, 
53-63 (1986).
\item{[vHvE]} J.L. van Hemmen and A.C.D.van Enter, 
``Chopper model for pattern recognition'',
Phys.Rev. {\bf A 34}, 2509-2512,(1986).
\item{[vHvEC]} J.L. van Hemmen, A.C.D. van Enter, and J. Canisius. 
``On a classical spin-glass
model'', Z. Phys. {\bf B 50}, 311-336 (1983).
\item{[K]} H. Koch, ``A free energy bound for the Hopfield
model'', J. Phys.   {\bf A 26}, L353-L355 (1993).
\item{[KP]}  H. Koch and J. Piasko, ``Some rigorous results on the Hopfield
neural network model'', J. Stat. Phys. {\bf 55},  903-928 (1989).
\item{[Ku]} Ch. K\"ulske, private communiction.
 \item{[LT]} M. Ledoux and M. Talagrand, ``Probability in Banach spaces'',
 Springer, Berlin-Heidelberg-New York  (1991).
\item{[Lu]} D. Loukianova, ``Two rigorous bounds in the Hopfield model 
of associative memory'', to appear in Probab. Theor. Rel. Fields (1996).
\item{[Lut]} J.M. Luttinger,  ``Exactly Soluble Spin-Glass Model'',
 Phys.Rev. Lett. {\bf 37}, 778-782 (1976).
\item{[Mar]} S. Martinez, ``Introduction to neural networks'', 
preprint, Temuco, (1992).
\item{[Ma]} D.C. Mattis, ``Solvable spin system with random interactions'', 
Phys. Lett.
{\bf 56A}, 421-422 (1976).
\item{[McE]} R.J. McEliece, E.C. Posner, E.R. Rodemich and S.S. Venkatesh,
``The capacity of the Hopfield associative memory'', IEEE Trans. Inform. 
Theory {\bf 33}, 461-482 (1987).
\item{[Mi]} Y. Miyashita, ``Neuronal correlate of visual associative long term 
memory in the primate temporal cortex'', Nature {\bf 335}, 817-819 (1988).
\item{[MPR]} E. Marinari, G. Parisi, and F. Ritort, ``On the 3D Ising 
spin glass'', J. Phys.  {\bf A 27}, 2687-2708 (1994).
\item{[MPV]} M. M\'ezard, G. Parisi, and M.A. Virasoro, 
``Spin-glass theory
and beyond'', { World Scientific}, Singapore (1988).
\item{[MR]} B. M\"uller and J. Reinhardt, ``Neural networks: an introduction'',
Springer Verlag, Berlin  (1990).
\item{[MS]} V.A. Malyshev and F.M. Spieksma, ``Dynamics of binary neural 
networks with a finite number of patterns''. Part 1: General picture of the 
asynchronous zero temperature dynamics'', MPEJ {\bf 3}, 1-36 (1997).
\item{[N]}  Ch.M. Newman, ``Memory capacity in neural network models:
Rigorous results'', Neural Networks {\bf 1}, 223-238 (1988).  
\item {[NS]} Ch.M. Newman and D.L. Stein, ``Non-mean-field behaviour in
realistic spin glasses'', Phys. Rev. Lett. {\bf 76}, 515-518 (1996);
``Spatial inhomogeneity and thermodynamic chaos'', Phys. Rev. Lett. {\bf 76},
4821-4824 (1996); ``Topics in disordered systems'', to appear in Birkh\"auser,
Boston (1997); ``Thermodynamic chaos and the structure of short range
spion glasses'', this volume.
 \item{[P]} D. Petritis, ``Thermodynamic formalism of neural 
 computing'', preprint Universit\'e de Rennes (1995)
\item{[PS]} L. Pastur and M. Shcherbina, ``Absence of self-averaging
of the order parameter in the Sher\-ring\-ton-Kirkpatrick model'', 
J. Stat. Phys. {\bf 62}, 1-19 (1991).
 \item{[PST]} L. Pastur, M. Shcherbina, and B. Tirozzi, ``The replica
 symmetric solution without the replica trick for the Hopfield model'',
 J. Stat. Phys. {\bf 74}, 1161-1183 (1994).
\item{[PN]} P. Peretto and J.J. Niez, ``Long term memory storage capacity
of multiconnected neural networks'', Biological Cybernetics {\bf 39},
53-63 (1986).
\item{[Ro]} R.T. Rockafellar, ``Convex analysis'', Princeton University Press,
Princeton  (1970).
\item{[RV]} A.W. Roberts and D.E. Varberg, ``Convex functions'', 
Academic Press, New York and London  (1973).
\item{[SK]} D. Sherrington and S. Kirkpatrick, ``Solvable model of a
spin glass'', { Phys. Rev. Lett.}
{\bf 35}, 1792-1796 (1972).
\item{[Si]} J. Silverstein, ``Eigenvalues and eigenvectors of large 
sample covariance matrices'', Contemporary Mathematics {\bf 50}, 
153-159 (1986).
\item{[Sn]} A.-S. Snitzman, ``Equations de type Boltzmann spatialement 
homog\`enes'', Z. Wahrscheinlichkeitstheorie verw. Gebiete {\bf 66}, 559-592
(1986).
\item{[ST]} M. Shcherbina and B. Tirozzi, ``The free energy for a class
of Hopfield models'', J. Stat. Phys. {\bf 72}, 113-125 (1992).
\item{[SW]} M. Schl\"uter and E. Wagner, Phys. Rev. {\bf E49}, 1690
(1994).
 \item{[Yu]} V.V. Yurinskii, ``Exponential inequalities for sums of
 random vectors'', J. Multivariate Anal. {\bf 6}, 473-499 (1976). 
\item{[T1]} M. Talagrand, ``Concentration of measure and isoperimetric
inequalities in product space'', Publ. Math. I.H.E.S., {\bf 81}, 73-205 
(1995). 
\item{[T2]} M. Talagrand, ``A new look at independence'', 
Ann. Probab. {\bf 24}, 1-34 (1996).
 \item{[T3]} M. Talagrand, ``R\'esultats rigoureux pour le mod\`ele de 
 Hopfield'', C. R. Acad. Sci. Paris, {\bf t. 321, S\'erie I}, 109-112 (1995).
\item{[T4]} M. Talagrand, ``Rigorous results for the Hopfield model with many
 patterns'', preprint 1996, to appear in Probab. Theor. Rel. Fields.
\item{[T5]} M. Talagrand, ``The Sherrington-Kirkpatrick model: A challenge for 
mathematicians'', preprint 1996, to appear in Prob. Theor. Rel. Fields.
\item{[TAP]} D.J. Thouless, P.W. Anderson, and R.G. Palmer, Phil. Mag. {\bf
35},593 (1977).
\item{[YBK]} Y.Q. Yin, Z.D. Bai, and P.R. Krishnaiah, ``On the limitof the 
largest eigenvalue of the large dimensional sample covariance matrix'',
Probab. Theor. Rel. Fields {\bf 78}, 509-521 (1988).
 \item{[Yu]} V.V. Yurinskii, ``Exponential inequalities for sums of
 random vectors'', J. Multivariate Anal. {\bf 6}, 473-499 (1976).

\end